\documentclass{article}

\usepackage{arxiv}

\usepackage[english]{babel}

\usepackage{amsmath,amsthm,amssymb}
\usepackage{mathrsfs}
\usepackage{graphicx}
\usepackage[colorlinks=true, allcolors=blue]{hyperref}
\newtheorem{definition}{Definition}
\newtheorem{theorem}{Theorem}

\usepackage[linesnumbered,ruled]{algorithm2e}
\usepackage{diagbox}
\usepackage{multirow}
\usepackage{subfigure}
\usepackage{bbm}
\usepackage{amsmath}
\usepackage{makecell}
\usepackage{enumitem}

\newcommand{\etal}{\textit{et al.}\xspace}
\newcommand{\eg}{\textit{e.g.}\xspace}
\newcommand{\ie}{\textit{i.e.}\xspace}
\newcommand{\etc}{\textit{etc.}\xspace}

\begin{document}
%
\title{Sub-optimal Learning in Meta-Classifier Attacks: A Study of Membership Inference on Differentially Private Location Aggregates}


\author{
\hspace{6em}{Yuhan Liu}\\
	\hspace{6em}Renmin University of China\\
	\hspace{6em}\texttt{liuyh2019@ruc.edu.cn} \\ \\
    \hspace{6em}{\textbf{Igor Shilov}} \\
	\hspace{6em}Imperial College London\\
	\hspace{6em}\texttt{i.shilov23@imperial.ac.uk} \\
	\And
    {Florent Guepin} \\
	Imperial College London\\
	\texttt{florent.guepin@imperial.ac.uk} \\\\
    {\textbf{Yves-Alexandre De Montjoye}} \\
	Imperial College London\\
	\texttt{demontjoye@imperial.ac.uk} \\
 	\And
}



%


\maketitle

\begin{abstract}
The widespread collection and sharing of location data, even in aggregated form, raises major privacy concerns.
Previous studies used meta-classifier-based membership inference attacks~(MIAs) with multi-layer perceptrons~(MLPs) to estimate privacy risks in location data, including when protected by differential privacy (DP).
In this work, however, we show that a significant gap exists between the expected attack accuracy given by DP and the empirical attack accuracy even with informed attackers (also known as DP attackers), indicating a potential underestimation of the privacy risk.
To explore the potential causes for the observed gap,
we first propose two new metric-based MIAs: the one-threshold attack and the two-threshold attack. We evaluate their performances on real-world location data and find that different data distributions require different attack strategies for optimal performance: the one-threshold attack is more effective with Gaussian DP noise, while the two-threshold attack performs better with Laplace DP noise.
Comparing their performance with one of the MLP-based attack models in previous works shows that the MLP only learns the one-threshold rule, leading to a suboptimal performance under the Laplace DP noise and an underestimation of the privacy risk.
Second, we theoretically prove that MLPs can encode complex rules~(\eg, the two-threshold attack rule), which can be learned when given a substantial amount of training data. We conclude by discussing the implications of our findings in practice, including broader applications extending beyond location aggregates to any differentially private datasets containing multiple observations per individual and 
how techniques such as synthetic data generation and pre-training might enable MLP to learn more complex optimal rules.

\end{abstract}


%

\section{Introduction}
\begin{figure}[tbp!]
    \centering
    \includegraphics[width=0.4\linewidth]{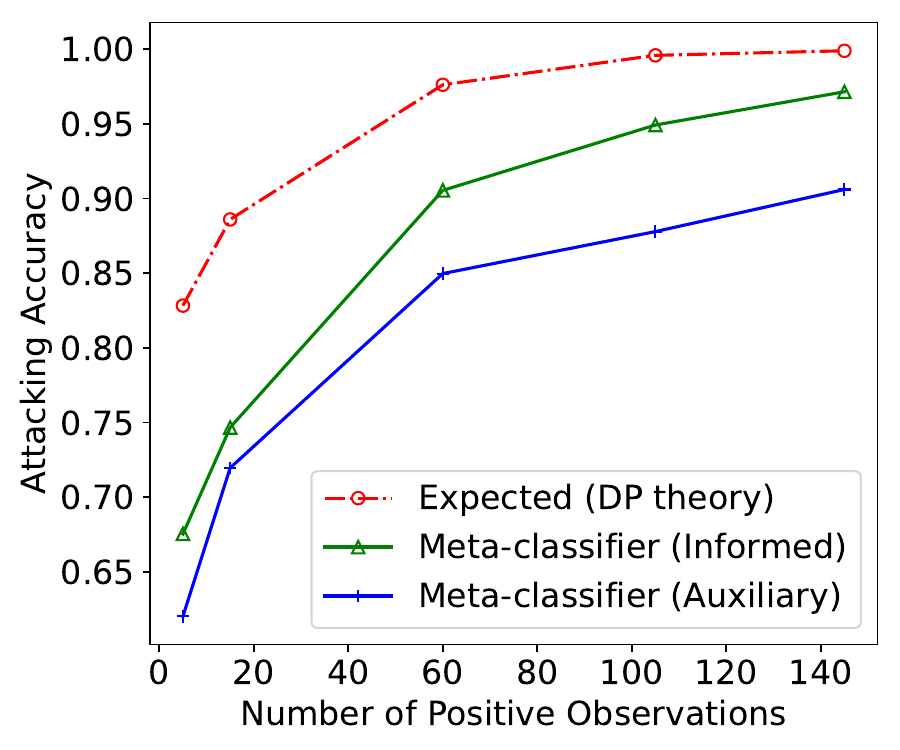}
    \caption{Comparison between the expected accuracy and attack accuracy of the typical meta-classifier-based attack~(MLP with one hidden layer and 100 nodes) with the informed attacker. Each observation is perturbed with the Laplace mechanism $\mathrm{Lap}(\frac{1}{0.5})$. To compute the expected attack accuracy, we first compute the expected false positive rate~($\alpha$) and false negative rate~($\beta$) of MIA given a DP mechanism under multiple observations as in \cite{kairouz2015composition}. Then, the expected attack accuracy is derived as $ACC=\frac{1-\alpha}{2}+\frac{1-\beta}{2}$.}
    \label{fig: gap}
\end{figure}
Location data, a form of time-series data, consists of individual traces represented as binary vectors that denote the presence of individuals at specific sites over a period. This data is crucial across a wide range of fields, from disease transmission monitoring~\cite{riley2007large,lin2022spatial} to urban planning~\cite{rathore2016urban}, and is extensively utilized by various organizations, including enterprises and governments. Despite its growing applications, the privacy concerns of the individuals involved are also escalating significantly. 

To mitigate the privacy risks, previous studies apply differential privacy~(DP) to location aggregates, providing a formal privacy guarantee for involved individuals~\cite{xiao2015protecting}. Meanwhile, membership inference attack~(MIA), a widely used auditing tool, is exploited before data releases to evaluate the lower bound of the empirical privacy risks~\cite{guan2024zero}.
In particular, meta-classifier-based MIAs are adopted in previous works, where a multi-layer perceptron~(MLP) is trained on a auxiliary location dataset as the attack model to distinguish members from non-members.

However, a significant gap is seen between the expected attack accuracy given by the DP theory and the empirical results given by the MLP-based MIAs adopted in previous works. As shown in Figure~\ref{fig: gap}, the gap exists not only in the auxiliary settings where the attacker knows no traces in the target aggregate, but also in the informed settings where the attacker knows every individual trace in the target aggregate except for the presence of the target trace. Since the informed setting matches the threat model assumed by DP, this discrepancy points to a sub-optimality in the previous MLP-based methods, potentially leading to an underestimation of the true privacy risks.
This raises a key question—\textit{can we reduce the gap by enhancing the MLP-based attacks on location aggregates?}

To answer this question, we first investigate potential causes that lead to the sub-optimal performance of the MLP-based MIAs. 
At a high level, we analyze the potential attacking rules learned by the attack model by comparing the learned parameters with two metric-based MIA rules newly proposed in this work, namely the one-threshold attack and the two-threshold attack.
Instead of casting the MIAs into a binary classification problem, both of our proposed metric-based attacks formalize it as a hypothesis testing problem.
Specifically, to distinguish members from non-members, the one-threshold attack performs a likelihood ratio test~(LRT) on a score function that sums all observations in the released aggregate. In contrast, the two-threshold attack performs an LRT on the sum of individual LRT results for each observation.

Our comparison reveals two main findings.
Firstly, different underlying data distribution requires different attacking strategies for optimal performance. For instance, the one-threshold attack yields an optimal performance for released aggregates perturbed with Gaussian DP noise, whereas the two-threshold attack outperforms when the released aggregates are perturbed with Laplace DP noise.
Secondly, the MLP-based attack model trained in previous works learns only the one-threshold attack rule, resulting in a sub-optimality under the Laplace DP noise and, consequently, an underestimation of the privacy risk.

Second, we theoretically prove that the MLP-based attack model can encode complex rules, such as the two-threshold attack rule. Then, our experiments conducted on the real-world dataset further demonstrate that such complex and better rules can be learned with substantial training data.

We conclude by discussing the implications of our findings in practice, including how techniques such as synthetic data generation and pre-training might enable MLP to learn more complex optimal rules, as well as how our findings in this work can potentially extend beyond location aggregates to any differentially private datasets containing multiple observations per individual.




To summarize, our main contributions are as follows:
\begin{itemize}
    \item We compare the performance of MLP-based membership inference attacks (MIAs) from previous works with two newly proposed metric-based MIAs, demonstrating that the MLP-based approach learns only a simple rule, regardless of the data distribution, leading to sub-optimal attack performance.
    \item We provide a theoretical proof that MLP-based attack models can encode complex attack rules, which, when properly learned, yield better performance compared to the simpler rule.
    \item Through experiments on a real-world dataset, we show that MLP-based models can learn more complex attack rules when given sufficient training data, resulting in a significant improvement in attack accuracy.
\end{itemize}
\section{Backgrounds}
In Section~\ref{subsec: location_data}, we provide the formal definitions of location data, including user traces and location aggregates. In Section~\ref{subsubsec: DP}, we introduce the definition and corresponding tools of differential privacy~(DP). In particular, we show in Section~\ref{subsubsec: dp_location} how DP is applied to location aggregates to provide a formal privacy guarantee for each individual.
\subsection{Location Data}\label{subsec: location_data}
While the formal definitions of user traces and location aggregates are presented in Section~\ref{subsubsec: traces} and Section~\ref{subsubsec: location_aggregates}, respectively, we also emphasize in Section~\ref{subsubsec: location_aggregates} that the more positive observations released for an individual, the higher privacy risks they face.
\begin{figure}
    \centering
    \subfigure[User traces~($x$).]{
    \includegraphics[width=0.25\linewidth]{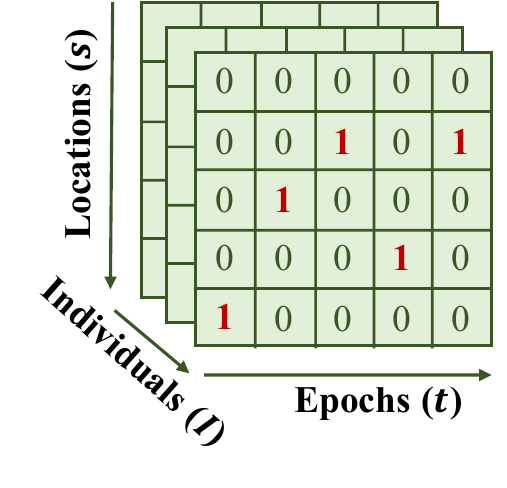}\label{subfig: traces}
    }
    \hspace{2cm}
    \subfigure[A location aggregate~($A$).]{
    \includegraphics[width=0.25\linewidth]{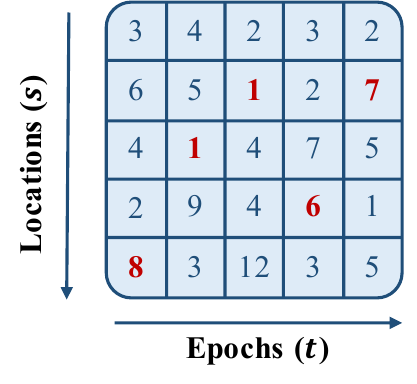}\label{subfig: aggregates}
    }
    \caption{An illustration of user traces and their aggregation. Figure~\ref{subfig: traces} demonstrates a set of individuals $I$, where $t\in E$ is the timestamps of interest~(also referred to as an Epoch) and $s\in S$ is the site of interest. Figure~\ref{subfig: aggregates} is an aggregate of user traces. If the shown trace in Figure~\ref{subfig: traces} is the target trace, the cells in \textcolor{red}{red} in $A$ are \textit{positive observations}.}
    \label{fig:location_data}
\end{figure}
\subsubsection{User Traces}\label{subsubsec: traces}
As shown in Figure~\ref{fig:location_data}, a location dataset $D=\{x^1,\ldots,x^n\}$ contains $n$ traces from distinct individuals.
Each trace $x^i$ for $i\in[n]$, following an underlying distribution $\pi$, is a binary matrix where each column represents a timestamp $e\in E$, and each row represents a distinct site $e\in E$. If an individual $I_i$ visits site $l$ at a timestamp $e$, it holds that $x^{i}_{le}=1$ for the corresponding cell in trace $x^{i}$. Otherwise, it holds that $x^{i}_{le}=0$. 


\subsubsection{Location Aggregates}\label{subsubsec: location_aggregates}
An aggregation matrix $A$ over $D$ is the sum of all traces in $D$ such that for each cell $A_{le}$ with $s\in S$ and $e\in E$, $A_{le}=\sum\limits_{x^i\in D}x^{i}_{le}$ holds true.
While a cell $A_{le}$ is referred to as an $\textit{observation}$,
the cell $A_{le}$ such that $z_{le}=1$ holds for the target trace $z$ is further referred to as a \textit{positive observation}.

\textbf{Remark.} Note that both traces and aggregates are time-series data, where data points from an individual are repeatedly collected and released at consistent intervals over time. Intuitively, each observation of an individual, especially the positive observations, can reveal the presence of the individual. Hence, the privacy risk for each individual increases along with the number of observations.

\subsection{Differential Privacy}\label{subsec: dp-background}
Differential privacy~(DP), first brought up by Dwork \etal~\cite{dwork2006differential,dwork2014algorithmic}, is a prevalent tool providing rigorous privacy guarantee for data releases, which is defined as follows:
\subsubsection{Definitions and Tools}\label{subsubsec: DP}
\begin{definition}[Differential Privacy~\cite{dwork2006differential}]\label{def:DP}
A randomized algorithm $\mathcal{M}:\mathbb{D}\rightarrow\mathbb{R}$ satisfies $(\varepsilon,\delta)$-differential privacy if for any two neighboring datasets $D$ and $D^{\prime}$, and any output $o\subseteq\mathbb{O}$
$$\Pr[\mathcal{M}(D)\in o]\leq e^{\varepsilon}\cdot\Pr[\mathcal{M}(D^{\prime})\in o]+\delta.$$
\end{definition}
When $\delta=0$, we say that $\mathcal{M}$ satisfies pure DP~(denoted by $\varepsilon$-DP). Otherwise, it guarantees an approximate DP.

The additive-noise mechanism is one of the most typical methods to achieve DP for real-valued data releases. Specifically, it corrupts and releases query results with additive noise randomly drawn from certain distributions~\cite{geng2019optimal,liu2024unleash}. That is, given a dataset $D$, to guarantee DP, a randomized algorithm $\mathcal{M}$ releases:
\begin{equation}\label{eq: dp_mechanism}
    \mathcal{M}(D) = q(D) + X,
\end{equation}
wherein $q(D)$ is the result of a query carried on $D$ and $X$ is the additive noise drawn from a probability distribution $\mathcal{N}$. The noise scale is calibrated to the sensitivity of $q(D)$ defined in the following:
\begin{definition}[$l_p$-sensitivity]
    For a real-valued query $q:\mathbb{D}\rightarrow \mathbb{R}$, the $l_p$-sensitivity of $q$ is defined as:
    $$\Delta_p=\max\limits_{D,D^{\prime}\in\mathbb{D}}\Vert q(D)-q(D^{\prime})\Vert_p,$$
    where $\Vert \cdot \Vert_{p}$ denotes the $l_p$ norm and $D, D^{\prime}$ is a pair of neighboring datasets that differ by one element.
\end{definition}

In particular, the \textit{Laplace} and \textit{Gaussian} mechanisms are two of the most widely adopted additive-noise mechanisms. The Laplace mechanism is achieved by drawing the random variable $X$ in Equation~\ref{eq: dp_mechanism} from Laplace distribution centered at $0$ denoted by $\mathrm{Lap}\left(\frac{\Delta_1} {\varepsilon}\right)$, where $\Delta_1$ is the $l_1$ sensitivity of the released data and $\frac{\Delta_1}{\varepsilon}$ is the noise scale denoted by $b$.
Similarly, the Gaussian mechanism is achieved by drawing $X$ from Gaussian distribution centered at $0$ denoted by $\mathrm{Gau}\left(\sigma\right)$, where $\sigma$ is the noise scale proportional to $\frac{\Delta_2}{\varepsilon}$.

DP provides a rigorous privacy guarantee mathematically and is straightforward to achieve. Additionally, it allows to combine multiple differentially private building blocks elegantly for designing more sophisticated algorithms, as described by the following theorem:
\begin{theorem}[Optimal Composition~\cite{kairouz2015composition}]\label{theo: composition}
    For any $\varepsilon\geq0$ and $\delta\in[0,1]$, the class of $\left(\varepsilon,\delta\right)$-differentially private mechanisms satisfies $\left(\left(k-2i\right)\varepsilon,1-\left(1-\delta\right)^{k}\left(1-\delta_{i}\right)\right)$-differential privacy under $k$-fold adaptive composition, for all $i=\left\{0,1,\ldots,\lfloor k/2\rfloor\right\}$, where
    \begin{equation}
        \delta_{i}=\frac{\sum\limits_{l=0}^{i-1}\binom{k}{l}\left(e^{(k-l)\varepsilon}-e^{(k-2i+l)\varepsilon}\right)}{\left(1+e^{\varepsilon}\right)^{k}}.
    \end{equation}
\end{theorem}

Theorem~\ref{theo: composition} indicates that the overall privacy budget $\varepsilon$ consumed by multiple differentially private mechanisms applied to one dataset increases with the accumulation of each privacy mechanism's privacy budget $\varepsilon_{i}$, which agrees with our intuition in Section~\ref{subsubsec: location_aggregates} that privacy risks of an individual increases along with the number of observations.

Aside from Theorem~\ref{theo: composition}, many other composition theorems are applied for different scenarios, such as the sequential composition theorem~\cite{dwork2006differential}, the advanced composition theorem~\cite{dwork2014algorithmic}, composition with R\'enyi differential privacy~(RDP)~\cite{mironov2017renyi}, and the composition with Gaussian differential privacy~(GDP)~\cite{dong2022gaussian}, \etc. As the application of different composition theorems does not affect the main takeaways of this work, we adopt theorem~\ref{theo: composition} for simplicity.

\subsubsection{Differential Privacy on Location Aggregate}\label{subsubsec: dp_location}
To protect individual privacy while releasing the aggregations over location data, DP is usually applied to each cell in matrix $A$.
Typically, three steps are incorporated. 

Firstly, to bound the amount of injected noise, each individual is made sure to contribute no more than $C$ ones at an epoch $e$. That is, $A_{el}$ is clipped to have a $\Delta_{p}$ sensitivity no larger than $C$ at each epoch.
While other clipping method can also be applied, for traces that have more than $C$ ones at epoch $e$ , we randomly flip $1$ to $0$ in vector $x^{i}_{[:,e]}$ so that $\sum\limits_{l\in L}x_{le}^{i}\leq C$ for each $i\in[n]$ and $t\in E$ for simplicity.

Secondly, additive random noise with scale in proportion to $\frac{\Delta_{p}}{\varepsilon}$ is added to each cell in the aggregation matrix $A$. The noisy aggregation, defined as $\tilde{A}=A+X$, where $X$ is the noise matrix with the same shape as $A$~(\ie, $\lvert
L\rvert$ rows and $\lvert E\rvert$ columns), is then publicized for downstream applications.

Finally, the overall privacy budget consumption $\varepsilon^{\prime}$ for $\lvert E\rvert$ epochs is computed according to Theorem~\ref{theo: composition}.

\subsection{Membership Inference Attack against Location Data}
Hereinafter, we first formally define the membership inference game in the context of location data in Section~\ref{subsubsec:mia_game}, where we emphasize that this work mainly focuses on studying the average privacy risk of one specific target over multiple location aggregates under the DP protection.

Then, we introduce evaluation metrics of MIA used in this work in Section~\ref{subsubsec: mia_metric}, including attack accuracy, AUC, and the ROC curve, especially under small false positive rates as suggested by Carlini \etal~\cite{carlini2022membership}.

Finally, we elaborate on the methodology of meta-classifier-based MIAs against location data adopted in previous works in Section~\ref{subsubsec: meta-classifer-attack}.
\subsubsection{Membership Inference Game}\label{subsubsec:mia_game}
Overall, given dataset $D=\{x^1,\ldots,x^{n}\}$, a membership inference attacker $\mathcal{A}$ aims to infer whether a specific target trace $z$ in $D$ was used by a released noisy aggregation matrix $\tilde{A}$.
\begin{figure}[tbp!]
    \centering
    \includegraphics[width=0.7\linewidth]{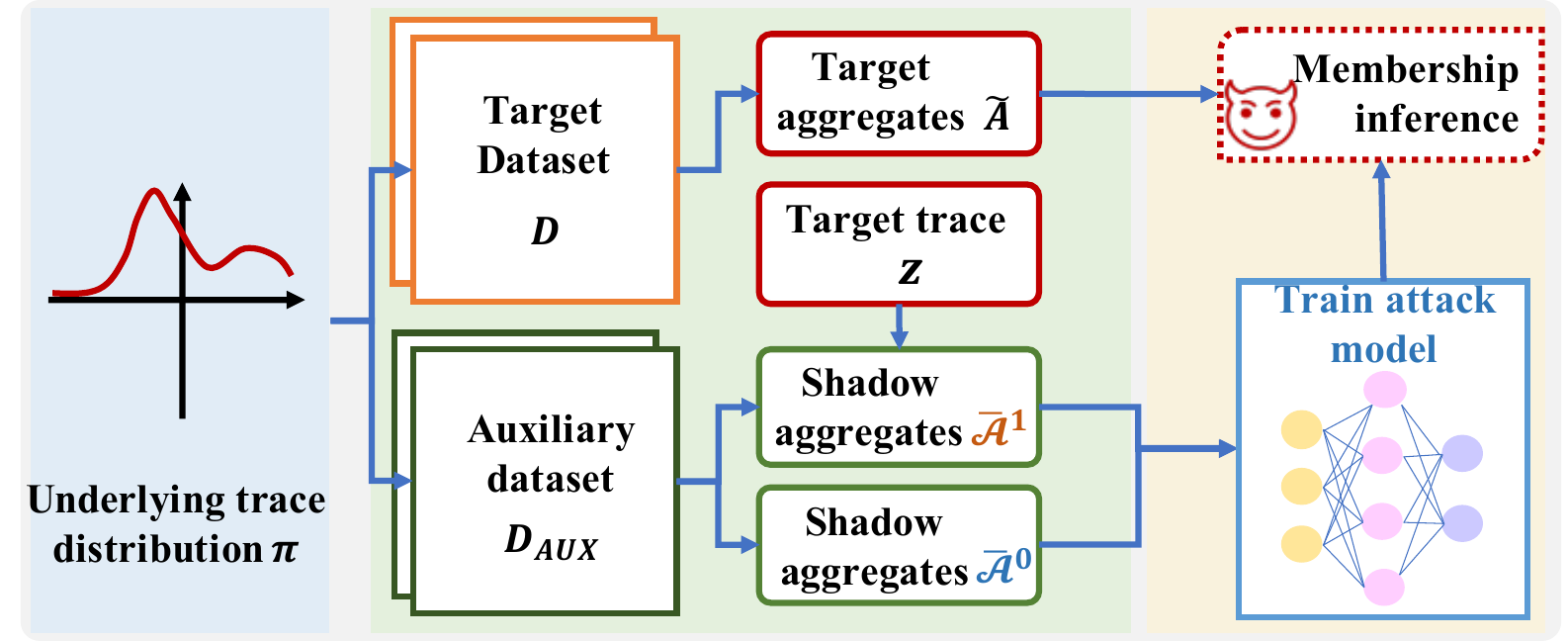}
    \caption{The overall pipeline of MIAs with meta-classifier: infer the membership of the target $z$ from a released aggregate $\tilde{A}$ derived from the target dataset $D$ by using an attack model trained on an auxiliary $D_{AUX}$ that follows the same underlying data distribution as $D$.}\label{fig: meta-attack}
\end{figure}
This is usually defined by an indistinguishability game between a challenger and an adversary. The game models random experiments related to the following two hypotheses:
\begin{equation*}
    \begin{aligned}
         &H_{IN}: \text{The target trace $z$ contributes to the released noisy aggregates $\tilde{A}$.}\\
         &H_{OUT}: \text{The target trace $z$ \textit{does not} contribute to the released noisy aggregates $\tilde{A}$.}
    \end{aligned}
\end{equation*}

\begin{definition}[Membership Inference Game~\cite{ye2022enhanced}]\label{def: mia}
    Let $\pi$ be the underlying distribution of traces, and $\tilde{A}$ is the released noisy aggregate.
    \begin{enumerate}
        \item The \textit{challenger} samples a set of traces $D\sim \pi$ via a fresh random seed $s_{D}$ and a trace $z\sim\pi$ with a fixed random seed $s_{z}$ such that $z\cap D=\emptyset$.
        \item The challenger flips a fair coin $b$. If $b=1$, it includes $z$ in $D$ and computes a noisy aggregation matrix $\tilde{A}$ over $D\cup\{z\}$. Otherwise, it computes a noisy aggregation matrix $\tilde{A}$ over $D$. Both $z$ and $\tilde{A}$ are sent to the adversary.
        \item The adversary, having access to an auxiliary dataset $D_{AUX}\sim \pi$ such that $D_{AUX}\cap D=\emptyset$ and a fraction $\theta$ of traces $D_{\theta}\subseteq {D}$ such that $D_{\theta}\cap \{z\}=\emptyset$, outputs a binary bit $\hat{b}\leftarrow\texttt{MIA}(\tilde{A},z)$.
        \item if $\hat{b}=b$, output $1$ to indicate a successful inference, otherwise, output $0$.
    \end{enumerate}
\end{definition}
There are two main takeaways from Theorem~\ref{def: mia}. For starters, we consider the average privacy risk of a specific target over multiple released aggregates by using a fresh random seed $s_{D}$ to select target aggregates and using a fixed random seed $s_{z}$ to select the target trace from the underlying data distribution.
Second of all, we consider attacks against noisy aggregates perturbed with DP noise.

In addition, we consider two types of adversary in this work, namely the \textit{informed attacker} and the \textit{auxiliary} attacker. An informed attacker knows the membership of all traces $x_i\neq z$ in $D$ as well as their values. That is to say $D_{\theta}=D$ in Definition~\ref{def: mia}. An auxiliary attacker, on the other hand, has no access to any data points in $D$.

\subsubsection{Evaluation Metric}\label{subsubsec: mia_metric}
For each target, the goal of an attacker is to infer its membership as accurately as possible.
Given a hypothesis testing experiment, two types of errors are involved, namely the type \uppercase\expandafter{\romannumeral1} error~($\alpha$) and the type \uppercase\expandafter{\romannumeral2} error~($\beta$).
While the type \uppercase\expandafter{\romannumeral1} error is the rejection of the null hypothesis~($H_{IN}$) when it is true, the type \uppercase\expandafter{\romannumeral2} error is the acceptance of the null hypothesis when it is false. Note that while $\alpha$ is equivalent to the false negative rate, $\beta$ is the false positive rate.
Assume that the attacker has no prior knowledge of the target's presence, we consider three MIA evaluation metrics, including accuracy, AUC, and the corresponding ROC~\cite{carlini2022membership}. While the definitions of AUC and ROC are straightforward under hypothesis testing, we present the definition of attack accuracy as follows:
\begin{equation}\label{eq:acc_prob}
    p = \frac{1-\alpha}{2}+\frac{1-\beta}{2},
\end{equation}
where $p$ is the probability of an attacker successfully performing the MIA on a balanced dataset where half of the aggregates are members. Particularly,
a higher probability $p$ suggests higher MIA accuracy, hence better performance.

\subsubsection{Meta-classifier-based MIAs}\label{subsubsec: meta-classifer-attack}
Previous works perform MIA against location data by casting it to a binary classification problem, where an attacking meta-model is trained on an auxiliary dataset to infer the membership of a target.
Specifically, As shown in Figure~\ref{fig: meta-attack}, given a set of auxiliary traces $D_{AUX}=\{\bar{x}^{1},\ldots,\bar{x}^{n_{AUX}}\}$ that follows the same underlying distribution $\pi$ as $D$, an attacker first creates a set of shadow aggregates $\mathcal{A}_{sd}=\{\bar{A}_{1},\ldots,\bar{A}_m\}$ by randomly sampling $n<n_{AUX}$ traces from $D_{AUX}$, aggregating sampled traces, and perturbing all aggregates with DP. Ideally, the shadow aggregates follow the same distribution as the released target aggregate $\tilde{A}$.

Then, the target trace $z$ is added to half of the created aggregates, which are then labeled with $1$, and the rest of the aggregates are further labeled with $0$.
By labeling half of the aggregates as members and the other half as non-members, we assume that the attacker has no prior knowledge of the membership of the target. That is to say, before seeing the released noisy aggregates, the best an attacker can do is to guess the presence of the target with probability $\frac{1}{2}$.

After that, the labeled aggregates and their corresponding labels are used to train an attacking model $\mathcal{C}$. The released aggregate $\tilde{A}$ are then fed into the attacking model $\mathcal{C}$ to infer the presence of the target $z$.

Most of the previous works train an MLP-based attack model with $100$ nodes in one hidden layer using $400$ shadow aggregates to infer targets with around $168$ observations~\cite{pyrgelis2018knock, guan2024zero}. As we aforementioned in Figure~\ref{fig: gap}, by implementing the previous MLP-based attacks, a large gap is observed between the theoretical attack accuracy given by DP theory and the MIA performance under the auxiliary attacker as well as the informed attacker, who shares a same threat model as DP. This further motivates us to explore the potential causes for the sub-optimality of MLP-based MIAs against location data and look for possible solutions.

\section{Metric-based Attack}
As we mentioned in Section~\ref{subsubsec: meta-classifer-attack}, the meta-classifier-based MIAs against location data cast the attack into a binary classification problem and distinguish members from non-members by training an attack model. The distinguishing rule of the attack model is automatically learned. Yet, it is hard for us to explain the learned rule explicitly. 

To better understand how the adversary knowledge is applied by the attacker to effectively launch MIAs against location aggregates, we propose two metric-based MIAs from the perspective of hypothesis testing as introduced in Section~\ref{subsubsec:mia_game}, which provides a better interpretability. By comparing the proposed metric-based rule and the distinguishing rule learned automatically by the attack model, we are able to explain the sub-optimality of the meta-classifier-based attack in previous settings.

In this section, we first provide an overview of our proposed metric-based methods, namely the one-threshold and two-threshold attack with two score functions, respectively, in Section~\ref{subsec: overview}, where we also summarize and compare the two proposed methods in the end to show more insights.

Then we elaborate on both \textbf{a)} the one-threshold attack, where the score function is the sum of all positive observations, and \textbf{b)} the two-threshold attack, where the score of the released aggregate is the sum of thresholding results of each positive observation, in Section~\ref{subsec: one-threshold} and Section~\ref{subsec: two-threshold}.

\subsection{Overview}\label{subsec: overview}
\begin{figure}[tbp!]
    \centering
    \includegraphics[width=0.7\linewidth]{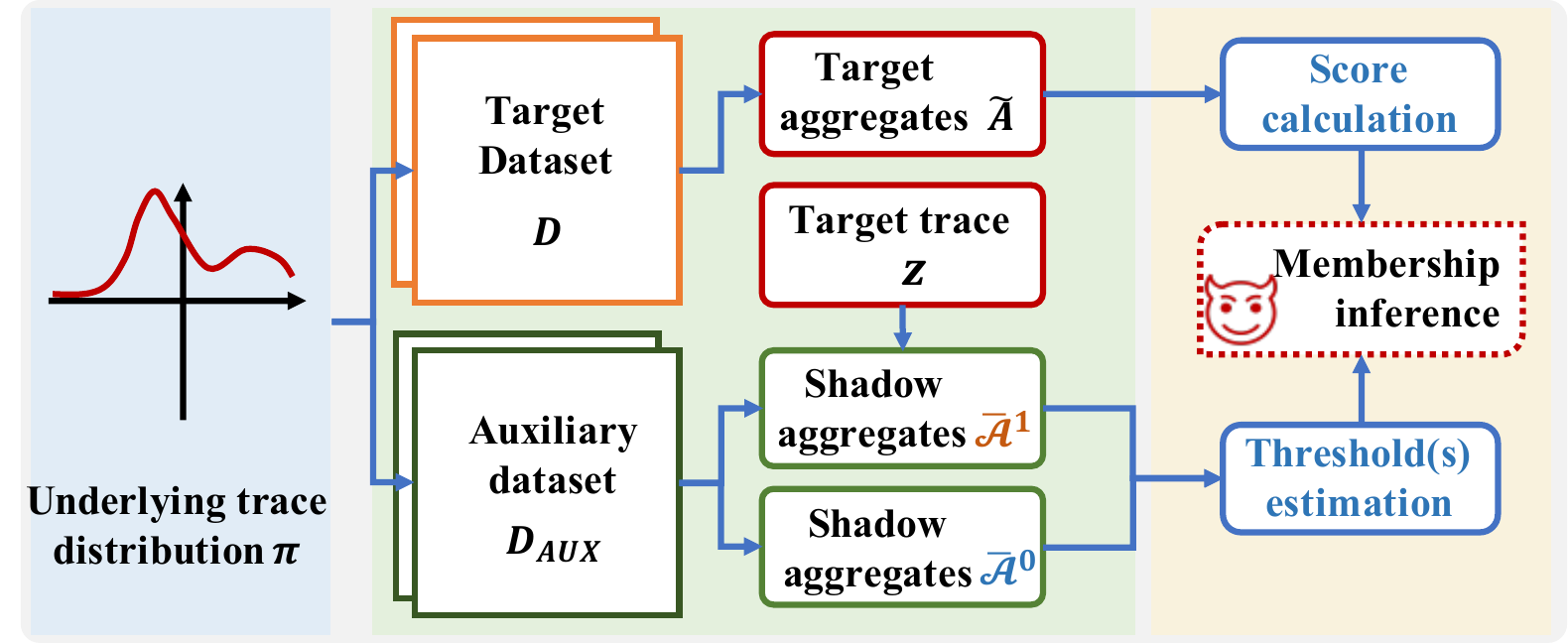}
    \caption{The overall pipeline of the metric-based MIAs: infer the membership of the target $z$ from a released aggregate $\tilde{A}$ derived from a target dataset $D$ by defining score functions and estimating thresholds using the auxiliary dataset $D_{AUX}$ that follows the same underlying data distribution as $D$. The membership of $z$ is then decided by comparing the score of the released aggregate and the estimated thresholds.}\label{fig: likelihood-ratio-attack}
\end{figure}

\begin{algorithm}
    \caption{Metric-based MIA against location aggregates.}\label{alg: overall}
    \LinesNumbered
    \SetKwInOut{Input}{Input}
    \SetKwInOut{Output}{Output}
    \Input{An auxiliary dataset $D_{AUX}$, the target aggregate $\tilde{A}$, the target trace $z$, number of non-target traces in the target aggregate $n$, an integer $m$, the background knowledge $A_{\theta}$, the score function indictor $I_{s}$, and privacy parameters $(\varepsilon,\delta)$.}
    \Output{A membership indicator $\hat{b}\in\{0,1\}$.}
    $\mathcal{A}_{sd}=\phi$\;
    \For{$i=1,2,\ldots,m$}{
    $\bar{A}^{i}=\texttt{SUM}(\texttt{Shuffle}(D_{AUX})[:n\left(1-\theta\right)])$\label{line: overall-shadowAgg}\tcp*{Generate shadow aggregates}

    \uIf{$i\leq\lfloor m/2\rfloor$\label{line: overall-label}}{
    $\bar{A}^{i}\leftarrow\bar{A}^{i}+z$\label{line: overall-label-end}\tcp*{Attach the target $z$ to half of the shadow aggregates}
    }
    $\bar{A}^{i}\leftarrow \texttt{Perturb}(\bar{A}^{i},\varepsilon,\delta)$\label{line: overall-perturb}\tcp*{Perturb with DP noise}
    $\mathcal{A}_{sd}\leftarrow\mathcal{A}_{sd}\cup\bar{A}^{i}$\;
    }
    \uIf{$I_{s}=1$}{
    $s\leftarrow\texttt{Score1}(\tilde{A},A_{\theta},z)$\;
    }
    \uElseIf{$I_{s}=2$}{$s\leftarrow\texttt{Score2}(\tilde{A},A_{\theta},z,\mathcal{A}_{sd})$
    \tcp*{Calculate the score of target aggregate}}
    $T\leftarrow\texttt{ThresholdEstimation}(\mathcal{A}_{sd},z)$\;
    \eIf{$s\leq T$}{$\hat{b}\leftarrow 0$\tcp*{Non-member}}
    {$\hat{b}\leftarrow 1$\tcp*{Member}}
    \Return{$\hat{b}$\;}
\end{algorithm}
\subsubsection{Overall Methodology}
The overall attack pipeline is presented in Figure~\ref{fig: likelihood-ratio-attack} and the corresponding algorithm is presented in Algorithm~\ref{alg: overall}, wherein four steps are involved.

\textbf{Step~1}: $m$ shadow aggregates are generated based on the auxiliary dataset $D_{AUX}$ and the attacker's background knowledge $A_{\theta}$. 

Specifically, suppose there are $n$ traces contributing to the released target aggregate $\tilde{A}$, the adversary then uniformly sample $n\left(1-\theta\right)$ traces without replacement from $D_{AUX}$ and sum them up at each cell to derive a shadow aggregate $\bar{A}^{i}$ for $i\in[m]$, where $\theta$ is the fraction of traces in $D$ that is known to the attacker, as shown in Line~\ref{line: overall-shadowAgg}.

Then, half of the generated shadow aggregates are attached with the target trace $z$ and labeled as members, while the other half are directly labeled as non-members, as shown from Line~\ref{line: overall-label} to Line~\ref{line: overall-label-end}. 

Finally, to mimic the underlying data distribution of $\tilde{A}$, the same differentially private mechanism as applied to $\tilde{A}$ is applied to all the generated shadow aggregates in Line~\ref{line: overall-perturb}.

\textbf{Step~2}: A score function that takes the target aggregate $\tilde{A}$ and the target trace $z$ as inputs is exploited to assign a score to $\tilde{A}$.

In particular, as shown by the following equations (informal), two score functions are defined in this work for the one-threshold attack and the two-threshold attack, respectively:
\begin{align}
    & \texttt{S}_{1}\left(\tilde{A},z\right)=\sum\limits_{l\in L}\sum\limits_{e\in E}\mathbbm{1}\left(z_{le}=1\right)\cdot \tilde{A}_{le},\\
    & \texttt{S}_{2}\left(\tilde{A},z\right)=\sum\limits_{l\in L}\sum\limits_{e\in E}\mathbbm{1}\left(z_{le}=1\right)\cdot\left(\mathbbm{1}\left(\tilde{A}_{le}\geq T_{le}
    \right)
    \right),
\end{align}
which we will further elaborate in Algorithm~\ref{alg: one-threshold-score} in Section~\ref{subsec: one-threshold} and in Algorithm~\ref{alg: two-threshold-score} in Section~\ref{subsec: two-threshold}.

At a high level, the score function of our one-threshold attack is defined as the sum of all positive-observation cells in $\tilde{A}$. 
On the other hand, the score function of our two-threshold attack is defined as the sum of sub-scores assigned to each positive observation, wherein the sub-scores are calculated based on individual comparison results between the value of each positive observation $\tilde{A}_{le}$ and an estimated threshold $T_{le}$.

\textbf{Step~3}: An overall threshold $T$ is estimated based on $\mathcal{A}_{sd}$ and the target trace $z$, which is also detailed in Algorithm~\ref{alg: one-threshold-threshold} in Section~\ref{subsec: one-threshold} and Algorithm~\ref{alg: one-threshold-threshold-fix-error} in Appendix~\ref{subsec: threshold-fixed-error}.

In particular, this work estimates $T$ by assuming a symmetric underlying data distribution and maximizing the attack accuracy for simplicity, which is shown by the following equation (informal):
\begin{align}
    T=\frac{1}{2}\left(
    \mathop{\texttt{MEAN}}\limits_{\bar{A}\in \mathcal{A}_{sd}^{0}}\left(\texttt{S}\left(\bar{A},z\right)\right)+
    \mathop{\texttt{MEAN}}\limits_{\bar{A}\in \mathcal{A}_{sd}^{1}}\left(\texttt{S}\left(\bar{A},z\right)\right)
    \right).
\end{align}

In more generic cases, $T$ is estimated based on the distribution of scores calculated from member shadow aggregates $\mathcal{A}_{sd}^{0}$ and its non-member counterpart $\mathcal{A}_{sd}^{1}$. For example, by first computing the empirical score histogram and then applying smooth techniques~\cite{ye2022enhanced}, one can estimate the cumulative distribution function of the score distributions for both members and non-members. After that, the threshold is computed either at a given level of attacking error or to maximize attacking accuracy.

\textbf{Step~4}:
The membership of the target $z$ is then inferred based on the comparison result between the threshold $T$ and the score $s$. Specifically, if $s$ surpasses $T$, $z$ is labeled as a member. Otherwise, it is labeled as a non-member.

\subsubsection{Comparison}\label{subsubsec: comparison}
\begin{figure}[tbp!]
    \centering
    \includegraphics[width=0.7\linewidth]{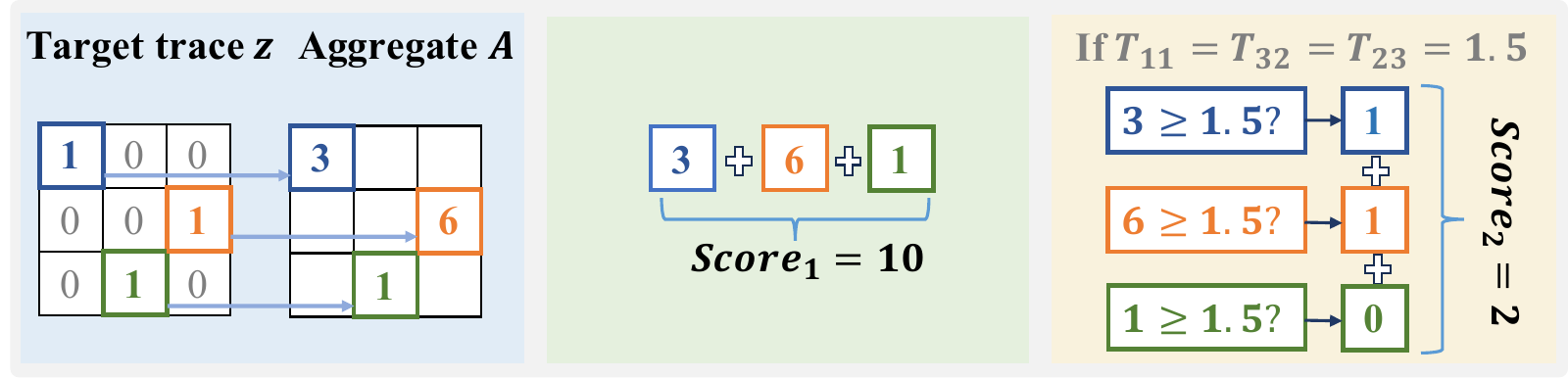}
    \caption{An illustration of two designed score functions for the one-threshold attack~(\ie, $\texttt{Score}_1$) and the two-threshold attack~(\ie, $\texttt{Score}_2$), respectively, given a target trace $z$ and an aggregate $A$. In particular, we suppose the sub-thresholds for each positive observation are $1.5$ without loss of generality.}\label{fig: score_function}
\end{figure}
The major difference between the one-threshold and the two-threshold attack lies in the design of the score function, which we summarize and compare in Figure~\ref{fig: score_function} for clarity.

\begin{figure}[tbp!]
    \centering
    \subfigure[Laplace mechanism.]{
    \includegraphics[width=0.4\linewidth]{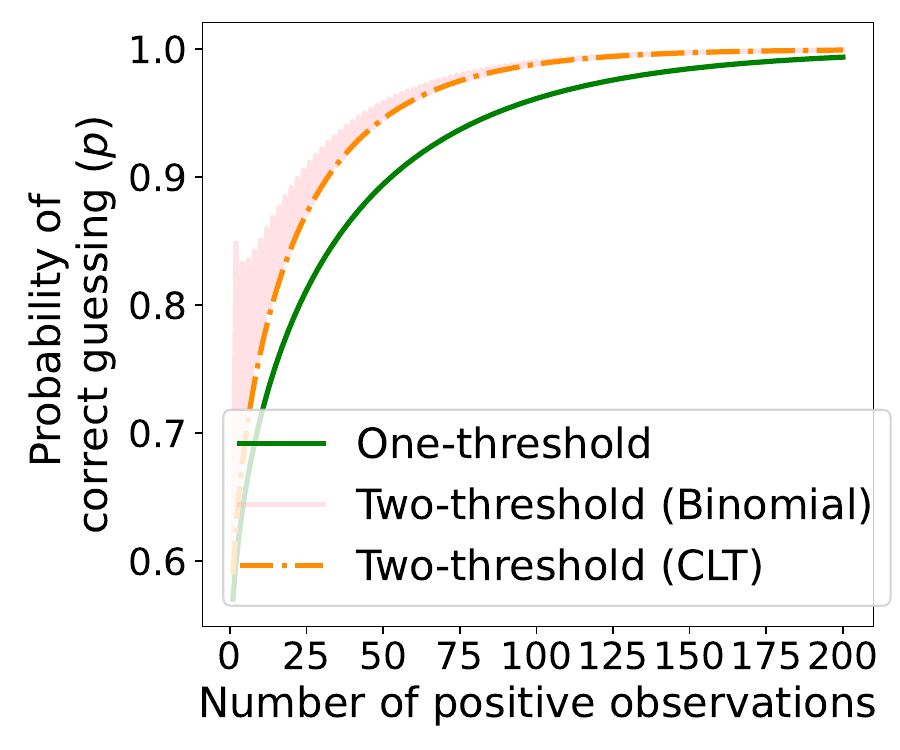}\label{subfig: compare_acc_lap}
    }
    \hspace{0.5cm}
    \subfigure[Gaussian mechanism.]{
    \includegraphics[width=0.4\linewidth]{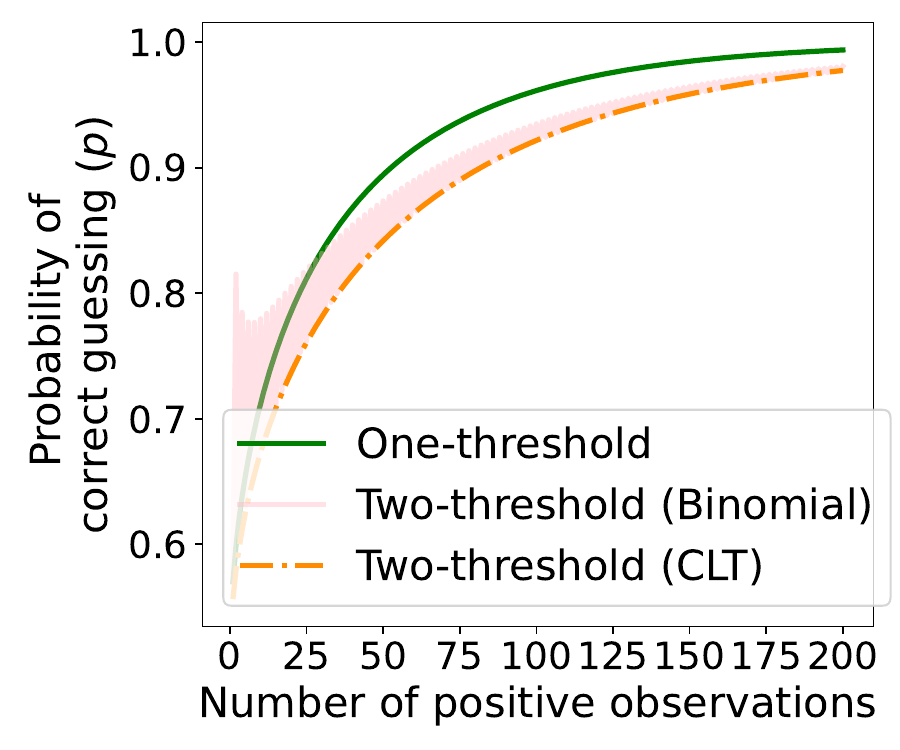}\label{subfig: compare_acc_gau}
    }
    \caption{Illustration of the attack accuracy~(\ie, probability of correct guessing $p$) with a varying number of positive observations for informed attacker~(\ie, DP attacker). While \textit{Two-threshold~(Binomial)} stands for the precise result, \textit{Two-threshold~(CLT)} is the result after approximating with the central limit theorem~(CLT), which is further explained in Section~\ref{subsec: two-threshold}.}
    \label{fig: compare-attack-acc}
\end{figure}

One of the main findings of this work is that the one-threshold attack and the two-threshold attack perform differently under different underlying data distributions. For instance, for the informed attacker, the one-threshold attack outperforms the two-threshold attack under Gaussian DP noise, whereas the other way around is under Laplace DP noise. We compare these two metric-based attacks under the informed attacker with different DP noise in terms of attack accuracy in Figure~\ref{fig: compare-attack-acc} for more insights. 

The intuition behind it is that, as the Laplace distribution has a relatively heavy tail, which accumulates after multiple hypothesis tests, hence the hypothesis testing error. Conducting a likelihood ratio test individually for each positive observation helps control the error accumulation, leading to a better performance.

\subsection{One-threshold Attack}\label{subsec: one-threshold}
In this section, we first define the \textit{score function} in Algorithm~\ref{alg: one-threshold-score}. Then, we analyze the \textit{score distribution} of both member aggregates and non-member aggregates in Theorem~\ref{theo: one-threshold-hypothesis} and Figure~\ref{fig: distribution-socre1} before we introduce our \textit{threshold estimation} method in Algorithm~\ref{alg: one-threshold-threshold}.

\begin{algorithm}
    \caption{\texttt{Score1}. The score function for the one-threshold attack.}\label{alg: one-threshold-score}
    \LinesNumbered
    \SetKwInOut{Input}{Input}
    \SetKwInOut{Output}{Output}
    \Input{The target aggregate $\tilde{A}$, the attacker's background knowledge $A_{\theta}$, and the target trace $z$.}
    \Output{Score of the target aggregate $s$.}
    $\tilde{A}^{r}\leftarrow\tilde{A}-A_{\theta}$\label{line: score-one-remove}\tcp*{Remove background knowledge from the target aggregate}
    $s\leftarrow \sum\limits_{l\in L}\sum\limits_{t\in T}\left(\mathbbm{1}\left(z_{le}=1\right)\cdot A^{r}_{le}\right)$\label{line: score-one-sum}\tcp*{Sum up all positive-observation cells}
    \Return{$s$}
\end{algorithm}

\textbf{Score function.}
As shown in Line~\ref{line: score-one-remove} in Algorithm~\ref{alg: one-threshold-score}, the aggregate of known traces to the attacker $A_{\theta}$ is first taken out from the target aggregate $\tilde{A}$ to reduce the uncertainty. Then, the score of $\tilde{A}$ is calculated by summing up all positive observations in the remaining aggregate $\tilde{A}^{r}$ as in Line~\ref{line: score-one-sum}.

The intuition behind Algorithm~\ref{alg: one-threshold-score} is straightforward. While the cell $\tilde{A}_{st}$ such that $z_{st}=0$ does not reveal much about the membership of $z$ in $\tilde{A}$, each of the cells $\tilde{A}_{st}$ such that $z_{st}=1$ leaks a bit about $z$'s presence. By summing up all cells that provide evidence, the difference between members and non-members grows larger, increasing the inference confidence level.

\textbf{Score distributions.}
As introduced in Section~\ref{subsubsec:mia_game}, Definition~\ref{def: mia} is considered as a hypothesis testing problem between the null hypothesis $H_{IN}$ and the alternative hypothesis $H_{OUT}$, which is further equivalent to distinguish between the score distribution over member aggregates $\mathbbm{Q}$ and that over non-member aggregates $\mathbbm{P}$. Specifically, $\mathbbm{Q}$ and $\mathbbm{P}$ are defined as follows:
\begin{equation}\label{eq: p_q}
    \begin{aligned}
    &\mathbbm{Q}=\left\{s\leftarrow\texttt{S}\left(\tilde{A},A_{\theta},z\right)\vert \tilde{A}\left(D\cup z,X\right), D\sim \pi, X\sim\mathcal{N}\right\},\\
    &\mathbbm{P}=\left\{s\leftarrow\texttt{S}\left(\tilde{A},A_{\theta},z\right)\vert \tilde{A}\left(D\backslash z,X\right), D\sim \pi, X\sim\mathcal{N}\right\},
    \end{aligned}
\end{equation}
where $D$ is a set of traces following an underlying distribution $\pi$ and contributing to the target aggregate $\tilde{A}$, $z$ is the target trace, $X$ is random DP noise drawn from distribution $\mathcal{N}$, and $\texttt{S}\left(\cdot\right)$ is the score function.

Specifically, the analytical forms of $\mathbbm{Q}$ and $\mathbbm{P}$ with Algorithm~\ref{alg: one-threshold-score} is given by the following theorem:
\begin{theorem}\label{theo: one-threshold-hypothesis}
    For any attacker that has access to a subset traces $D_{\theta}$ of the target trace dataset $D$ such that $D_{\theta}\subseteq D\backslash z$,
    to infer the membership of a target trace $z$ based on a released aggregate $\tilde{A}$ with Algorithm~\ref{alg: overall} and Algorithm~\ref{alg: one-threshold-score} is equivalent to distinguish between the following two hypotheses:
    \begin{equation*}
        \begin{aligned}
            &H_{IN}: s \text{ follows the distribution }\mathbbm{Q}:= N\left(\mu_{1},\sigma\right),\\
            &H_{OUT}: s \text{ follows the distribution }\mathbbm{P}:=N\left(\mu_{0},\sigma\right),
        \end{aligned}
    \end{equation*}
where $\mu_{b}=\sum\limits_{l\in L}\sum\limits_{e\in E}\mathbbm{1}\left(z_{le}=1\right)\cdot\left(\mu_{le}+bz_{le}\right)$ for $b\in[0,1]$,
$\sigma=\sum\limits_{l\in L}\sum\limits_{e\in E}\mathbbm{1}\left(z_{le}=1\right)\cdot\left(\sigma_{le}^{r}+\sigma_{X}\right)$, $\sigma_{X}$ is the standard deviation of the injected DP noise, and $\mu_{le}$, $\sigma_{le}^{r}$ are the mean and the standard deviation of the clean location aggregation at cell $A_{le}^{r}$ after subtracting the attacker background knowledge $A_{\theta}$, respectively.
\end{theorem}

Theorem~\ref{theo: one-threshold-hypothesis} is derived by first looking into the per-cell distribution of $A^{r}$, then applying the central limit theorem to the sum of all positive observations.

Since given a clean aggregate $A$, $\tilde{A}^{r}=A+X-A_{\theta}$ holds, we have that for any non-member aggregates at $A_{le}^{r,0}$ such that $l\in L$ and $e\in E$, it holds that
\begin{align}\label{eq: distribution-percell}
    \tilde{A}_{le}^{r,0}\sim \pi_{le}^{n\left(1-\theta\right)}\otimes\mathcal{N}_{le},
\end{align}
where $\pi_{le}^{n\left(1-\theta\right)}$ is the distribution of the sum of $n\left(1-\theta\right)$ randomly drawn from $\pi_{le}$, $\mathcal{N}_{le}$ is the noise distribution, and $\otimes$ denotes for convolution of two distributions.

Similarly, for any member aggregates at $A_{lt}^{r,1}$ such that $l\in L$ and $t\in T$, we have that
\begin{align}\label{eq: distribution-percell-mem}
    A_{lt}^{r,1}\sim z_{lt}+\pi_{lt}^{n\left(1-\theta\right)}\otimes\mathcal{N}_{lt}.
\end{align}

Without loss of generality, we assume the additive DP noise centers at $0$. Then, it is derived that the mean of $A_{le}^{r,0}$, denoted as $\mu_{le}^{0}$, is equivalent to the mean of of distribution $\pi_{le}^{n(1-\theta)}$, while the mean of $A_{le}^{r,1}$, denoted as $\mu_{le}^{1}$, is equivalent to $z_{le}+\mu_{le}^{0}$. In addition, the standard deviation of $A_{le}^{r,0}$ and $A_{le}^{r,1}$ is denoted by $\sigma_{le}^{r}$, such that $\sigma=\sigma_{le}+\sigma_{X}$, where $\sigma_{le}$ and $\sigma_{X}$ are the standard deviation of distribution $\pi_{le}^{n\left(1-\theta\right)}$ and the DP noise, respectively.
Then, by applying the central limit theorem, the score $s$ approximately follows Gaussian distribution, as stated in Theorem~\ref{theo: one-threshold-hypothesis}.

\textbf{Examples.}
\begin{figure}[tbp!]
    \centering
    \subfigure[Informed attacker.]{
    \includegraphics[width=0.4\linewidth]{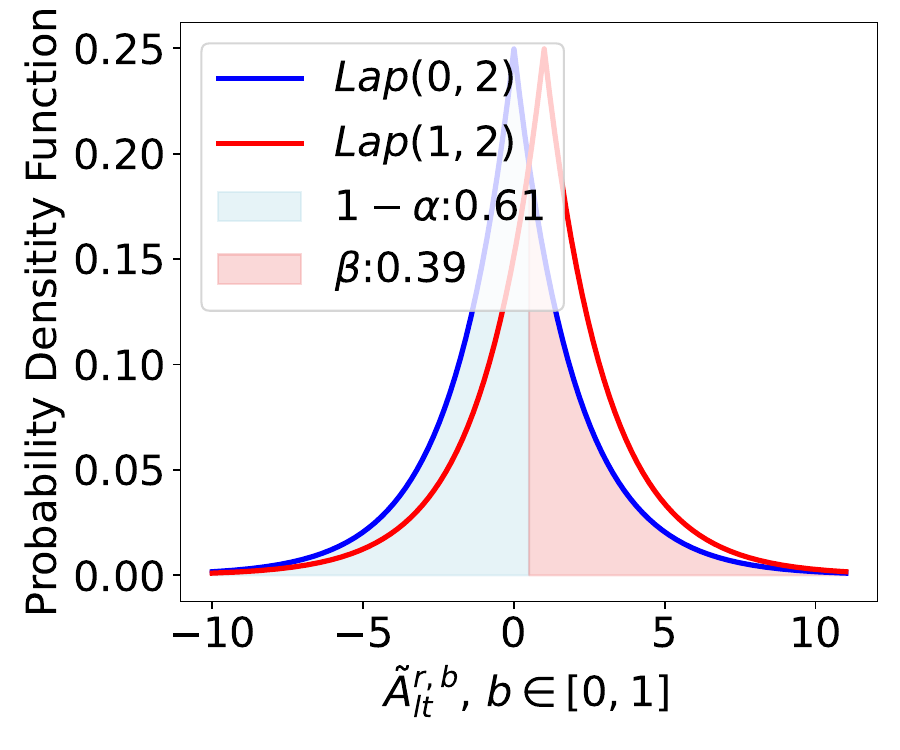}\label{subfig: per-cell-full-distribution}
    }
    \hspace{0.5cm}
    \subfigure[Auxiliary attacker.]{
    \includegraphics[width=0.4\linewidth]{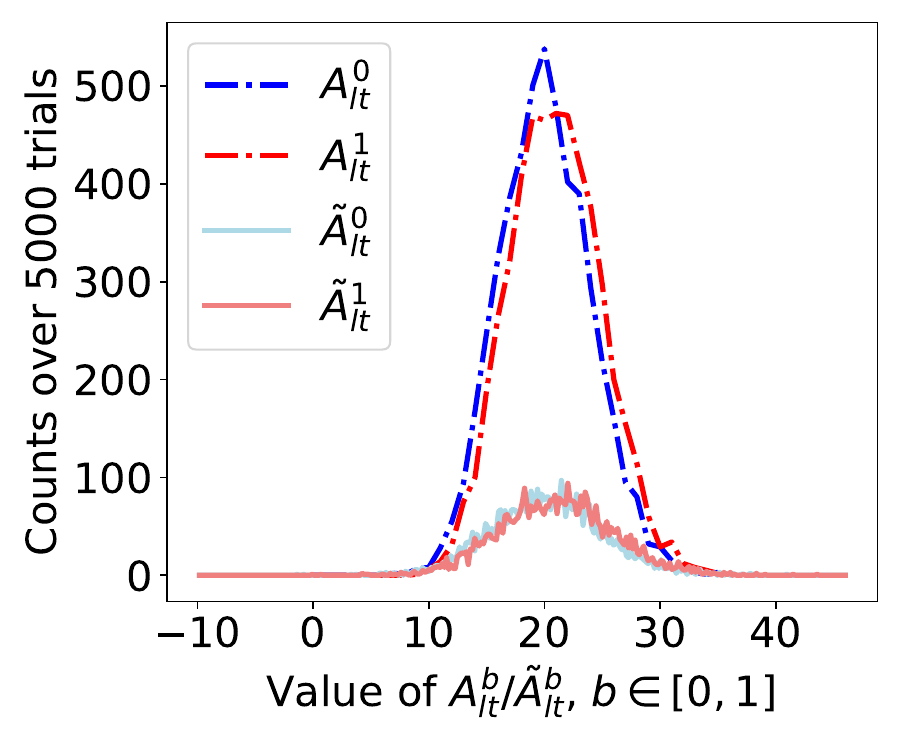}\label{subfig: per-cell-partial-distribution}
    }
    \caption{An example of data distribution at cell $\tilde{A}_{le}^{r}$ for both informed attacker and auxiliary attacker. In particular, for the auxiliary attacker in Figure~\ref{subfig: per-cell-partial-distribution}, we assume that the $2,000$ traces contributing to $\tilde{A}^{r,0}_{le}$ are uniformly selected over $10,000$ traces, $1\%$ of which have one at cell $A_{le}$. Additionally, noise randomly drawn from $\mathrm{Lap}\left(\frac{1}{0.5}\right)$ is injected to $A_{le}$.}
    \label{fig: one-threshold-distribution-percell}
\end{figure}
\begin{figure}[tbp!]
    \centering
    \subfigure[Informed attacker.]{
    \includegraphics[width=0.4\linewidth]{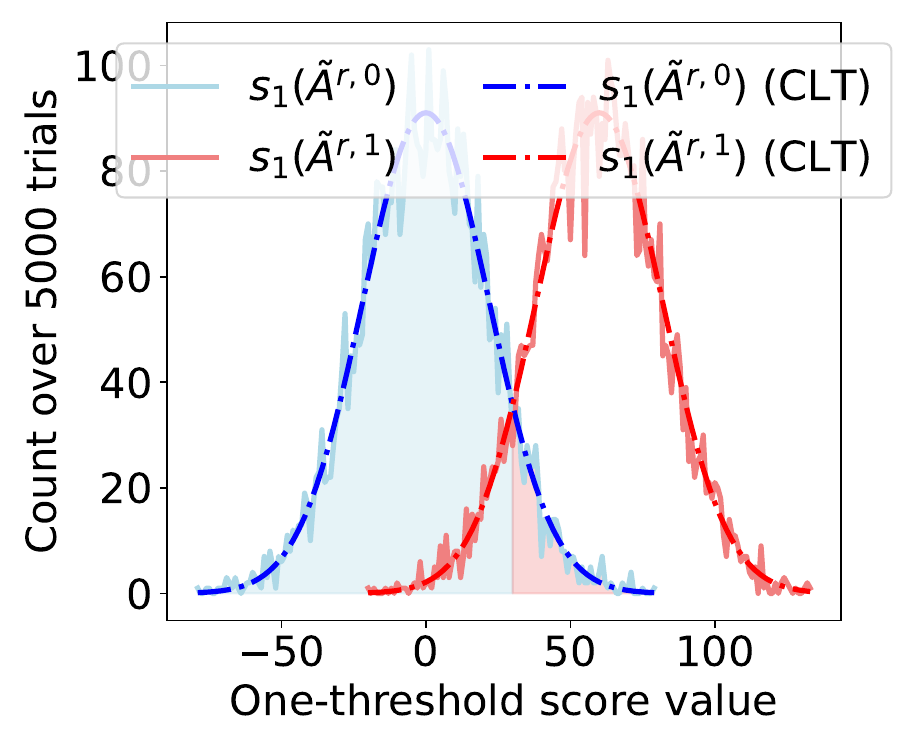}\label{subfig: multi-cell-full-distribution}
    }
    \hspace{0.5cm}
    \subfigure[Auxiliary attacker.]{\includegraphics[width=0.4\linewidth]{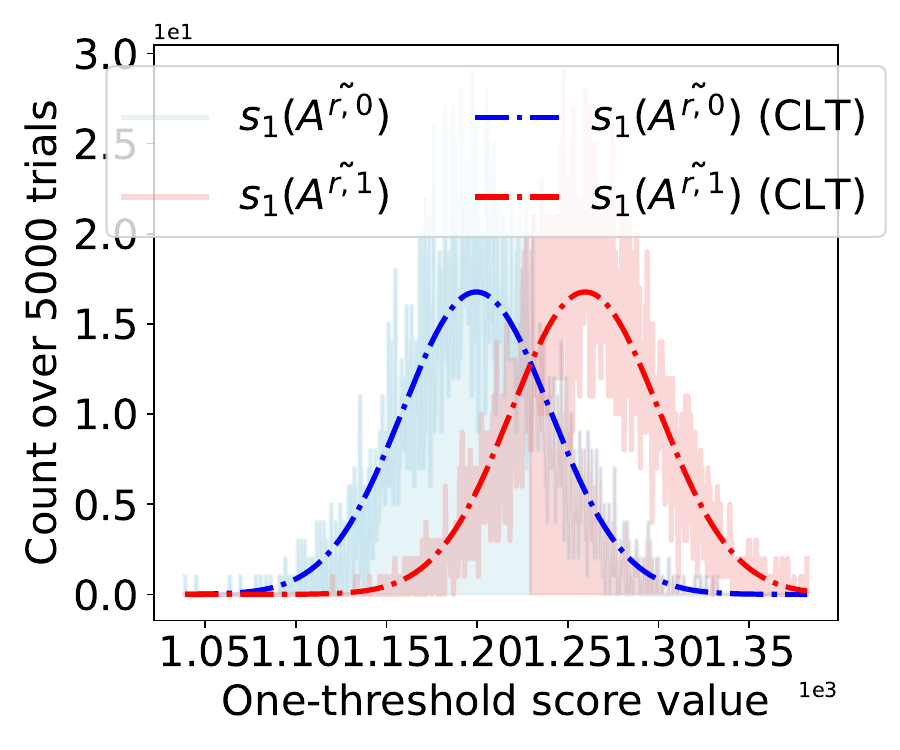}\label{subfig: multi-cell-partial-distribution}}
    \caption{The corresponding one-threshold score distributions of 60 positive observations with the individual cell distribution as shown in Figure~\ref{fig: one-threshold-distribution-percell} for both non-member aggregates~(\ie, $\texttt{s}_{1}\left(\tilde{A}^{r,0}\right)$ and $\texttt{s}_{1}\left(\tilde{A}^{r,0}\right)$  (CLT)) and member aggregates~(\ie, $\texttt{s}_{1}\left(\tilde{A}^{r,1}\right)$ and $\texttt{s}_{1}\left(\tilde{A}^{r,1}\right)$  (CLT)). While \textit{$\texttt{s}_{1}\left(\tilde{A}^{r,b}\right)$} for $b\in[0,1]$ is the empirical histogram over $5,000$ trials, \textit{$\texttt{s}_{1}\left(\tilde{A}^{r,b}\right)$ (CLT)} is the approximation with the central limit theorem.}
    \label{fig: distribution-socre1}
\end{figure}
In Figure~\ref{fig: one-threshold-distribution-percell}, we first present examples of the distribution at each cell $\tilde{A}_{le}^{r,b}$ for $l\in L$, $e\in E$, and $b\in[0,1]$ under both informed attackers and auxiliary attackers. Then, we show the corresponding score distributions in Figure~\ref{fig: distribution-socre1} for more insights.

For the informed attackers, as shown in Figure~\ref{subfig: per-cell-full-distribution}, each observation follows either a Laplace distribution $\mathrm{Lap}\left(\frac{C}{\varepsilon}\right)$ for non-members or a Laplace distribution $\mathrm{Lap}\left(C,\frac{C}{\varepsilon}\right)$ for non-members. That is to say, for informed attackers, the only uncertainty of the membership inference comes from the DP noise.
Then, the one-threshold score distribution, which is essentially the sum of all positive observations, approximately follows a Gaussian distribution as presented in Figure~\ref{subfig: multi-cell-full-distribution}.

For auxiliary attackers, as shown in Figure~\ref{subfig: per-cell-partial-distribution}, the distribution of each cell in the released aggregate follows a convoluted distribution between the underlying data distribution and the DP noise distribution.
Then, analogously, the sum of all positive observations follows a Gaussian distribution as presented in Figure~\ref{subfig: multi-cell-partial-distribution}.

The major difference between the informed attacker and the auxiliary attacker in terms of the one-threshold score lies in the uncertainty of the underlying data distribution before perturbation.
As aforementioned, for informed attackers, the uncertainty comes solely from the injected DP noise, whereas the underlying data distribution before perturbation also provides obfuscation for auxiliary attackers, leading to score distributions with larger variance and less distinguishability as shown in Figure~\ref{fig: distribution-socre1}.

\textbf{Threshold estimation.}
As suggested by the Neyman-Pearson lemma~\cite{dong2022gaussian}, given any level of error, the most powerful way to distinguish between two distributions is to perform a likelihood ratio test, where the probability ratio of two distributions outputting the same outcome is thresholded. In our case, this is equivalent to the threshold of the score $s$~\cite{lrt}. That is to say, to distinguish between $H_{IN}$ and $H_{OUT}$ as stated in Theorem~\ref{theo: one-threshold-hypothesis}, one can use a threshold $T$, such that
\begin{equation}
    \begin{aligned}
        \hat{b}=\begin{cases}
        0 & s<T,\\
        1 & s\geq T.
        \end{cases}
    \end{aligned}
\end{equation}

\begin{figure}[tbp!]
    \centering
    \subfigure[ROC.]{\includegraphics[width=0.4\linewidth]{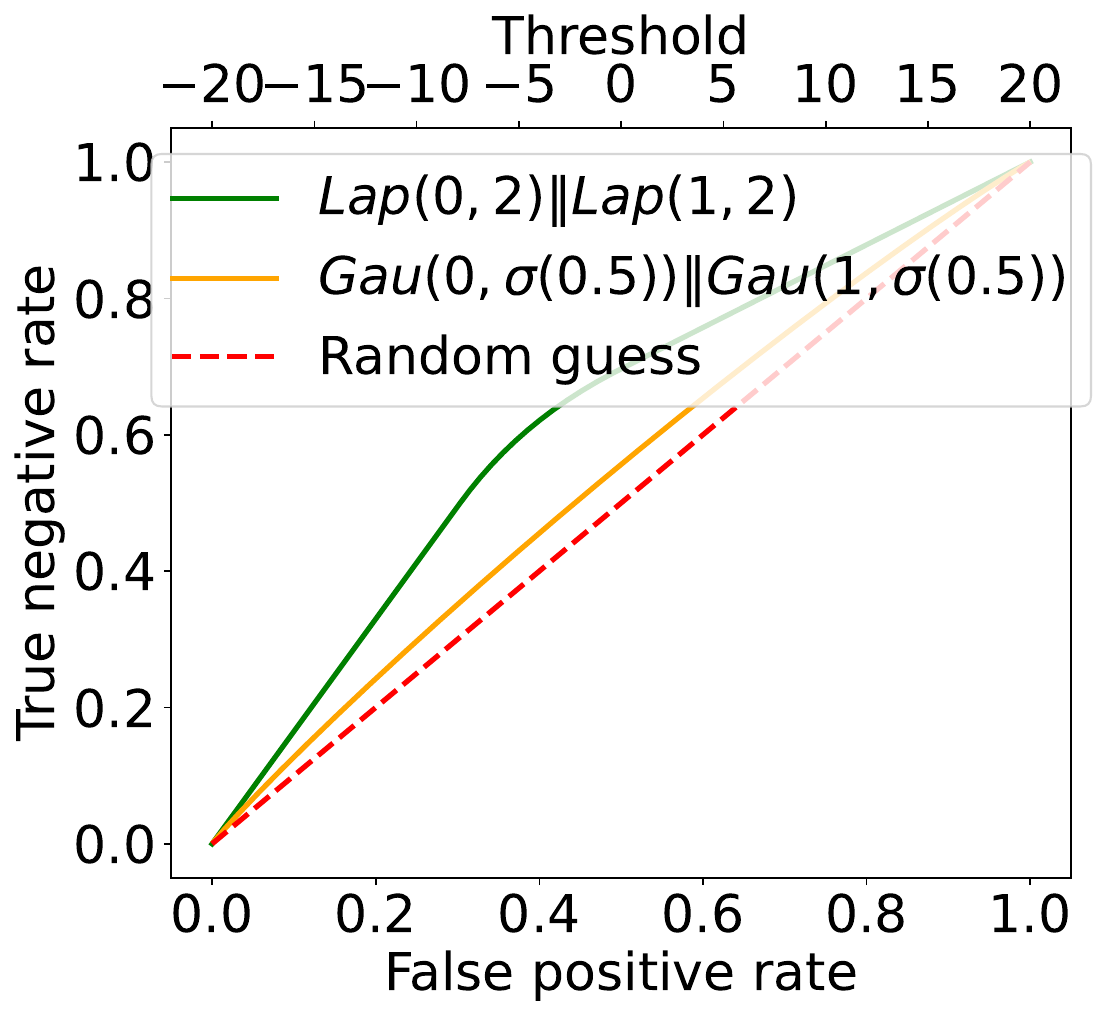}\label{subfig: roc_lap_gau}}
    \hspace{0.5cm}
    \subfigure[Attack accuracy.]{
    \includegraphics[width=0.4\linewidth]{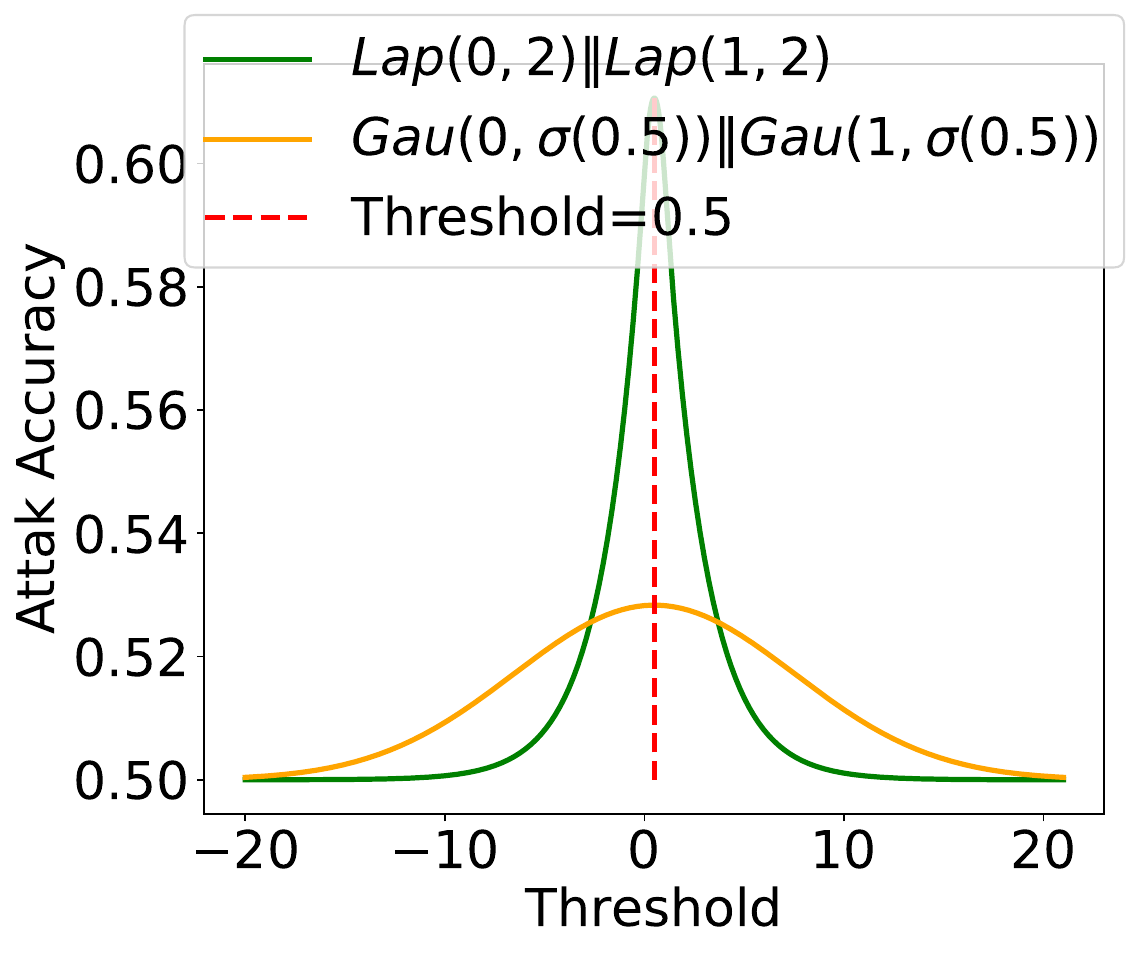}\label{subfig: acc_lap_gau}
    }
    \caption{An example of attack accuracy and the corresponding ROC with varying thresholds under informed attackers.}
    \label{fig: thresholds}
\end{figure}

The selection of $T$ is crucial as it decides the attacking power of an adversary as well as its error. As shown in Figure~\ref{subfig: roc_lap_gau}, a larger $T$ usually indicates a larger FPR and, therefore, a larger TPR. In cases where the underlying data distribution after the perturbation is symmetric, setting the threshold at the intersection point of the member and non-member score distribution leads to the highest attacking accuracy under balanced dataset settings as demonstrated in Figure~\ref{subfig: acc_lap_gau}.

For the attackers who are aware of the mean and the variance of each positive observation, the analytical form of distribution $\mathbbm{P}$ and $\mathbbm{Q}$ in Theorem~\ref{theo: one-threshold-hypothesis} is able to be derived as stated. The value of $T$ is then set based on $\mathbbm{P}$ and $\mathbbm{Q}$ given a fixed FPR or to maximize the attacking accuracy. 

On the other hand, for the attackers who have no prior of either the mean or the variance of the underlying data distribution,
the threshold $T$ is able to be set based on the empirical estimated distributions using generated shadow aggregates $\mathcal{A}_{sd}$, the concrete steps of which are summarized in Algorithm~\ref{alg: one-threshold-threshold}.
\begin{algorithm}
    \caption{\texttt{ThresholdEstimation}. Estimate the threshold $T$ for the one-threshold attack to maximize attack accuracy on the balanced dataset.}\label{alg: one-threshold-threshold}
    \LinesNumbered
    \SetKwInOut{Input}{Input}
    \SetKwInOut{Output}{Output}
    \Input{The shadow aggregates $\mathcal{A}_{sd}$ and the target trace $z$.}
    \Output{A threshold $T$.}
    $\mathcal{A}_{sd}^{1}\leftarrow\{\bar{A}^{1},\bar{A}^{2},\ldots,\bar{A}^{\lfloor m/2\rfloor}\}\subset\mathcal{A}_{sd}$\label{line: member-threshold-one}\tcp*{Get all member shadow aggregates}
    $\mathcal{A}_{sd}^{0}\leftarrow\mathcal{A}_{sd}\backslash \mathcal{A}_{sd}^{1}$\label{line: non-member-threshold-one}\tcp*{Get all non-member shadow aggregates}
   
    $\hat{\mu}_{1}\leftarrow\texttt{MEAN}_{\bar{A}^{i}\in\mathcal{A}_{sd}^{1}}\left(
    \texttt{Score1}\left(
    \bar{A}^{i},A_{\theta}=0^{\lvert L\rvert\times \lvert E\rvert},z
    \right)
    \right)$\;\label{line: one-threshold-estimate-mean-of-p}
    $\hat{\mu}_{0}\leftarrow\texttt{MEAN}_{\bar{A}^{i}\in\mathcal{A}_{sd}^{0}}\left(
   \texttt{Score1}\left(
    \bar{A}^{i},A_{\theta}=0^{\lvert L\rvert\times \lvert E\rvert},z
    \right)
    \right)$\label{line: one-threshold-estimate-mean-of-q}\tcp*{Mean estimation with Algorithm~\ref{alg: one-threshold-score}}
    $T\leftarrow \frac{1}{2}\left(\hat{\mu}_{0}+\hat{\mu}_{1}\right)$\;\label{line: one-threshold-threshold}
    \Return{T}
\end{algorithm}

We first divide the shadow aggregates into members~(\ie, $\mathcal{A}^{1}_{sd}$) and non-members~(\ie, $\mathcal{A}^{0}_{sd}$) as shown in Line~\ref{line: member-threshold-one}
and Line~\ref{line: non-member-threshold-one}.
Then, for each shadow aggregate $\bar{A}^{i}$ in $\mathcal{A}_{sd}^{0}$ and $\mathcal{A}_{sd}^{1}$, we calculate its score with Algorithm~\ref{alg: one-threshold-score}. 
After that, $\hat{\mu}_{1}$ and $\hat{\mu}_{0}$, which are the estimated mean of $\mathbbm{P}$ and $\mathbbm{Q}$,
are calculated by averaging out $\mathcal{A}_{sd}^{1}$ and $\mathcal{A}_{sd}^{0}$, respectively,
as shown in Line~\ref{line: one-threshold-estimate-mean-of-p} and Line~\ref{line: one-threshold-estimate-mean-of-q}.

In Algorithm~\ref{alg: one-threshold-threshold}, we aim to estimate a threshold that maximizes the attacking accuracy. As both $\mathbbm{P}$ and $\mathbbm{Q}$ are symmetric distributions, the threshold $T$ is set to the midpoint of $\hat{\mu}_{0}$ and $\hat{\mu}_{1}$, as depicted by Line~\ref{line: one-threshold-threshold}.
Note that we can also estimate the threshold given a fixed level of attack error~(\ie, FPR), the corresponding method of which is deferred into Algorithm~\ref{alg: one-threshold-threshold-fix-error} in Appendix~\ref{subsec: threshold-fixed-error}.

\subsection{Two-threshold Attack}\label{subsec: two-threshold}
\begin{figure}[tbp!]
    \centering
    \subfigure[Informed attacker.]{
    \includegraphics[width=0.4\linewidth]{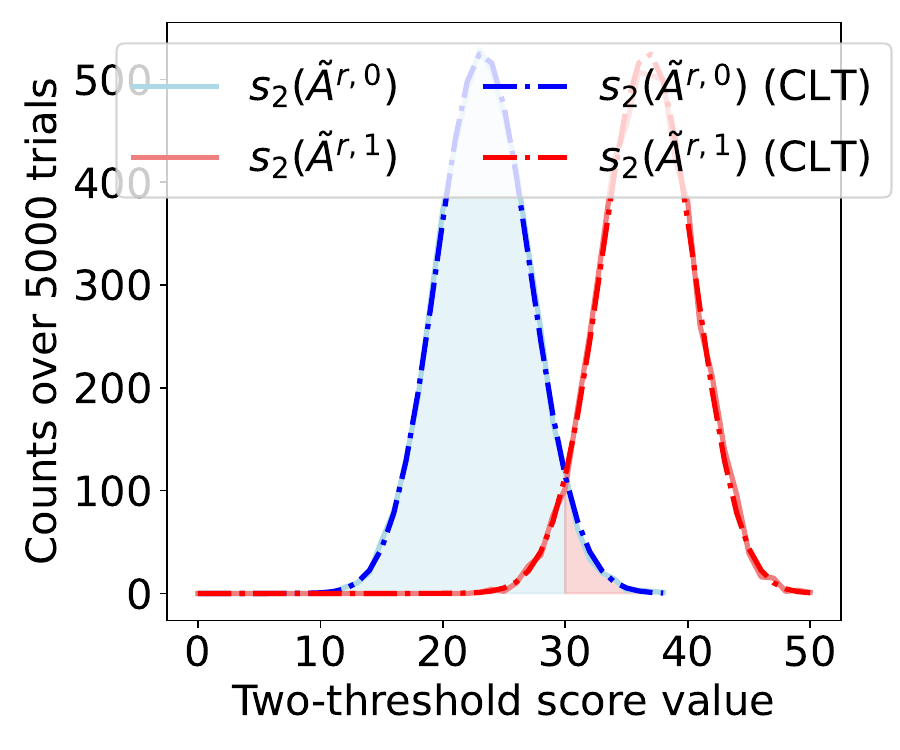}\label{subfig: multi-cell-full-distribution-score2}
    }
    \hspace{0.5cm}
    \subfigure[Auxiliary attacker.]{\includegraphics[width=0.4\linewidth]{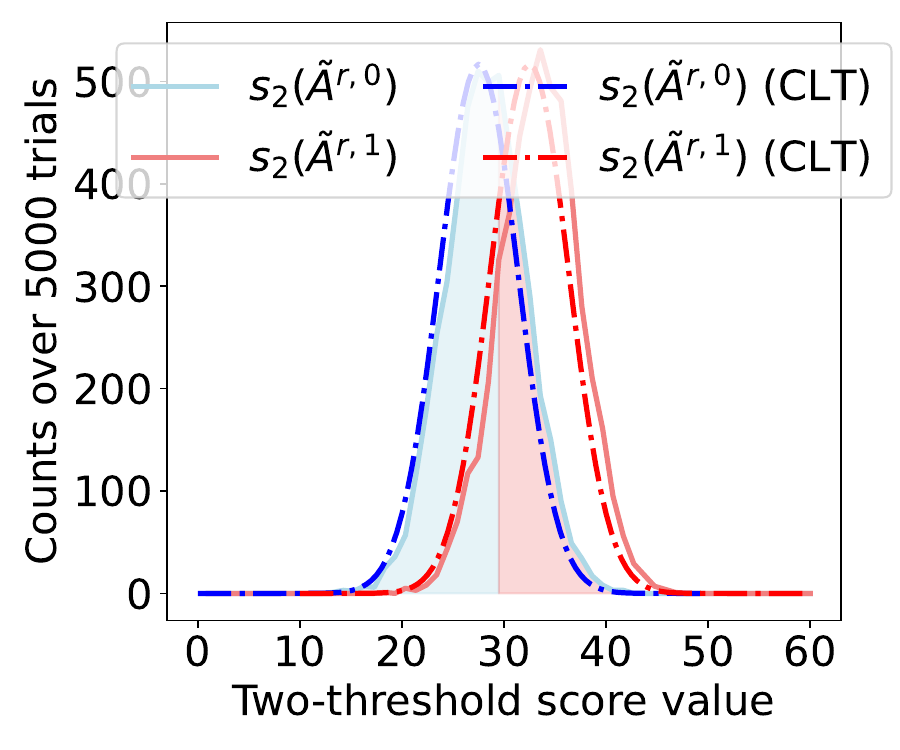}\label{subfig: multi-cell-partial-distribution-score2}}
    \caption{The corresponding two-threshold score distribution over $60$ positive observations with the individual cell distribution as shown in Figure~\ref{fig: one-threshold-distribution-percell} for both non-member and member aggregates.}
    \label{fig: distribution-score2}
\end{figure}

Unlike the one-threshold attack, our two-threshold attack defines a different score function, which performs the likelihood ratio test two times, leading to different score distributions.  
The definition of the two-threshold score function is summarized in Algorithm~\ref{alg: two-threshold-score}, the score distributions are shown in Figure~\ref{fig: distribution-score2}, and the corresponding threshold estimation is presented in Algorithm~\ref{alg: two-threshold-subthreshold}.

\textbf{Score function.}
\begin{algorithm}[tbp!]
    \caption{\texttt{Score2}. The score function for the two-threshold attack.}\label{alg: two-threshold-score}
    \LinesNumbered
    \SetKwInOut{Input}{Input}
    \SetKwInOut{Output}{Output}
    \Input{The target aggregate $\tilde{A}$, the attacker's background knowledge $A_{\theta}$, the target trace $z$, and the shadow aggregates $\mathcal{A}_{sd}$.}
    \Output{Score of the target aggregate $s$.}
    $\tilde{A}^{r}\leftarrow\tilde{A}-A_{\theta}$\label{line: score-two-remove}\tcp*{Remove background knowledge from the target aggregate}
    $\hat{\mathcal{T}}_{1}\leftarrow\texttt{PerCellThresholdEstimation}\left(\mathcal{A}_{sd},z\right)$\;
    $s\leftarrow \sum\limits_{l\in L}\sum\limits_{e\in E}
    \mathbbm{1}\left(
    z_{le}=1
    \right)\cdot
    \mathbbm{1}\left(
    \tilde{A}^{r}_{le}\geq \hat{T}_{le}^{\prime}
    \right)$\label{line: score-two-sum}\tcp*{Sum up all positive-observation cells}
    \Return{s}
\end{algorithm}
\begin{algorithm}[tbp!]
    \caption{\texttt{PerCellThresholdEstimation}. Estimate the threshold for each positive observation.}\label{alg: two-threshold-subthreshold}
    \LinesNumbered
    \SetKwInOut{Input}{Input}
    \SetKwInOut{Output}{Output}
    \Input{The shadow aggregate $\mathcal{A}_{sd}$ and the target trace $z$.}
    \Output{A threshold set $\mathcal{T}^{\prime}$.}
    $\mathcal{T}^{\prime}\leftarrow\phi$\;
    $\mathcal{A}_{sd}^{1}\leftarrow\{\bar{A}^{1},\bar{A}^{2},\ldots,\bar{A}^{\lfloor m/2\rfloor}\}\subset\mathcal{A}_{sd}$\label{line: member-threshold-two}\tcp*{Get all member shadow aggregates}
    $\mathcal{A}_{sd}^{0}\leftarrow\mathcal{A}_{sd}\backslash \mathcal{A}_{sd}^{1}$\label{line: non-member-threshold-two}\tcp*{Get all non-member shadow aggregates}
    \For{$l\in L$}{
    \For{$e\in E$}{
    $\hat{\mu}_{le}^{0}\leftarrow\texttt{MEAN}_{\bar{A}^{i}\in\mathcal{A}_{sd}^{0}}\left(
    \bar{A}^{i}_{le}
    \right)$\;\label{line: two-subthrehold-mem-t}
    $\hat{\mu}_{le}^{1}\leftarrow\texttt{MEAN}_{\bar{A}^{i}\in\mathcal{A}_{sd}^{1}}\left(
    \bar{A}^{i}_{le}
    \right)$\;\label{line: two-subthreshold-nmem-t}
    $T_{le}\leftarrow \frac{1}{2}\left(
    \hat{\mu}_{le}^{0}+\hat{\mu}_{le}^{1}
    \right)$\;\label{line: two-subthreshold-est}
    $\mathcal{T}^{\prime}\leftarrow \mathcal{T}^{\prime}\cup\hat{T}_{le}$\;
    }
    }
    \Return{$\mathcal{T}^{\prime}$}
\end{algorithm}
As summarized in Algorithm~\ref{alg: two-threshold-score} and Algorithm~\ref{alg: two-threshold-subthreshold}, in general, two steps are involved in the score computation, namely the sub-thresholding and the sub-results summation.

Instead of directly summing up all released positive observations as in Algorithm~\ref{alg: one-threshold-score}, Algorithm~\ref{alg: two-threshold-score} first estimates a sub-threshold for each cell using Algorithm~\ref{alg: two-threshold-subthreshold}.
Specifically, the sub-thresholds are estimated by first estimating the underlying distribution of each released positive observation ~(Line~\ref{line: two-subthrehold-mem-t}-\ref{line: two-subthreshold-nmem-t}), and then searching for the thresholds that maximize the attack accuracy~(Line~\ref{line: two-subthreshold-est}) following the same principle as Algorithm~\ref{alg: one-threshold-threshold}.
Also note that we can estimate the threshold based on given FPRs for each cell, which we defer to Algorithm~\ref{alg: two-threshold-subthreshold-fixed-error} in Appendix~\ref{subsec: threshold-fixed-error}.

For each positive observation $\tilde{A}^{r}_{le}$, we compare it with the estimated threshold $T_{le}$ as shown in Line~\ref{line: score-two-sum}. If $\tilde{A}^{r}_{lt}\geq T_{le}$ holds, we assign a score $1$ to $\tilde{A}_{le}^{e}$. Otherwise, a score of zero is assigned. 
Afterward, we sum up all assigned scores to the positive observations in aggregate $\tilde{A}^{r}$ as the final score $s$.

\textbf{Score distributions.}
Similar to the one-threshold attack, we analyze the analytical form of the distributions $\mathbbm{Q}$ and $\mathbbm{P}$ shown in Equation~\ref{eq: p_q}.
For a released aggregate with $n$ positive observations, the score $s$ computed with Algorithm~\ref{alg: two-threshold-score} follows a binomial distribution, which is summarized by Theorem~\ref{theo: two-threshold-hypothesis}. The intuition behind this is that, as thresholding each cell is equivalent to flipping an uneven coin as shown in Figure~\ref{fig: one-threshold-distribution-percell}, the overall flipping result over $n$ positive observations follows a (Poisson) binomial distribution with $n$ Bernoulli experiments.
\begin{theorem}\label{theo: two-threshold-hypothesis}
    For any attacker that has access to a subset traces $D_{\theta}$ of the target trace dataset $D$ such that $D_{\theta}\subseteq D\backslash z$,
    to infer the membership of a target trace $z$ based on a released aggregate $\tilde{A}$ with Algorithm~\ref{alg: overall} and Algorithm~\ref{alg: two-threshold-score} is equivalent to distinguish between the following two hypotheses:
    \begin{equation*}
        \begin{aligned}
            &H_{IN}: s \text{ follows the distribution }\mathbbm{Q}:= B\left(n,1-\beta^{\prime}\right),\\
            &H_{OUT}: s \text{ follows the distribution }\mathbbm{P}:=B\left(n,\alpha^{\prime}\right),
        \end{aligned}
    \end{equation*}
    where $\beta^{\prime}$ is the false negative rate of thresholding each positive observation and $\alpha^{\prime}$ is the false positive rate of thresholding each positive observation.
\end{theorem}

Analogously to that for the one-threshold attack, $\alpha^{\prime}$ and $\beta^{\prime}$ can either be directly obtained by Equation~\ref{eq: distribution-percell}, or estimated by first computing the histogram of each cell $\tilde{A}^{r}_{lt}$ for $l\in L$ and $t\in T$ with the shadow aggregates, then its CDF~\cite{ye2022enhanced}.
Particularly, for $n$ that is large enough~($n\geq5$), we can also use the Gaussian distribution with mean $\mu_{1}=n\left(1-\beta^{\prime}\right)$~($\mu_{0}=n\alpha^{\prime}$) and standard deviation $\sigma_{1}=\sqrt{n\beta^{\prime}\left(1-\beta^{\prime}\right)}$~($\sigma_{0}=\sqrt{n\alpha^{\prime}\left(1-\alpha^{\prime}\right)}$) to approximate the Binomial distribution in Theorem~\ref{theo: two-threshold-hypothesis}.

\textbf{Examples.}
In Figure~\ref{fig: distribution-score2}, we show the two-threshold score distribution for both informed attackers and auxiliary attackers.
Specifically, the Laplace DP noise is injected for perturbation. Therefore, for informed attackers, with a probability $1-\beta^{\prime}$ where $\beta^{\prime}=1-\mathrm{CDF}_{Lap\left(C,\frac{C}{\varepsilon}\right)}\left(\frac{C}{2}\right)$, $\tilde{A}^{r,1}_{le}$ is labeled as $1$. Similarly, with a probability $\alpha^{\prime}$ where $\alpha^{\prime}=\mathrm{CDF}_{Lap\left(C,\frac{C}{\varepsilon}\right)}\left(\frac{C}{2}\right)$, $\tilde{A}^{r,0}_{le}$ is labeled as $1$. The two-threshold score of an aggregate with $n$ positive observations thus follows a binomial distribution with parameter $n$.

For auxiliary attackers, we assume that the shadow aggregates are generated by aggregating $m$ traces after uniformly sampling them from the auxiliary dataset. Suppose there are $n_{AUX}^{le,1}$ out of $n_{AUX}$ ones at the $le$-th cell, the clean data distribution of the cell follows a binomial distribution $B\left(m,\frac{n_{AUX}^{le,1}}{n_{AUX}}\right)$.
After perturbation, each cell might have distinct distributions. Hence, the two-threshold score follows a Poisson binomial distribution~\cite{daskalakis2012learning}. 
For simplicity, we further approximate it with a Gaussian distribution $N\left(1+\mu_{0},\sigma\right)$ for members, and $N\left(\mu_{0},\sigma\right)$ for non-members, where $\mu_{0}=n\sum\limits_{l\in L}\sum\limits_{e\in E}\frac{n_{AUX}^{le,1}}{n_{AUX}}$ and $\sigma=\sum\limits_{l\in L}\sum\limits_{e\in E}\left(
\left(2\left(\frac{C}{\varepsilon}\right)^{2}\right)+n\frac{n_{AUX}^{le,1}}{n_{AUX}}\left(1-\frac{n_{AUX}^{le,1}}{n_{AUX}}\right)
\right)$.

\section{Attack Model Interpretation}
To explore the potential causes leading to the gap between the empirical results of MLP-based MIAs and the expected attack accuracy given by DP as shown in Figure~\ref{fig: gap}, as well as its inferiority to metric-based attacks~(Cf. Figure~\ref{fig: metric-based-acc-main}), we study and interpret the rules learned by the MLP-based attack model in previous settings.

Particularly, we compare the learned MLP-based attack model parameters and metric-based attack rules proposed in this work. Specifically, in Section~\ref{subsec: approximate-thresholding}, we show MLP can approximate the thresholding operation with its nonlinear layers~(activations). Then, in Section~\ref{subsec: parameter-analysis}, we show that the MLP-based attack models in previous works learns only the one-threshold attack rule, leading to a sub-optimality when it comes to scenarios where the released aggregate is perturbed by the Laplace mechanism.

\subsection{Approximate Metric-based attack with MLP}\label{subsec: approximate-thresholding}
In this section, we first use MLP to approximate the proposed metric-based attacks, proving the capability of MLP to learn different attack rules automatically in theory. In particular, we take MLP with Sigmoid function as an example, which is one of the most frequently used activations in the literature to provide nonlinearity.

\begin{figure}[tbp!]
    \centering
    \subfigure[MLP with one node in the hidden layer.]{
    \includegraphics[width=0.4\linewidth]{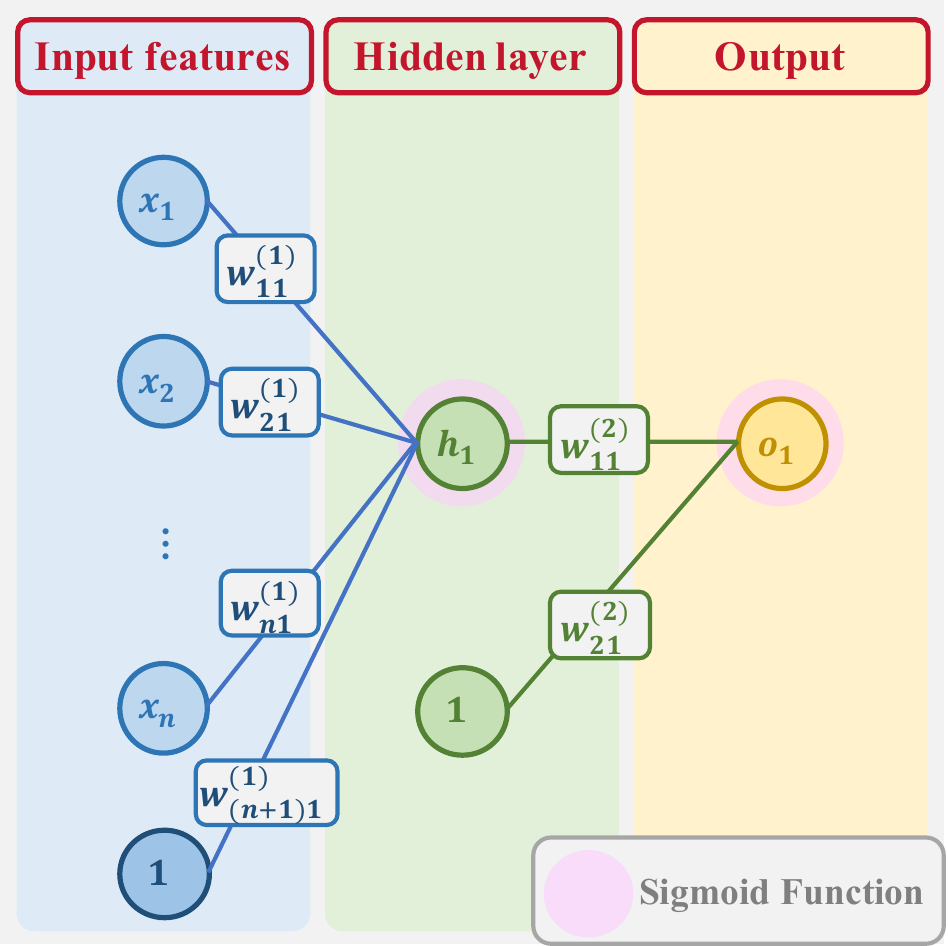}\label{subfig: mlp-one-node}
    }
    \hspace{0.5cm}
    \subfigure[MLP with $n$ nodes in the hidden layer.]{\includegraphics[width=0.4\linewidth]{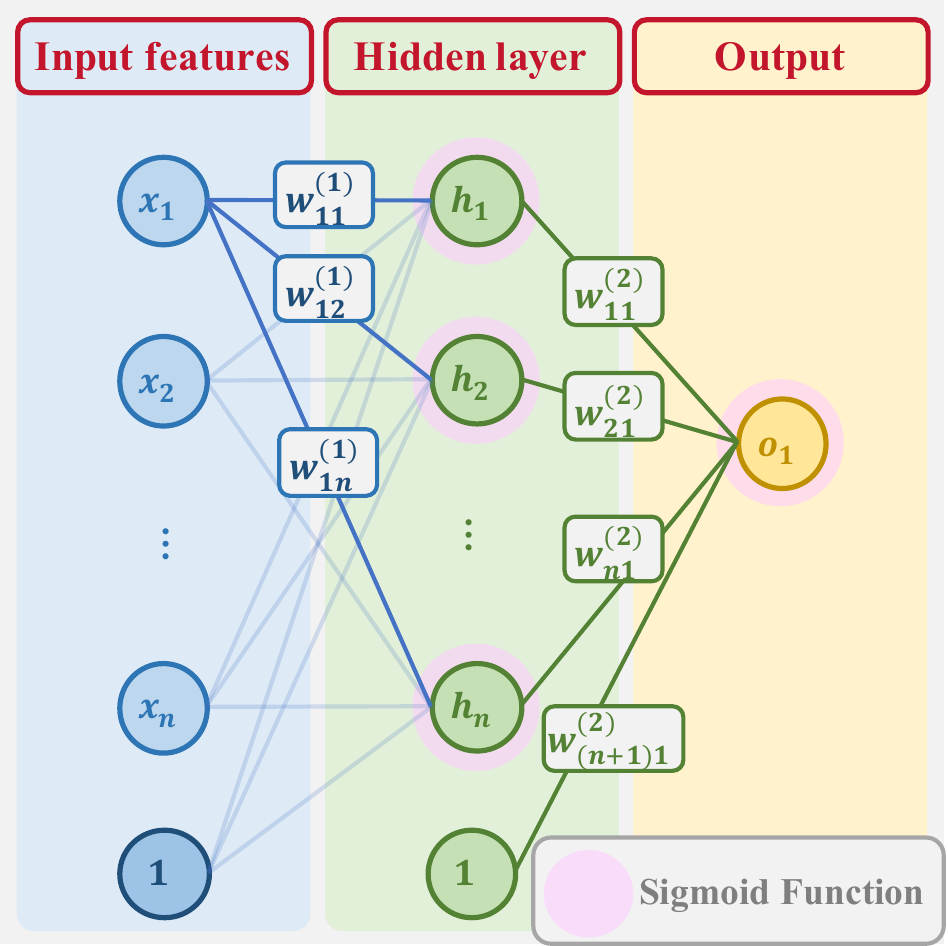}\label{subfig: mlp-multi-node}}
    \caption{Illustration of the MLP architecture considered in this work.}
    \label{fig: MLP-architecture}
\end{figure}

We start by considering a simple case where there is only one hidden node with bias in one hidden layer, as shown in Figure~\ref{subfig: mlp-one-node}. Note that each input feature corresponds to one positive observation in the input location aggregates.

\begin{theorem}\label{theo: one-threshold-approximation}
    For an input with $n$ features $\{x_1,\ldots,x_n\}$, given a real number $T$, the following step function
    \begin{equation}\label{eq: one-threshold-function}
        \begin{aligned}
            f\left(x\right)=\begin{cases}
                0 & \text{if }\sum\limits_{i=1}^{n}x_{i}<T,\\
                1 & \text{if }\sum\limits_{i=1}^{n}x_{i}\geq T
            \end{cases}
        \end{aligned}
    \end{equation}
    can be approximated by
    \begin{equation}
        \hat{f}(x)=\sigma\left(w^{(2)}_{11}
        \left(
        \sum\limits_{i=1}^{n}w_{i1}^{(1)}x_{i}+w_{n+1}^{(1)}
        \right)
        +w_{21}^{(2)}
        \right),
    \end{equation}
    where $w_{i1}^{(1)}=a$ for $i\in[n]$ is a real number, $w_{n+1}^{(1)}=aT$, $w_{11}^{(2)}=b$ is a real number $w_{21}=-\frac{1}{2}b$, and $\sigma\left(x\right)=\frac{e^{x}}{1+e^{x}}$ is the Sigmoid function.
\end{theorem}

Note that Equation~\ref{eq: one-threshold-function} is the one-threshold attack rule. Therefore, Theorem~\ref{theo: one-threshold-approximation} further indicates that the MLP-based attack can at least learn the one-threshold rule to distinguish members from non-members.

In the following theorem, we further show that with a slightly larger architecture, MLP is able to approximate the two-threshold rule. Particularly, for an input with $n$ features, we assume an architecture with with at least $n$ nodes in the hidden layer as shown in Figure~\ref{subfig: mlp-multi-node}.

\begin{theorem}\label{theo: two-threshold-approximation}
     For an input with $n$ features $\{x_1,\ldots,x_n\}$, given a set of real numbers $\{T_1,\ldots,T_{n}\}$ and a real number $T$, the following step function
     \begin{equation}\label{eq: two-threshold-function}
         \begin{aligned}
             f(x)=\begin{cases}
                 0 & \text{if }\left(
                 \sum\limits_{i=1}^{n}\mathbbm{1}\left(x_i\geq T_i\right)
                 \right)<T,\\
                 1 & \text{if }\left(
                 \sum\limits_{i=1}^{n}\mathbbm{1}\left(x_i\geq T_i\right)
                 \right)\geq T
             \end{cases}
         \end{aligned}
     \end{equation}
     can be approximated by
     \begin{equation}
         \begin{aligned}
        &\hat{f}\left(x\right)=\\
         &\sigma\left(
          \sum\limits_{j=1}^{n}\left(w_{j1}^{(2)}\sigma
          \left(
          \sum\limits_{i=1}^{n}w_{ii}^{(1)}x_i+w_{(n+1)i}^{(1)}
          \right)+w_{(n+1)1}^{(2)}
          \right)
         \right),
         \end{aligned}
     \end{equation}
     where $w_{ii}^{(1)}=a$ for $i\in[n]$, $w_{j1}^{(2)}=b$ for $j\in[n]$, $w_{(n+1)i}^{(1)}=-aT_{i}$ for $i\in[n]$, and $w_{j1}^{(2)}=b\frac{n}{2}$ for $j\in[n]$ hold for real numbers $a$ and $b$, $\sigma\left(x\right)=\frac{e^{x}}{1+e^{x}}$ is the Sigmoid function, and $\mathbbm{1}\left(\cdot\right)$ is the indicator function.
\end{theorem}

As Equation~\ref{eq: two-threshold-function} describes the two-threshold attack rule, Theorem~\ref{theo: two-threshold-approximation} indicates that the MLP-attack model with the architecture shown in Figure~\ref{subfig: mlp-multi-node} is able to learn the two-threshold rule to distinguish members from non-members.

In addition, to provide more insights, we study how the scale of the parameters~(\ie, the value of $a$ and $b$) affects the approximation error in Appendix~\ref{subsec: mlp-approximation}.

\subsection{Parameter Analysis}\label{subsec: parameter-analysis}
Now, we are ready to analyze the learned parameters of the MLP-based attack model in previous works. Particularly, we adopt a similar setting as in previous works, where an MLP with $n$ nodes in one hidden layer is trained with $2,000$ shadow aggregates as the attack model~\cite{pyrgelis2018knock,guan2024zero}\footnote{Previous works~\cite{pyrgelis2018knock,guan2024zero} use $400$ shadow aggregates to train the attack model, which has a fixed architecture for input with any sizes. We show in Appendix~\ref{subsec: fixed-nodes-num} that this setting demonstrates similar model behaviors to ours.}. 

\begin{figure}[tbp!]
    \centering
    \includegraphics[width=0.45\linewidth]{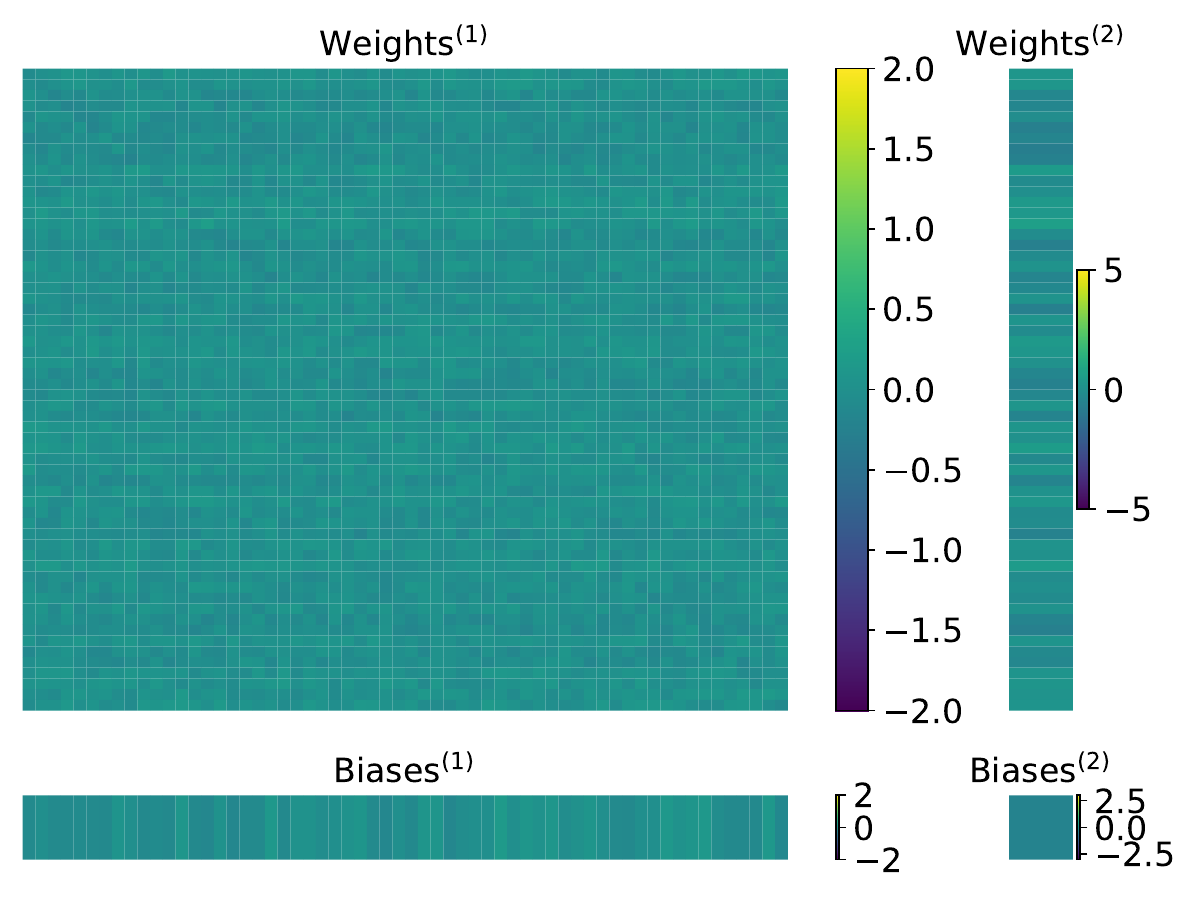}
    \caption{The learned weights of the MLP-based attacker model with an architecture as shown in Figure~\ref{subfig: mlp-multi-node} with $60$ input features.}
    \label{fig: lap_weights_2k}
\end{figure}

The learned weights of the MLP-based attacker model with an architecture as shown in Figure~\ref{subfig: mlp-multi-node} with $60$ input features. As shown in Figure~\ref{fig: lap_weights_2k},
while the learned weights $w_{ij}^{(1)}$ for $i,j\in[n]$ are roughly equivalent to each other, the learned biases are all around $0$. By comparing the learned parameters with Theorem~\ref{theo: one-threshold-approximation}, we have that, in essence, each node in the hidden layer learns the sum of all input features. Afterward, the weights are further summed up and thresholded by the learned bias $w_{(n+1)1}^{(2)}$, which is roughly equivalent to half of the weighted sum of all hidden nodes.

To summarize, an MLP-based attack model with $n$ neurons in the hidden layer trained in the aforementioned settings learns the following rule to distinguish members from non-members:
\begin{equation}
    \begin{aligned}
        f(x)=\begin{cases}
            0 & \text{if } n\sum\limits_{i=1}^{n}x_{i}<T,\\
            1 & \text{if }n\sum\limits_{i=1}^{n}x_{i}\geq T,
        \end{cases}
    \end{aligned}
\end{equation}
which, in essence, is our one-threshold attack rule. Therefore, as discussed in Section~\ref{subsubsec: comparison}, for cases such as Laplace noise perturbed aggregate releases, the trained MLP-based attack model demonstrates inferiority compared to the two-threshold attack, as well as the expected attack accuracy given by the DP theory.
\section{Evaluations}\label{sec: evaluation}
In this section, we verify the effectiveness of our proposed metric-based MIAs against location aggregates in Section~\ref{subsubsec: metric-based-attack}, where we confirm that both our one-threshold and two threshold attack is able to generalize to auxiliary knowledge settings for real-world datasets.
In Section~\ref{subsubsec: mlp-based-attack}, we further demonstrate that by increasing the training dataset size, the MLP-based attack model can learn both the one-threshold and two-threshold attack automatically for the optimal performance.
\subsection{Baselines and Dataset}\label{subsec: baseline-dataset}
\textbf{Baselines.}
In this work, we compare five different MIAs against location aggregates in total, namely the \textit{one-threshold attack} described in Algorithm~\ref{alg: overall}, the \textit{two-threshold attack} also described in Algorithm~\ref{alg: overall}, the MLP-based attack trained with 2,000 shadow aggregates, which we referred as \textit{meta-classifier~(2k)}, the MLP-based attack trained with 200,000 shadow aggregates, which we referred as \textit{meta-classifier~(200k)}, and the metric-based attack proposed by Dwork\etal~\cite{dwork2015robust}, which we referred as the \textit{reference attack}. In particular, for the meta-classifier-based attack, we adopt the MLP architecture shown in Figure~\ref{subfig: mlp-multi-node}.

\textbf{Dataset.}
We test the aforementioned attacks on the \textit{Milano} dataset, which contains data derived from an analysis of time-stamped and geo-localized tweets originating from Milan during November and December~\cite{barlacchi2015multi, DVN/9IZALB_2015}.
In particular, we grid the entire area into $96$~(\ie,$12\times 8$) sites in total. In addition, we consider the data from roughly the last $5$ weeks, leading to $800$ hourly epochs overall.

\subsection{Set-ups}\label{subsec: set-ups}
First, we randomly select and remove one trace from the dataset as the target trace $z$. Then, we evenly split the dataset into halves as the target~($7,896$ traces) and auxiliary dataset~($7,896$ traces), respectively. For the MLP-based attack methods, we further split the auxiliary dataset into halves evenly as the training dataset~($3,948$ traces) and the validation dataset~($3,948$ traces), respectively.
To generate a shadow aggregate, we randomly sample $m=1,999$ traces from the auxiliary dataset~(or the train and the validating dataset, respectively) before aggregating. To generate test aggregates, we repeat the same procedure over the target dataset. In particular, we test all baseline methods over $50,000$ test aggregates and average the results over 10 runs. Note that we adopt the balanced-dataset setting where half of the aggregates are added by the target for both the shadow aggregates and the test aggregates.

To provide DP guarantees, both the Laplace mechanism and Gaussian mechanism with a privacy budget $\varepsilon=0.5$ per observation are adopted. For the Gaussian mechanism, we further set the parameter $\delta$ to $\frac{1}{2m}$ by following the convention~\cite{dwork2014algorithmic,liu2022collecting,liu2024edge}. Note that we also clip the dataset as described in Section~\ref{subsubsec: dp_location} by setting $C=1$. Therefore, both the $l_1$ and $l_2$ sensitivity of released aggregates are $1$.
\subsection{Main results}\label{subsec: main-rersults}
\subsubsection{Effectiveness of the proposed metric-based attack}\label{subsubsec: metric-based-attack}
\begin{figure}[tbp!]
    \centering
    \subfigure[Informed attacker.]{
    \includegraphics[width=0.4\linewidth]{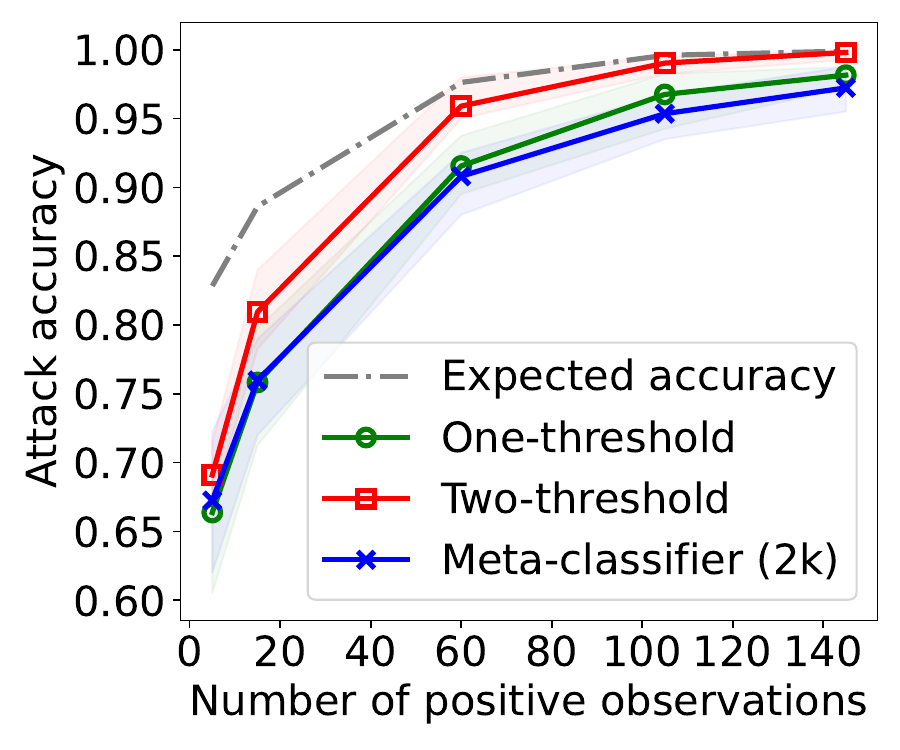}\label{subfig: metric-based-acc-fk}
    }
    \hspace{0.5cm}
    \subfigure[Auxiliary attacker.]{
    \includegraphics[width=0.4\linewidth]{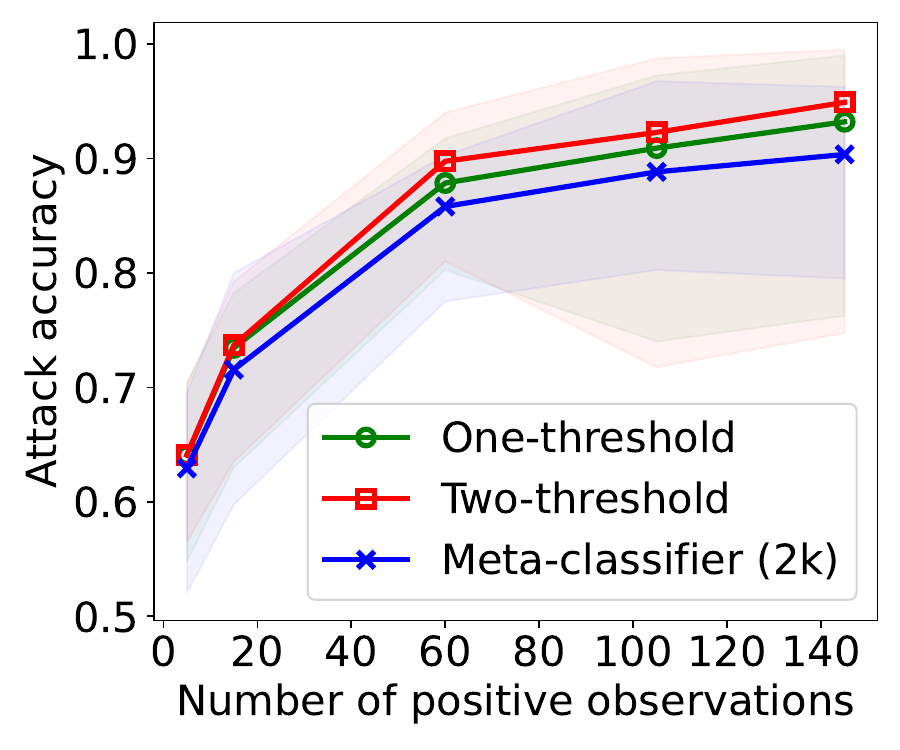}\label{subfig: metric-based-acc-pk}
    }
    \caption{Attack accuracy with an increasing number of positive observations under the Laplace mechanism. The attack models are trained with $2k$ shadow aggregates.}\label{fig: metric-based-acc-main}
\end{figure}
Figure~\ref{fig: metric-based-acc-main} demonstrates the effectiveness of our proposed metric-based MIAs under the \textit{Laplace} mechanism. The results for other metrics, including the AUC and the ROC, and for the Gaussian mechanism are deferred to Appendix~\ref{subsec: auc-roc} and Appendix~\ref{subsec: gau-performance}, respectively, due to the space limitation\footnote{As the \textit{reference attack} does not provide a method to achieve the maximized attack accuracy in their work, we compare it with other baselines in terms of AUC and ROC in the appendix.}.

First of all, non-surprisingly, for all compared MIAs, the attack accuracy increases along with the number of positive observations. In addition, by comparing Figure~\ref{subfig: metric-based-acc-fk} and Figure~\ref{subfig: metric-based-acc-pk}, it is obvious that for the same MIA method, the informed attackers benefit from a much higher attack accuracy compared to auxiliary attackers. This further confirms that the clean underlying data distribution provides privacy protection against auxiliary attackers to some extent, which agrees with our analysis in Section~\ref{subsec: one-threshold} and Section~\ref{subsec: two-threshold}.

Second of all, for both informed attackers and auxiliary attackers, the proposed metric-based methods, including the one-threshold and the two-threshold attack, demonstrate better performance compared to the meta-classifier-based method whose attack model is trained over $2,000$ shadow aggregates as in previous works~\cite{pyrgelis2018knock,guan2024zero}. 

While for both types of attackers, the two-threshold attack shows a great advantage over the one-threshold attack, which agrees with our theoretical analysis in Section~\ref{subsubsec: comparison}, the one-threshold attacker demonstrates a relevant smaller advantage over the meta-classifier-based attack~(around $1\%\sim 2\%$ in terms of attack accuracy). The reason is that, in essence, the meta-classifier-based attack learns the one-threshold rule to distinguish members from non-members as shown in Figure~\ref{fig: lap_weights_2k}. Yet, previous works use too little training data~(\eg, at most $2,000\times 60$) to train a model with at least $60\times 60$ parameters, leading to a slightly overfitted attack model.

\subsubsection{Effectiveness of the MLP-based attack with increased training dataset size}\label{subsubsec: mlp-based-attack}
\begin{figure}[tbp!]
    \centering
    \subfigure[Informed attacker.]{
    \includegraphics[width=0.4\linewidth]{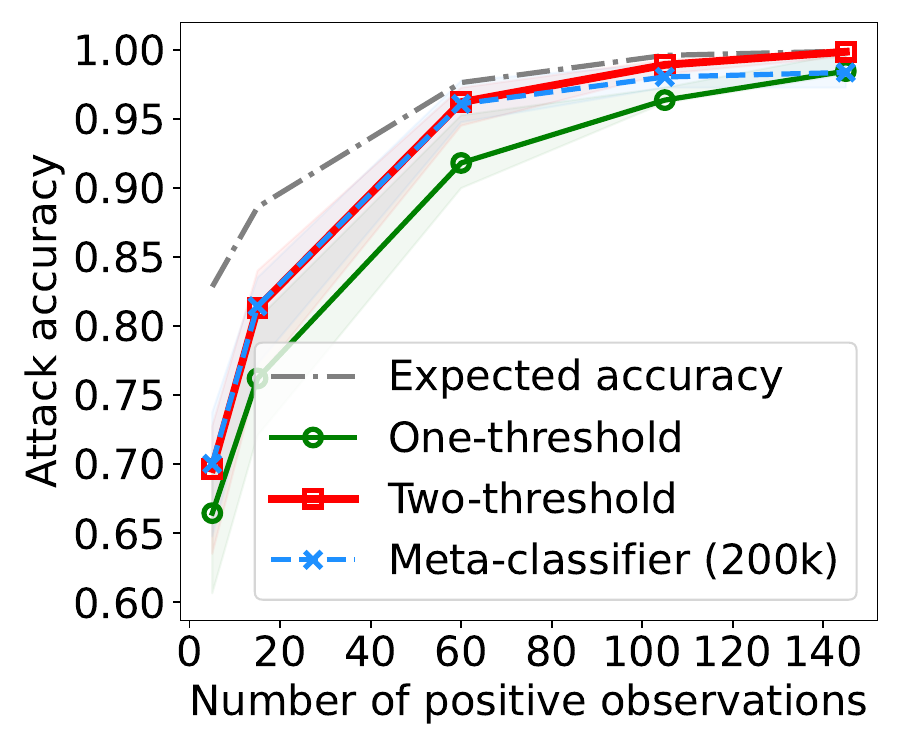}\label{subfig: metric-based-acc-fk-200k}
    }
    \hspace{0.5cm}
    \subfigure[Auxiliary attacker.]{
    \includegraphics[width=0.4\linewidth]{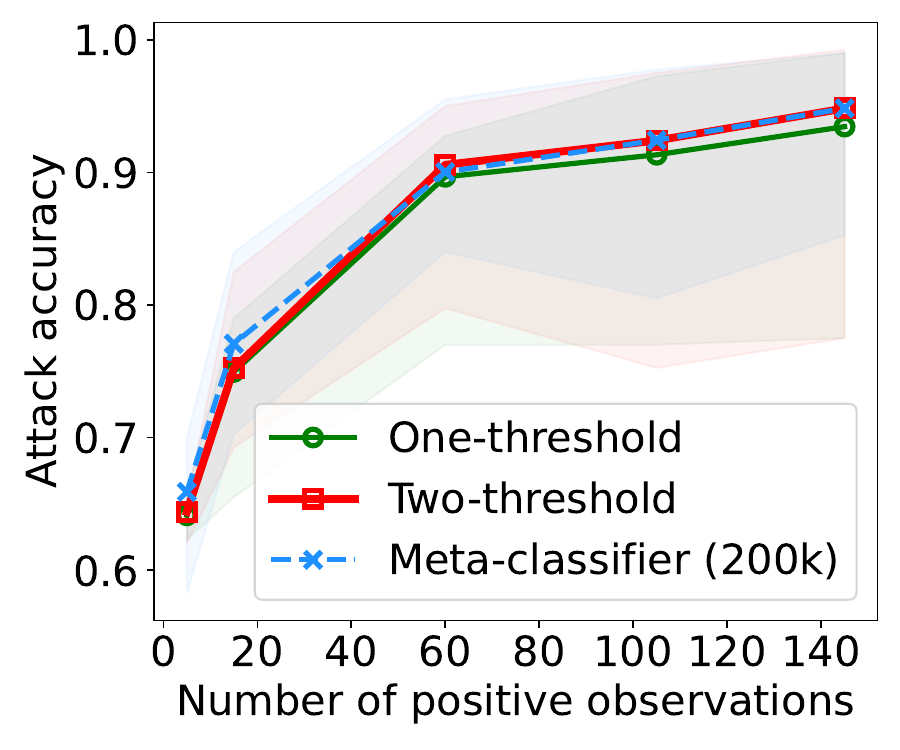}\label{subfig: metric-based-acc-pk-200k}
    }
    \caption{Attack accuracy with an increasing number of positive observations under the Laplace mechanism. The attack models are trained on $200k$ shadow aggregates.}\label{fig: metric-based-acc-main-200k}
\end{figure}

Figure~\ref{fig: metric-based-acc-main-200k} shows the attack results with $200k$ shadow aggregates. The results for other metrics and DP mechanisms are deferred to the Appendix~\ref{subsec: auc-roc} and Appendix~\ref{subsec: gau-performance} due to the space limitation.

First of all, it is shown that the metric-based attacks are not affected much by the increased shadow aggregates. In other words, relatively fewer shadow aggregates are required for the metric-based MIAs to achieve a satisfying result.

By contrast, after increasing the total amount of the training aggregates from $2k$ to $200k$, the attack accuracy of the meta-classifier-based attack boosts to the same level as our two-threshold attack for both informed attackers and auxiliary attackers. This further suggests that, while it is relatively easy for the MLP-based attack model to learn the one-threshold rule with a small number of shadow aggregates, which is the local optima for the released aggregates perturbed by the Laplace mechanism, $100$ times or more, shadow aggregates are required to learn the two-threshold rule for membership inference.

\begin{figure}[htbp!]
    \centering
    \includegraphics[width=0.45\linewidth]{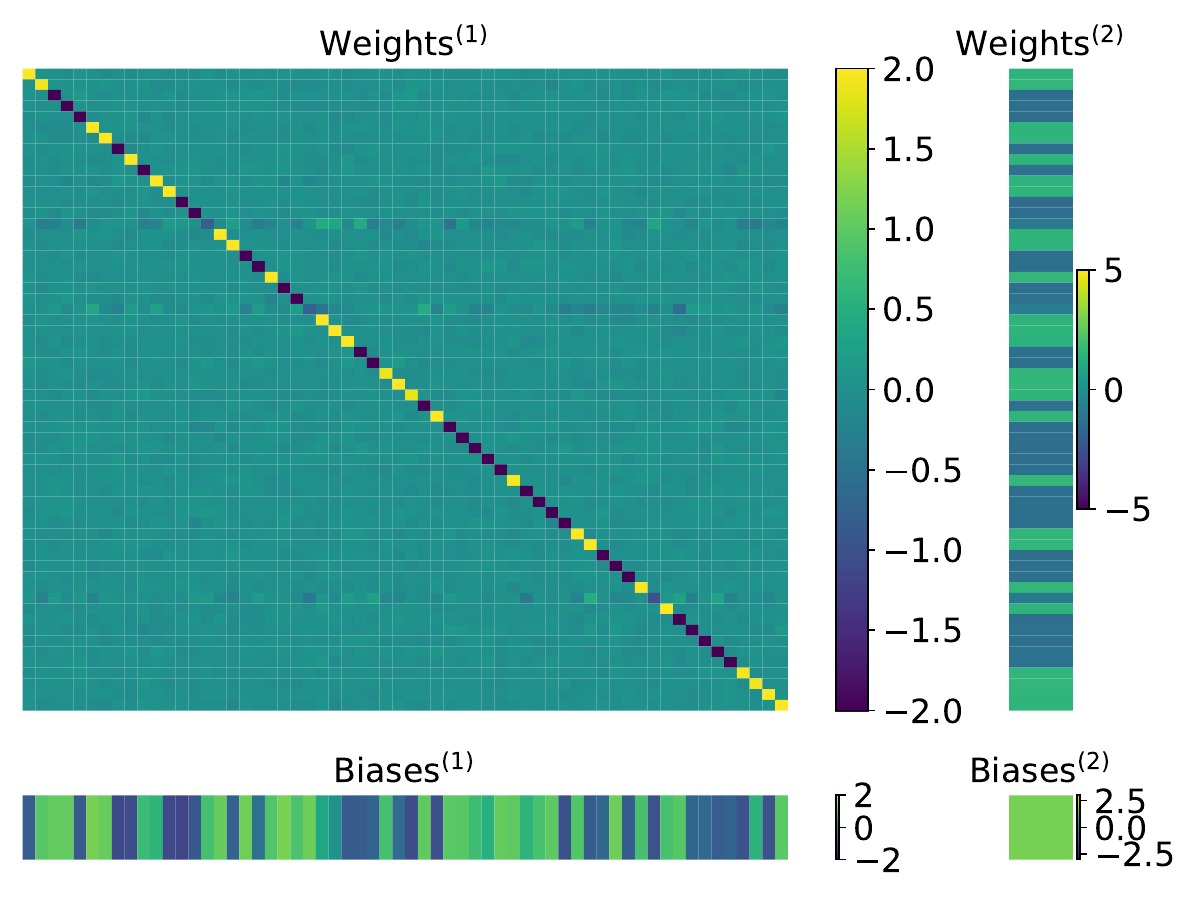}
    \caption{The learned weights of the MLP-based attacker model with an architecture as shown in Figure~\ref{subfig: mlp-multi-node} with $60$ input features. The MLP-based attack model is trained with $200k$ shadow aggregates under the informed attacker.}
    \label{fig: lap_weights_200k}
\end{figure}

To further prove that the MLP-based attack model indeed learns the two-threshold rule in the end, we present the learned weights and biases in Figure~\ref{fig: lap_weights_200k} to show more insights.
For starters, it is obvious that except for the diagonal values in the weight matrix $w_{ij}^{(1)}$ for $ I,j\in[n]$, all weights at other positions are roughly equivalent to $0$. Meanwhile, the bias at the corresponding dimensions~(\ie, $w_{(n+1)j}^{(1)}$ for $j\in[n]$) has the opposite sign as the weight~(\ie, $w_{jj}^{(1)}$ for $j\in[n]$). By comparing the learned parameters with Algorithm~\ref{alg: two-threshold-score} and Theorem~\ref{theo: two-threshold-approximation}, we have that the MLP-based attack model is able to learn the two-threshold rule to infer the membership, which is a better solution in terms of the released aggregates perturbed by the Laplace mechanism, only if large enough amount of training aggregates are provided. Otherwise, it is stuck at the local optima~(one-threshold rule) regardless of the underlying data distributions~(\eg, aggregates perturbed with Laplace noise or Gaussian noise).

In Appendix~\ref{subsec: performance-with-varying-san}, we also study the attack performance with varying size of training shadow aggregates and the corresponding learned weights to show more insights.

\section{Discussion}
Our experiment results shown in Figure~\ref{fig: metric-based-acc-main-200k} and Appendix~\ref{subsec: performance-with-varying-san} indicate that to enable the MLP to learn complex rules, such as the two-threshold attack rule, a large amount of training data is necessary. However, such a large amount of training data is not always available in practice. One of the potential solutions is using generated synthetic data. For instance, Guan \etal~\cite{guan2024zero} have shown that the synthetic location aggregates yield comparable performance in terms of MIAs compared with auxiliary aggregates directly drawn from the same underlying data distributions. Another potentially feasible option is pre-training. For instance, by first pre-training the MLP-based attack model on public data perturbed with the same DP mechanism as the targets and then fine-tuning it with a small amount of auxiliary data, one would achieve optimal performance.

Moreover, we emphasize that our findings in this work is not limited to locations aggregates and potentially apply broadly in scenarios where multiple DP-protected observations are released per individual, such as the allele frequencies as in \cite{dwork2015robust} and the loss trajectories as in \cite{liu2022membershipinferenceattacksexploiting}, by, for example, directly applying the score functions to the allele frequency vectors or concatenation of the loss trajectory from all distilled shadow models.
\section{Related Work}
Various types of membership inference attacks~(MIA) are used under scenarios from summary statistics~\cite{homer2008resolving,dwork2015robust,pyrgelis2018knock,pyrgelis2020measuringmembershipprivacyaggregate, guan2024zero} to machine learning~(ML) models~\cite{shokri2017membership,yeom2018privacy,ye2022enhanced,liu2022membershipinferenceattacksexploiting,zarifzadeh2024low} for privacy auditing purposes. Our work falls into evaluating the average privacy risks of a specific individuals trace over different location data aggregates.
Particularly, two generic methodologies are widely adopted in the literature to perform MIAs, namely the meta-classifier-based attack and the metric-based attack.

The meta-classifier-based attacks cast the membership inference problem into a binary classification problem, where an attack is trained to automatically distinguish non-member targets from member targets. First brought up by Shokri~\etal~\cite{shokri2017membership}, this type of attacks
applies not only to ML models~\cite{yeom2018privacy,salem2018ml,long2020pragmatic,shokri2021privacy,he2021stealing,zhang2022inference,liu2022membershipinferenceattacksexploiting,wu2024link} but also to statistical aggregates releases. One of the typical scenarios is the location aggregates release. \cite{pyrgelis2018knock} uses a multi-layer perceptron trained on shadow aggregates derived from an auxiliary location dataset as the attack model to infer the membership of a target trace. Later on, \cite{guan2024zero} removes the requirement of an auxiliary dataset in \cite{pyrgelis2018knock} by adopting a synthetic data generator and marginal target dataset statistics. Yet, MLP with the same architecture is still leveraged as the attack model. Different from the previous works, this work pinpoints that the typical MLP architecture as an attack model may easily get stuck at the local optima and thus yield a sub-optimal performance, especially when applied to statistical aggregates where multi-observations per individual are released.

Unlike the meta-classifier-based attacks, the metric-based attacks make inferences by calculating and thresholding a score for each target on certain metrics.
Salem~\etal~\cite{salem2018ml} first mentioned in their work that one can distinguish a member record in the training dataset from a non-member one for ML model with a high probability by threshold the posterior of the record. Song~\etal~\cite{song2021systematic} then studied the privacy risks in machine learning models systematically with MIAs that decides the membership of a record by thresholding its prediction entropy. In addition, Yeom~\etal~\cite{yeom2018privacy}, followed by Carlini~\etal~\cite{carlini2022membership}, Ye~\etal~\cite{ye2022enhanced}, and Zarifzadeh~\etal~\cite{zarifzadeh2024low}, perform MIAs against ML models by thresholding the loss of a given target record. Different from their works, this work focuses on developing high power MIAs against location aggregates where multiple observations are sequentially revealed for each individual involved. This multi-observation feature of location aggregates poses extra challenges on how these observations are combined to maximize an adversary's attacking power.

The most relevant work to ours is from Dwork~\etal~\cite{dwork2015robust}, who proposed a metric-based MIA against genomics statistics. Assuming that multiple allele frequencies from one individual are released, an attacker first calculates the vector similarity between the target and the released frequencies. Then, they randomly draw a reference record from the underlying data distribution and compute its similarity to the released frequencies. The target record is labeled as a non-member if the two similarity scores are close enough. Otherwise, it is regarded as a member. Different from their work, first of all, we use multiple reference records rather than one, which benefits the attack performance with a smaller variance. Second of all, rather than a privacy mechanism agnostic approach, we apply different attack methodologies to different underlying data distributions~(\eg, the two-threshold attack for the Laplace mechanism and the one-threshold attack for the Gaussian mechanism), providing better performance.


\section{Conclusion}
MLP-based MIA has been widely adopted as a default privacy auditing tool in numerous scenarios.
In this work, we have proposed two metric-based attacks against DP-protected dataset containing multiple observations per individual, \ie, the one-threshold attack and the two-threshold attack, and compared their performance with one of the previous MLP-based attack models adopted in the literature. It has been revealed by the comparison results that the previous MLP-based attack models only learned the one-threshold attack rule, leading to a sub-optimality for the Laplace DP noise.
After theoretically proving that the MLPs are capable of encoding the more complex and better rules, such as the two-threshold attack rule, we confirmed empirically that MLPs can learn these rules when substantial training data is given.
For cases where such a large amount of training data is unavailable, we further suggest adopting techniques, such as pretraining or synthetic data generation, as practical solutions to ensure the MLP works at its best.
\bibliographystyle{IEEEtran}
\bibliography{sample}
\appendix
\section{Threshold Estimation with Fixed Error}\label{subsec: threshold-fixed-error}
\begin{algorithm}
    \caption{\texttt{ThresholdEstimation}. Estimate the threshold $T$ for the one-threshold attack given a fixed attack error $\alpha$.}\label{alg: one-threshold-threshold-fix-error}
    \LinesNumbered
    \SetKwInOut{Input}{Input}
    \SetKwInOut{Output}{Output}
    \Input{The shadow aggregates $\mathcal{A}_{sd}$, the target trace $z$, and a given attack error $\alpha$.}
    \Output{A threshold $T$.}
    $\mathcal{A}_{sd}^{1}\leftarrow\{\bar{A}^{1},\bar{A}^{2},\ldots,\bar{A}^{\lfloor m/2\rfloor}\}\subset\mathcal{A}_{sd}$\tcp*{Get all member shadow aggregates}
    $\mathcal{A}_{sd}^{0}\leftarrow\mathcal{A}_{sd}\backslash \mathcal{A}_{sd}^{1}$\tcp*{Get all non-member shadow aggregates}
   
    $\hat{\mu}_{1}\leftarrow\texttt{MEAN}_{\bar{A}^{i}\in\mathcal{A}_{sd}^{1}}\left(
    \texttt{Score}\left(
    \bar{A}^{i},A_{\beta}=0^{u\times w},z
    \right)
    \right)$\;\label{line: params_estimation_start}
    $\hat{\mu}_{0}\leftarrow\texttt{MEAN}_{\bar{A}^{i}\in\mathcal{A}_{sd}^{0}}\left(
   \texttt{Score}\left(
    \bar{A}^{i},A_{\theta}=0^{u\times w},z
    \right)
    \right)$\tcp*{Mean estimation with Algorithm~\ref{alg: one-threshold-score}}

    $\hat{\sigma}_{1}\leftarrow\texttt{VAR}_{\bar{A}^{i}\in\mathcal{A}_{sd}^{1}}\left(
    \texttt{Score}\left(
    \bar{A}^{i},A_{\theta}=0^{u\times w},z
    \right)
    \right)$\;
    $\hat{\sigma}_{0}\leftarrow\texttt{VAR}_{\bar{A}^{i}\in\mathcal{A}_{sd}^{0}}\left(
   \texttt{Score}\left(
    \bar{A}^{i},A_{\theta}=0^{u\times w},z
    \right)
    \right)$\tcp*{Variance estimation with Algorithm~\ref{alg: one-threshold-score}}
    
    $\hat{\sigma}=\frac{1}{2}\left(\hat{\sigma}_0+\hat{\sigma}_1\right)$\tcp*{Variance estimation}
    $\bar{\mathbbm{P}}:=\bar{N}\left(\bar{\mu}_{0},\bar{\sigma}\right)$\;
    $\bar{\mathbbm{Q}}:=\bar{N}\left(\bar{\mu}_{1},\bar{\sigma}\right)$\tcp*{Score distribution estimation}\label{line: params_estimation_end}
    $T\leftarrow\texttt{CDF}_{\bar{\mathbbm{P}}}^{-1}\left(1-\alpha\right)$\tcp*{Calculate the threshold given $\alpha$}
    \Return{T}
\end{algorithm}

\begin{algorithm}
    \caption{\texttt{PerCellThresholdEstimation}. Estimate the threshold for each positive observation given a set of fixed attack errors $\alpha=\{\alpha_{le}\vert l\in L,e\in E\}$.}\label{alg: two-threshold-subthreshold-fixed-error}
    \LinesNumbered
    \SetKwInOut{Input}{Input}
    \SetKwInOut{Output}{Output}
    \Input{The shadow aggregate $\mathcal{A}_{sd}$ and the target trace $z$.}
    \Output{A threshold set $\hat{\mathcal{T}}^{\prime}$.}
    $\hat{\mathcal{T}}^{\prime}\leftarrow\phi$\;
    $\mathcal{A}_{sd}^{1}\leftarrow\{\bar{A}^{1},\bar{A}^{2},\ldots,\bar{A}^{\lfloor m/2\rfloor}\}\subset\mathcal{A}_{sd}$\label{line: member-threshold-one-fixed}\tcp*{Get all member shadow aggregates}
    $\mathcal{A}_{sd}^{0}\leftarrow\mathcal{A}_{sd}\backslash \mathcal{A}_{sd}^{1}$\label{line: non-member-threshold-one-fixed}\tcp*{Get all non-member shadow aggregates}
    \For{$l\in L$}{
    \For{$e\in E$}{
    $\bar{\mathbbm{P}}:=\texttt{Hist}_{\bar{A}^{i}\in\mathcal{A}_{sd}^{0}}\left(\bar{A}^{i}_{le}\right)$\;\label{line: two-histogram-nmem}
    $\bar{\mathbbm{Q}}:=\texttt{Hist}_{\bar{A}^{i}\in\mathcal{A}_{sd}^{1}}\left(\bar{A}^{i}_{le}\right)$\;\label{line: two-histogram-mem}
    $T_{le}\leftarrow\texttt{CDF}_{\bar{P}}^{-1}\left(1-\alpha_{le}\right)$\;\label{line: two-cdf}
    $\mathcal{T}^{\prime}\leftarrow\mathcal{T}^{\prime}\cup T_{le}$\;
    }
    }
    \Return{$\mathcal{T}^{\prime}$}
\end{algorithm}
Algorithm~\ref{alg: one-threshold-threshold-fix-error} and Algorithm~\ref{alg: two-threshold-subthreshold-fixed-error} show the method of estimating threshold given a fixed error for the one-threshold attack and the sub-threshold for the two-threshold attack, respectively.

As mentioned in Section~\ref{subsec: one-threshold}, Algorithm~\ref{alg: one-threshold-threshold-fix-error} first estimates the means and the standard deviations for both the member and non-member shadow aggregates. Then it fits the estimated parameters into the Gaussian distributions $\mathbbm{P}$ and $\mathbbm{Q}$, respectively, as shown from Line~\ref{line: params_estimation_start} to Line~\ref{line: params_estimation_end}. After estimating the CDF of $\mathbbm{P}$, we are able to derive the threshold given a fixed error $\alpha$ accordingly.

Similarly, Algorithm~\ref{alg: two-threshold-subthreshold-fixed-error} derives $\mathbbm{P}$ and $\mathbbm{Q}$ by estimating histograms first. Then, smooth techniques are exploited as also described in \cite{ye2022enhanced} to estimate the CDF as shown from Line~\ref{line: two-histogram-nmem} to Line~\ref{line: two-cdf}.
\section{MLP Approximation}\label{subsec: mlp-approximation}
\begin{figure}[htbp!]
    \centering
    \subfigure[Impact of $a$.]{
    \includegraphics[width=0.4\linewidth]{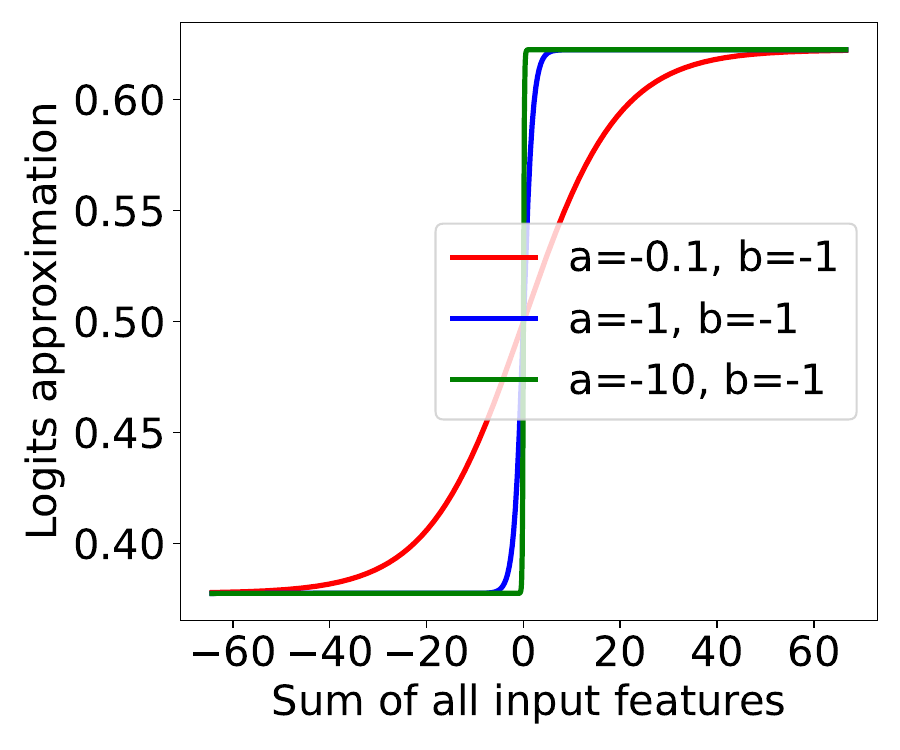}\label{subfig: impact-of-a}
    }
    \hspace{0.5cm}
    \subfigure[Impact of $b$.]{
    \includegraphics[width=0.4\linewidth]{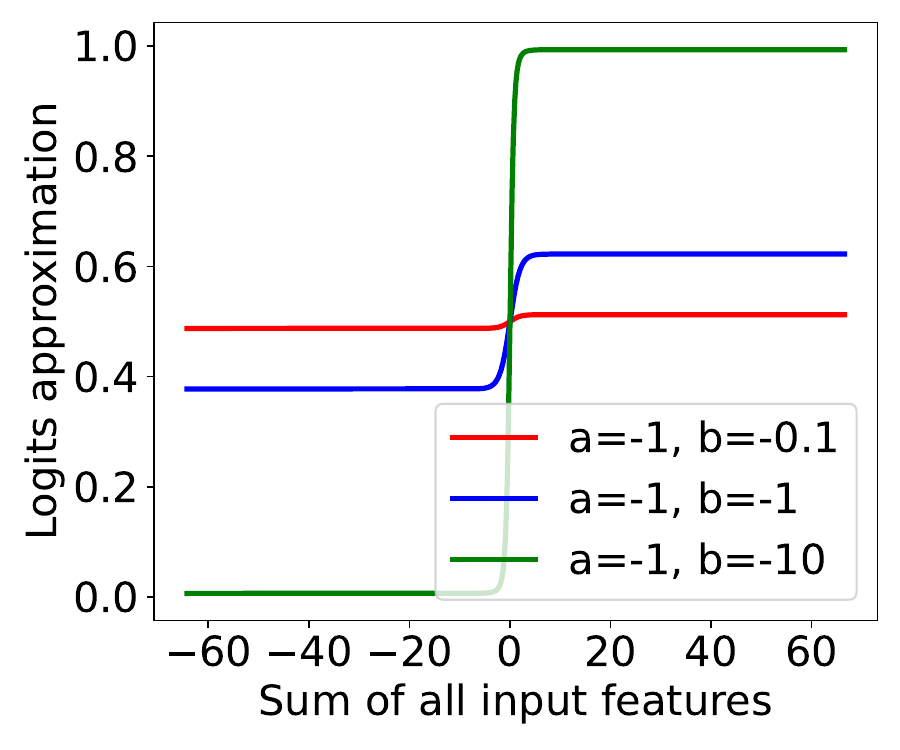}\label{subfig: impact-of-b}
    }
    \caption{The impact of value $a$ and $b$ in Theorem~\ref{theo: one-threshold-approximation}.}\label{fig: mlp-approximation}
\end{figure}

In Figure~\ref{fig: mlp-approximation}, we demonstrate the impact of the value $a$ and $b$ on the approximation result in Theorem~\ref{theo: one-threshold-approximation}. As the impact of the values on the approximation results in Theorem~\ref{theo: two-threshold-approximation} demonstrates a similar pattern as in Theorem\ref{theo: one-threshold-approximation}, we omit the corresponding results here. In general, large $a$ and $b$ lead to a better approximation result~(\ie, a smaller loss). Though the approximation loss keeps decreasing as the increase of $a$ and $b$, by setting an appropriate threshold, the accuracy will not increase much after reaching a certain level.
\section{AUC and ROC for Laplace Mechanism}\label{subsec: auc-roc}
\begin{figure}[htbp!]
    \centering
    \subfigure[Informed attacker with $2k$ shadow aggregates.]{
    \includegraphics[width=0.4\linewidth]{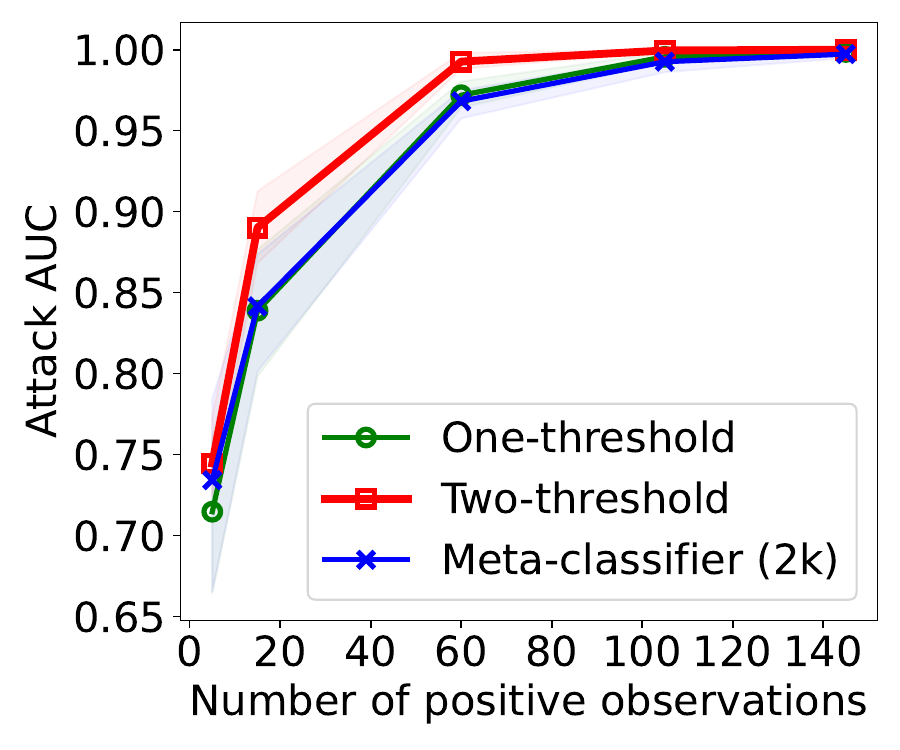}\label{subfig: metric-based-auc-fk-2k}
    }
    \subfigure[Auxiliary attacker with $2k$ shadow aggregates.]{
    \includegraphics[width=0.4\linewidth]{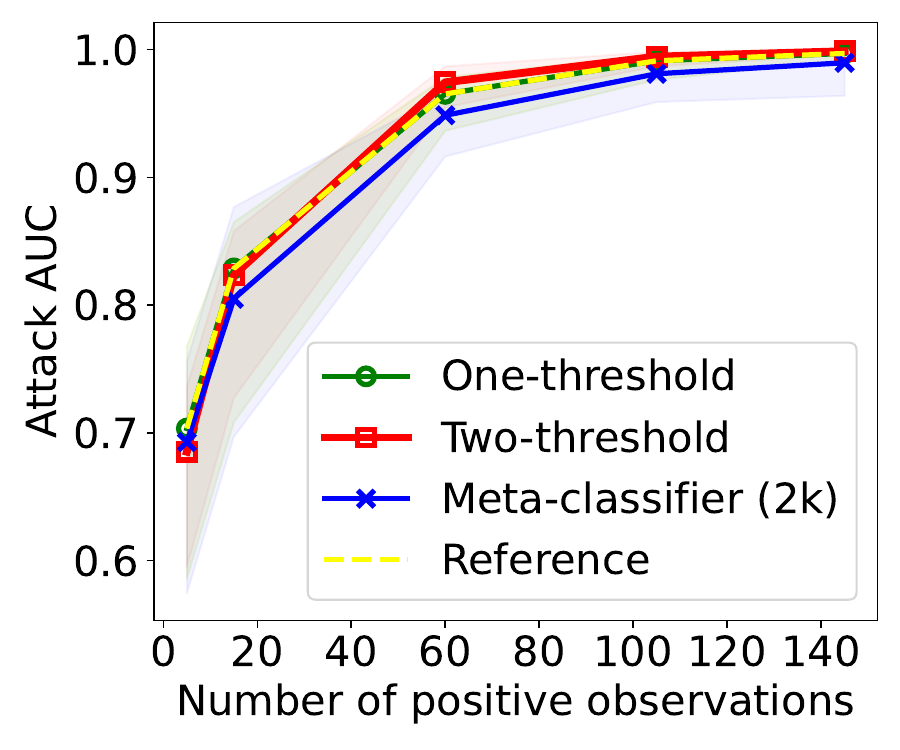}\label{subfig: metric-based-auc-pk-2k}
    }
    \subfigure[Informed attacker with $200k$ shadow aggregates.]{
    \includegraphics[width=0.4\linewidth]{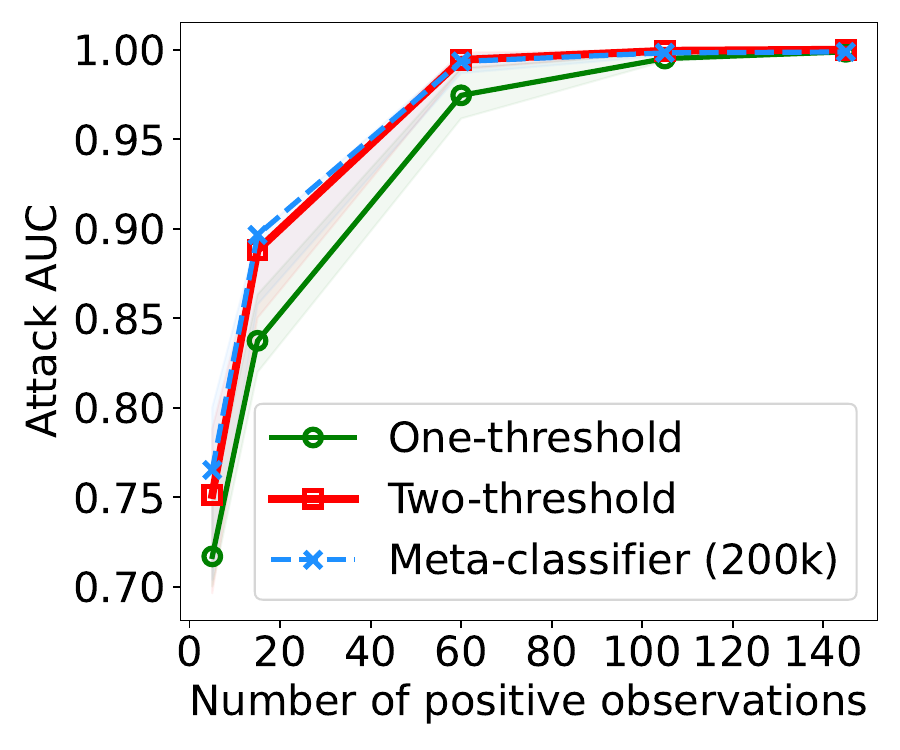}\label{subfig: metric-based-auc-fk-200k}
    }
    \subfigure[Auxiliary attacker with $200k$ shadow aggregates.]{
    \includegraphics[width=0.4\linewidth]{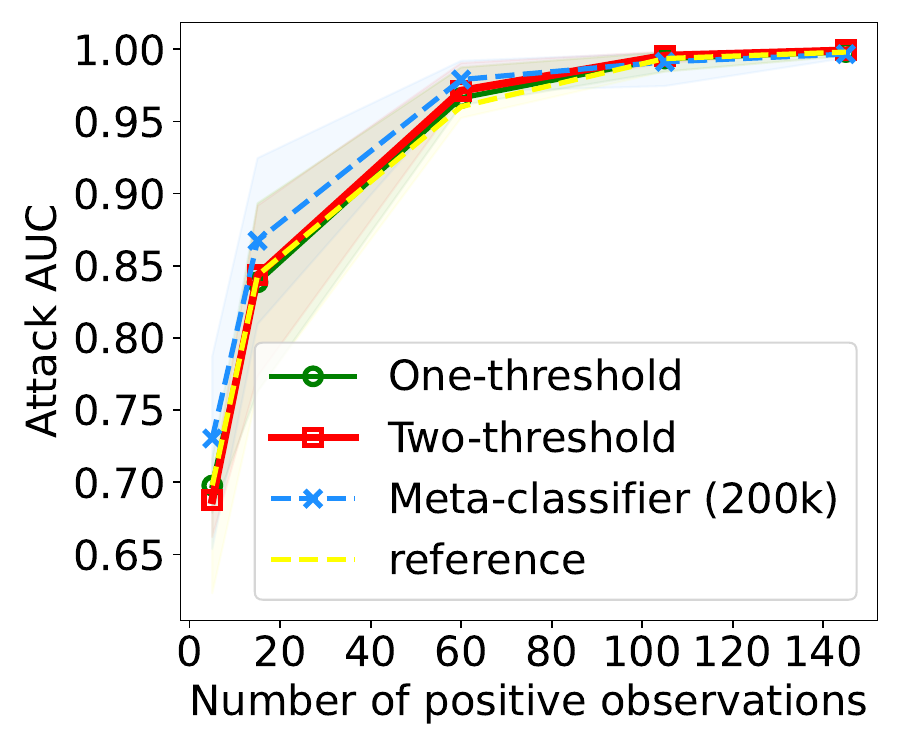}\label{subfig: metric-based-auc-pk-200k}
    }
    \caption{Attack AUC with increasing number of positive observations under the Laplace mechanism.}\label{fig: metric-based-auc-lap}
\end{figure}

\begin{figure}[htbp!]
    \centering
    \subfigure[Informed attacker with $2k$ shadow aggregates.]{
    \includegraphics[width=0.4\linewidth]{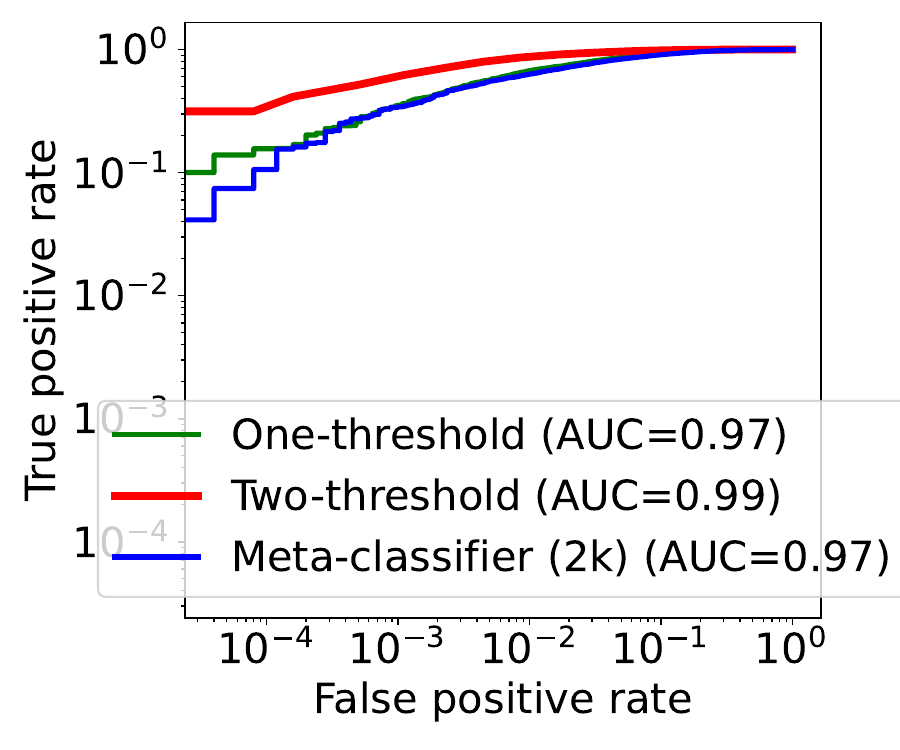}\label{subfig: metric-based-roc-fk-2k}
    }
    \subfigure[Auxiliary attacker with $2k$ shadow aggregates.]{
    \includegraphics[width=0.4\linewidth]{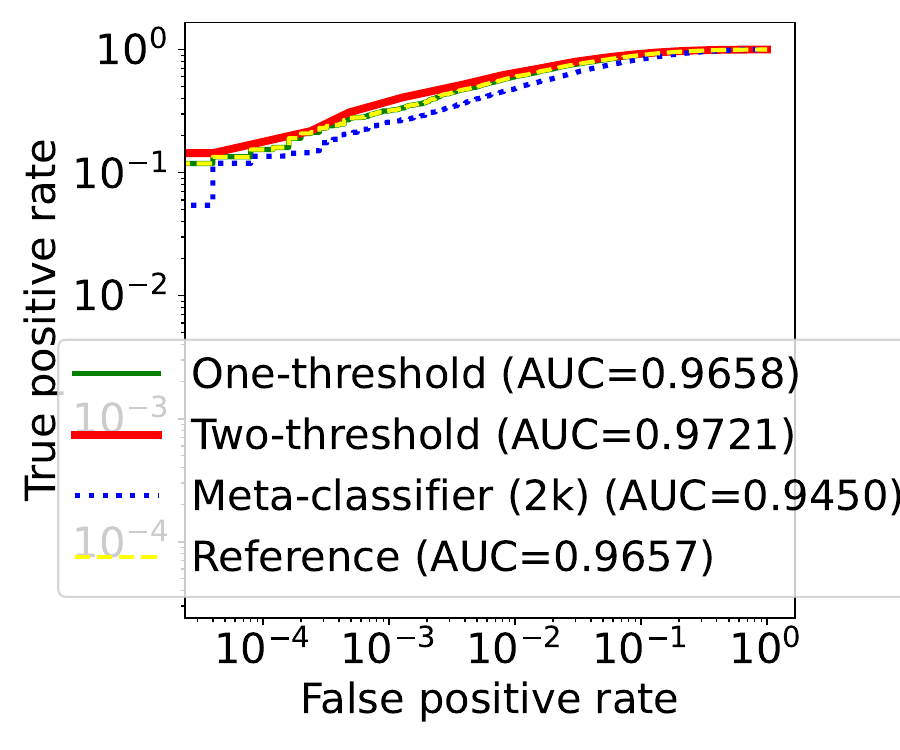}\label{subfig: metric-based-roc-pk-2k}
    }
    \subfigure[Informed attacker with $200k$ shadow aggregates.]{
    \includegraphics[width=0.4\linewidth]{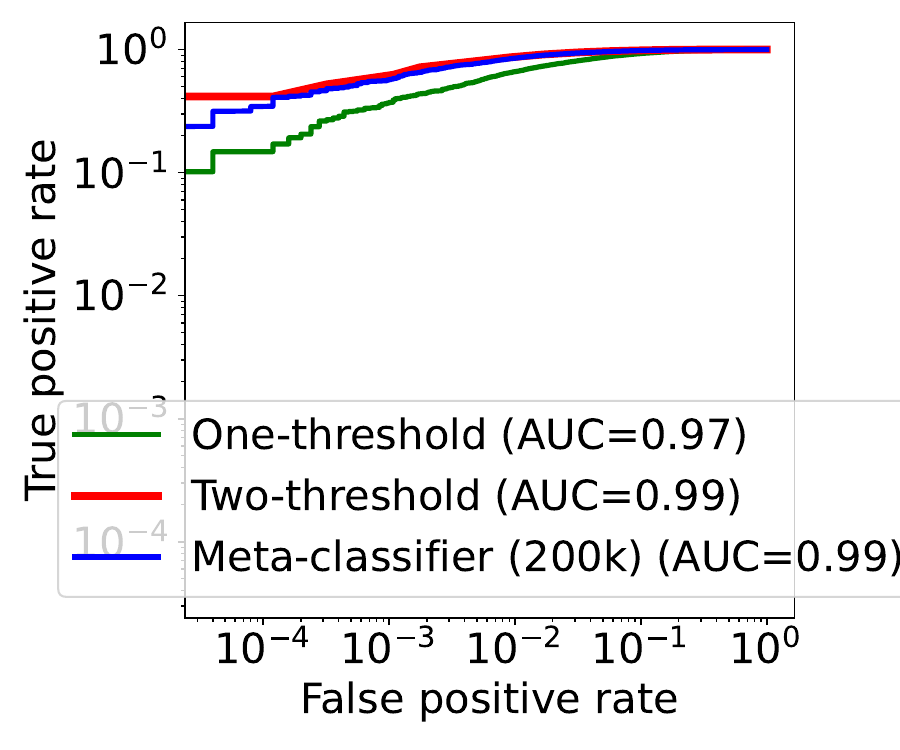}\label{subfig: metric-based-roc-fk-200k}
    }
    \subfigure[Auxiliary attacker with $200k$ shadow aggregates.]{
    \includegraphics[width=0.4\linewidth]{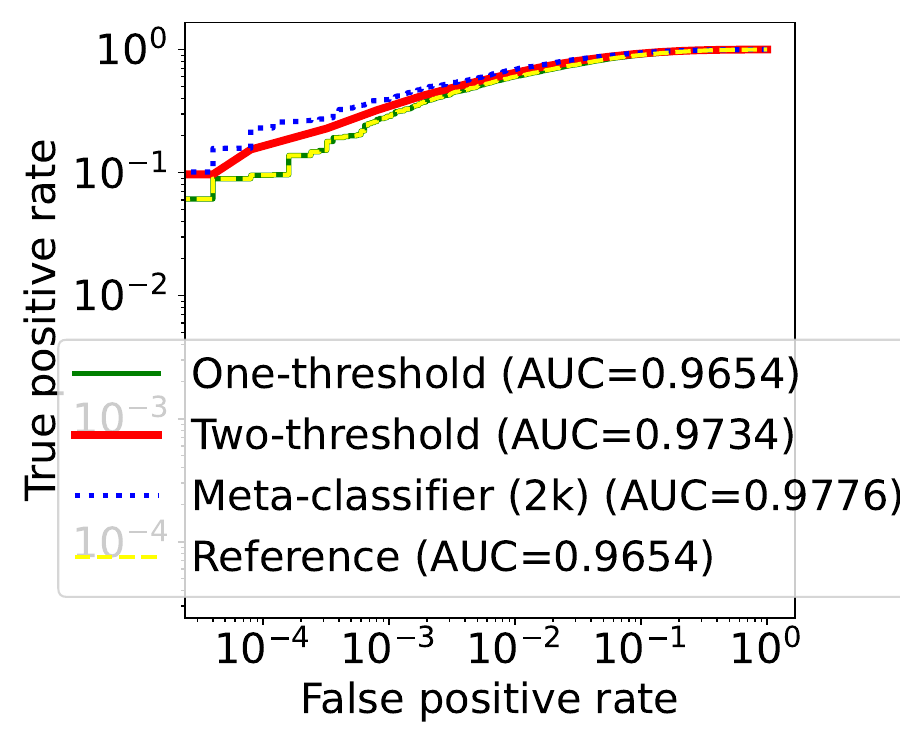}\label{subfig: metric-based-roc-pk-200k}
    }
    \caption{Attack ROC with increasing number of positive observations under the Laplace mechanism.}\label{fig: metric-based-roc-lap}
\end{figure}

Figure~\ref{fig: metric-based-auc-lap} and Figure~\ref{fig: metric-based-roc-lap} present the attack AUC and the ROC for both informed attackers and auxiliary attackers under the Laplace mechanism.

While the AUC demonstrates the same trends as the accuracy shown in Figure~\ref{fig: compare-attack-acc}, the ROC further demonstrates the advantage of the two-threshold attack over the one-threshold attacker for the Laplace mechanism under small false positive rates. In addition, we further prove that by increasing the training dataset size from $2k$ to $200k$, the MLP-based attack model is able to learn the two-threshold rule.

Aside from this, we compare the reference attack~\cite{dwork2015robust} with others in terms of AUC and ROC. In essence, the referenced attack distinguishes members from non-members by first computing two inner products, the target vector, and the released aggregate, and the non-member~(reference) vector and the released aggregate. Then, the distance between the two inner products is computed. If the distance is large, the target vector is inferred as a member. Otherwise, it is inferred as a non-member.

When the reference vector is orthogonal with the target vector, the reference attack is equivalent to our proposed one-threshold attack. Otherwise, it is harder to distinguish members and non-members as the target vector and the reference vector growing similarly to each other compared to the one-threshold attack. Therefore, the reference attack demonstrates a very slight disadvantage over the one-threshold attack.

\section{Attack Performance for Gaussian Mechanism}\label{subsec: gau-performance}
\begin{figure}[htbp!]
    \centering
    \subfigure[Informed attacker with $2k$ shadow aggregates.]{
    \includegraphics[width=0.4\linewidth]{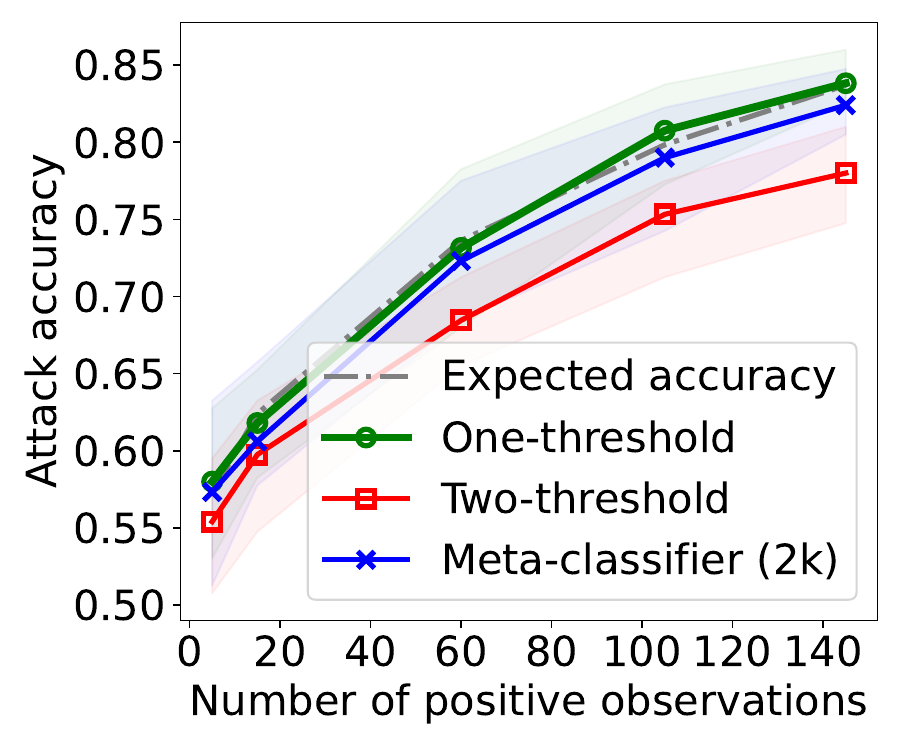}\label{subfig: metric-based-acc-fk-2k-gau}
    }
    \subfigure[Auxiliary attacker with $2k$ shadow aggregates.]{
    \includegraphics[width=0.4\linewidth]{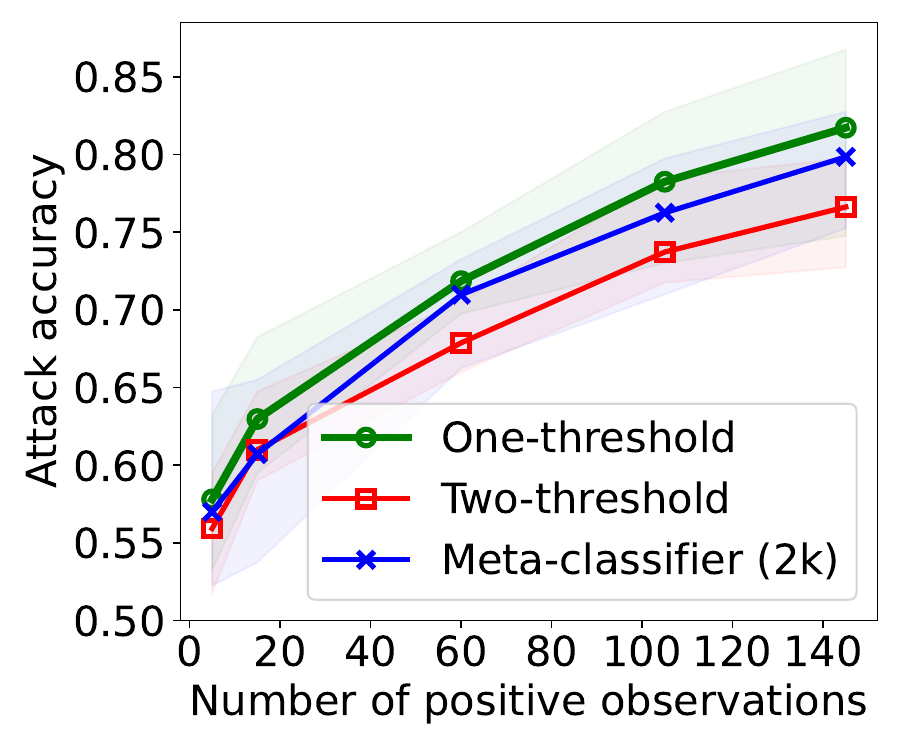}\label{subfig: metric-based-acc-pk-2k-gau}
    }
    \subfigure[Informed attacker with $200k$ shadow aggregates.]{
    \includegraphics[width=0.4\linewidth]{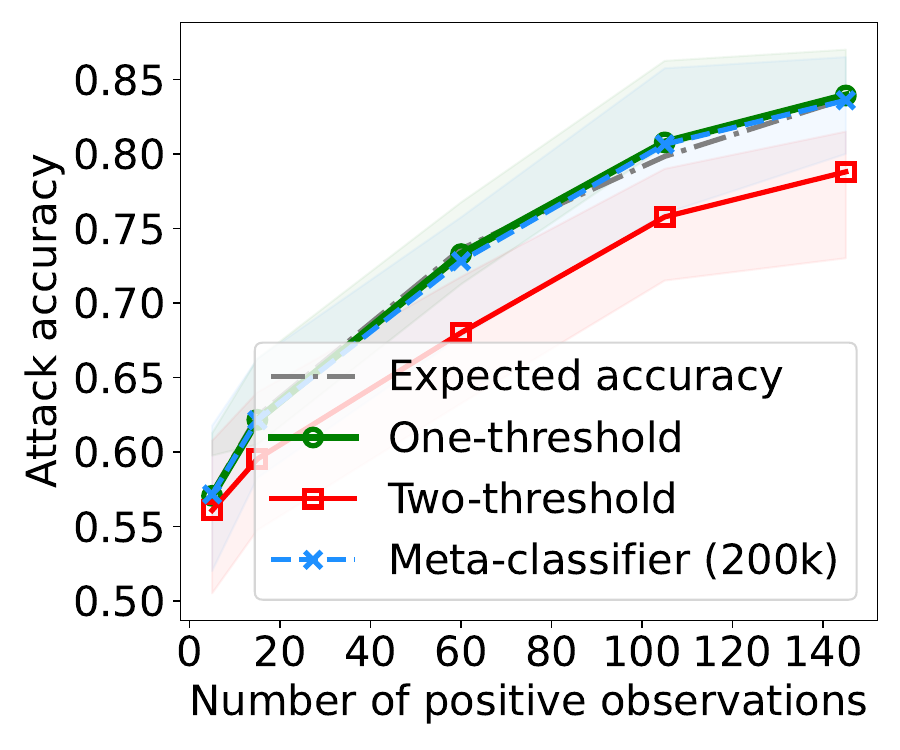}\label{subfig: metric-based-acc-fk-200k-gau}
    }
    \subfigure[Auxiliary attacker with $200k$ shadow aggregates.]{
    \includegraphics[width=0.4\linewidth]{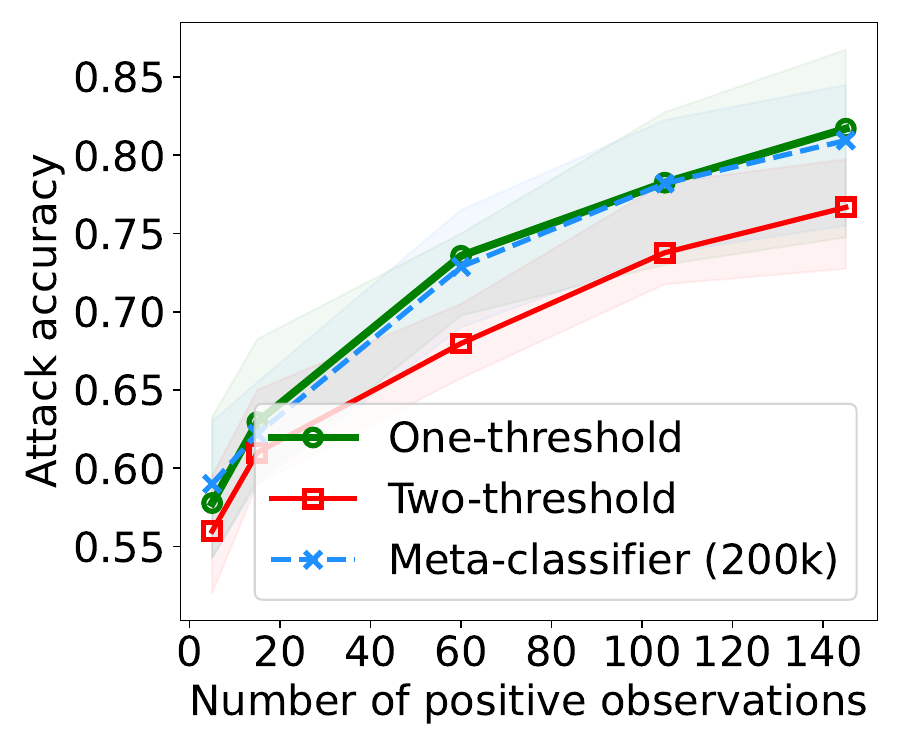}\label{subfig: metric-based-acc-pk-200k-gau}
    }
    \caption{Attack accuracy with increasing number of positive observations under the Gaussian mechanism.}\label{fig: metric-based-acc-gau}
\end{figure}

\begin{figure}[htbp!]
    \centering
    \subfigure[Informed attacker with $2k$ shadow aggregates.]{
    \includegraphics[width=0.4\linewidth]{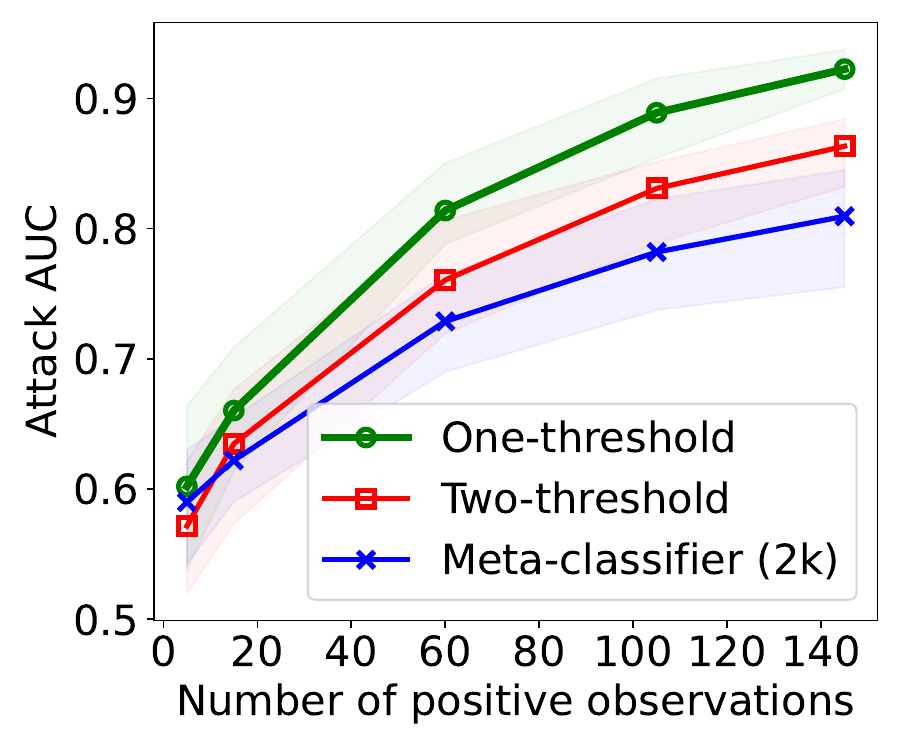}\label{subfig: metric-based-auc-fk-2k-gau}
    }
    \subfigure[Auxiliary attacker with $2k$ shadow aggregates.]{
    \includegraphics[width=0.4\linewidth]{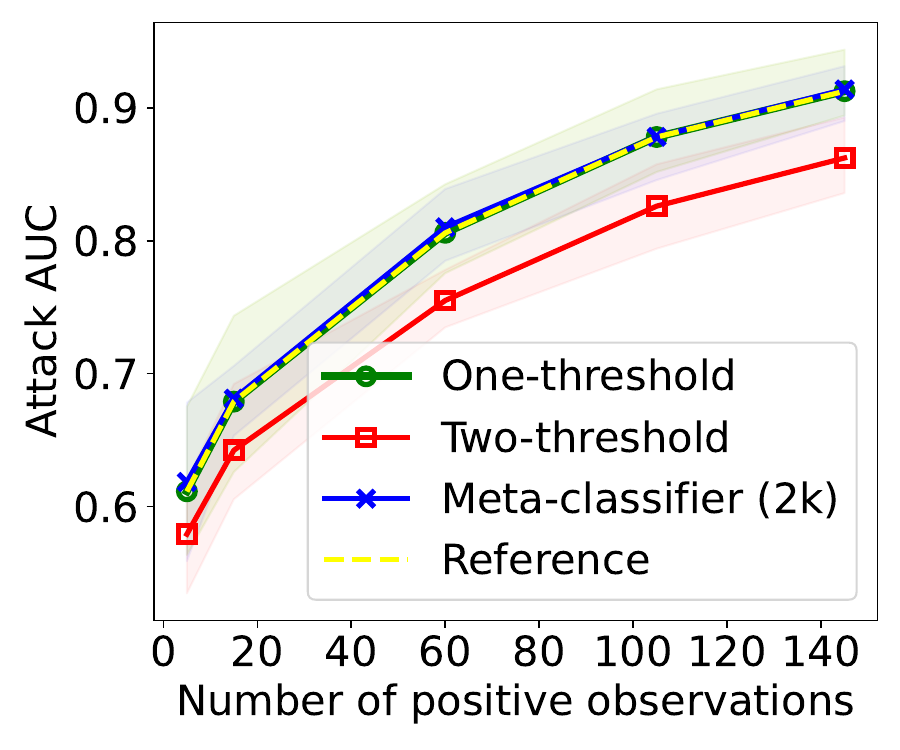}\label{subfig: metric-based-auc-pk-2k-gau}
    }
    \subfigure[Informed attacker with $200k$ shadow aggregates.]{
    \includegraphics[width=0.4\linewidth]{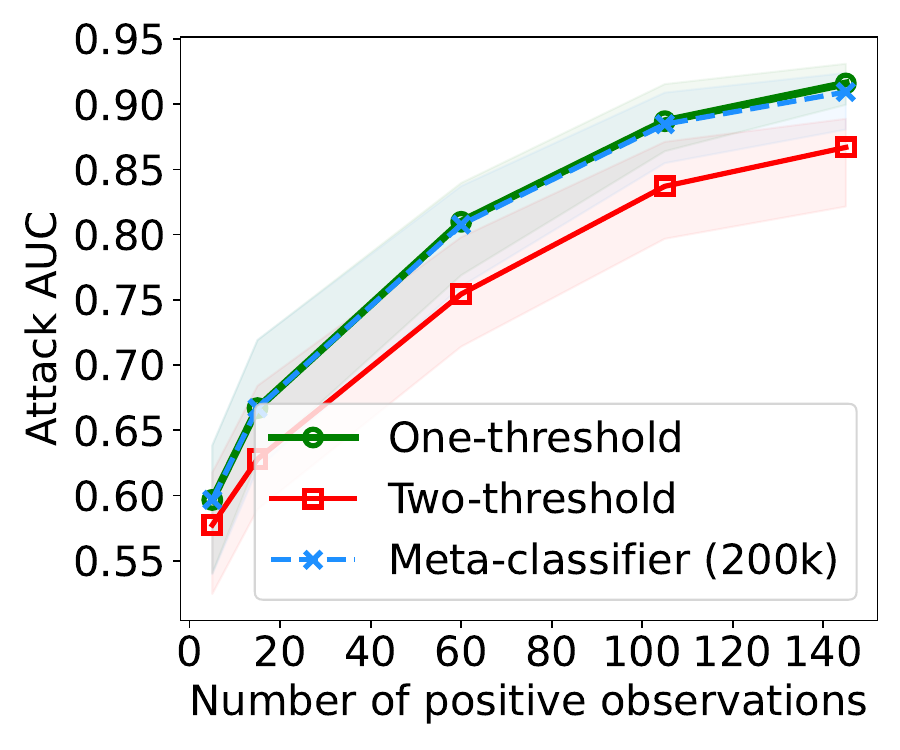}\label{subfig: metric-based-auc-fk-200k-gau}
    }
    \subfigure[Auxiliary attacker with $200k$ shadow aggregates.]{
    \includegraphics[width=0.4\linewidth]{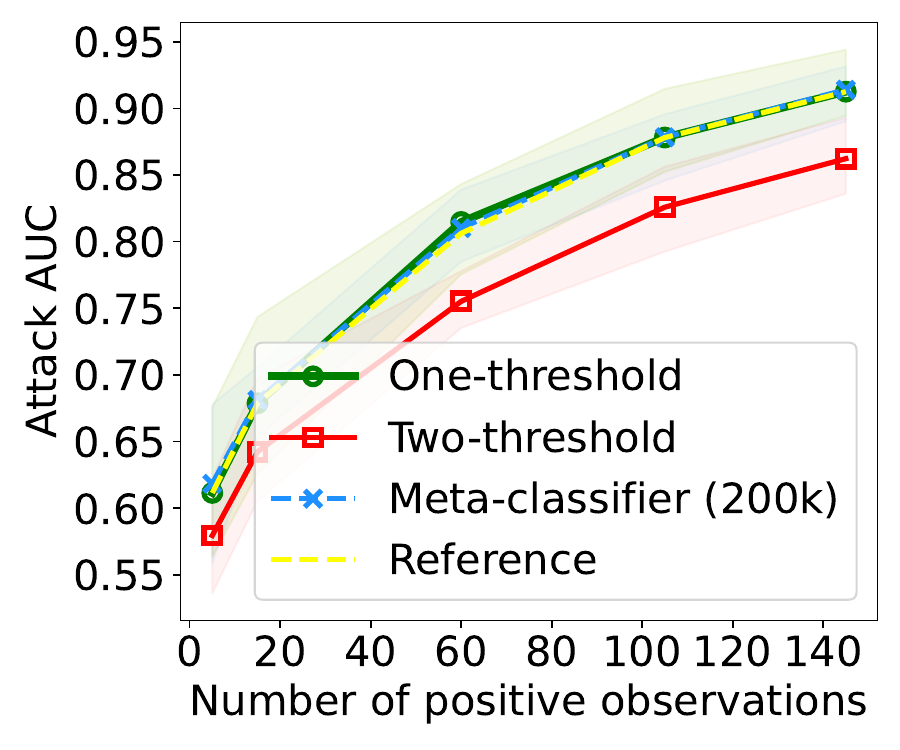}\label{subfig: metric-based-auc-pk-200k-gau}
    }
    \caption{Attack AUC with increasing number of positive observations under the Gaussian mechanism.}\label{fig: metric-based-auc-gau}
\end{figure}

\begin{figure}[htbp!]
    \centering
    \subfigure[Informed attacker with $2k$ shadow aggregates.]{
    \includegraphics[width=0.4\linewidth]{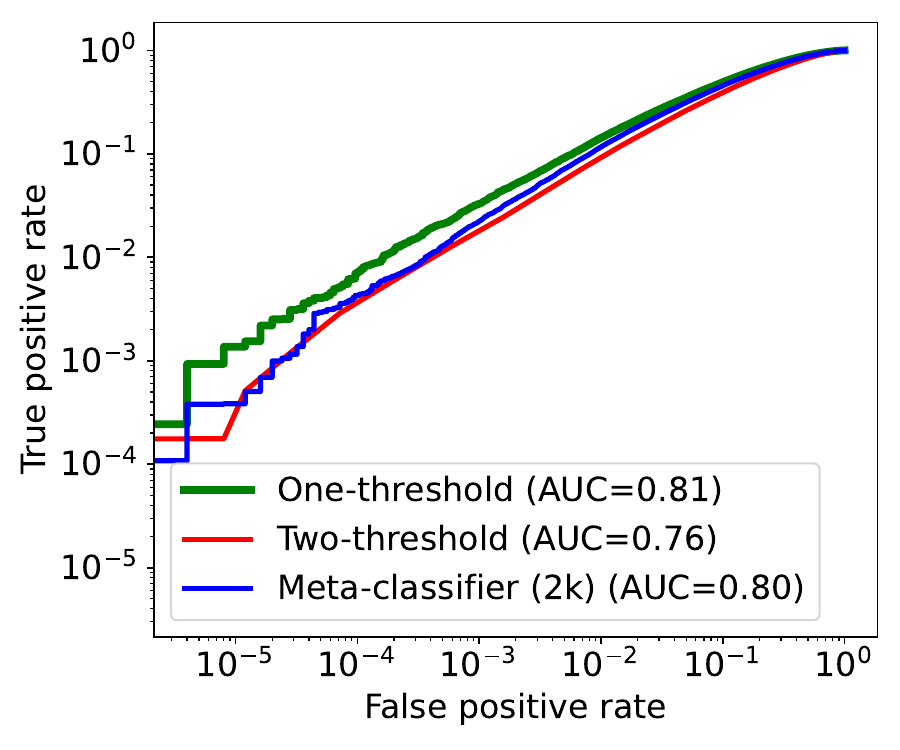}\label{subfig: metric-based-roc-fk-2k-gau}
    }
    \subfigure[Auxiliary attacker with $2k$ shadow aggregates.]{
    \includegraphics[width=0.4\linewidth]{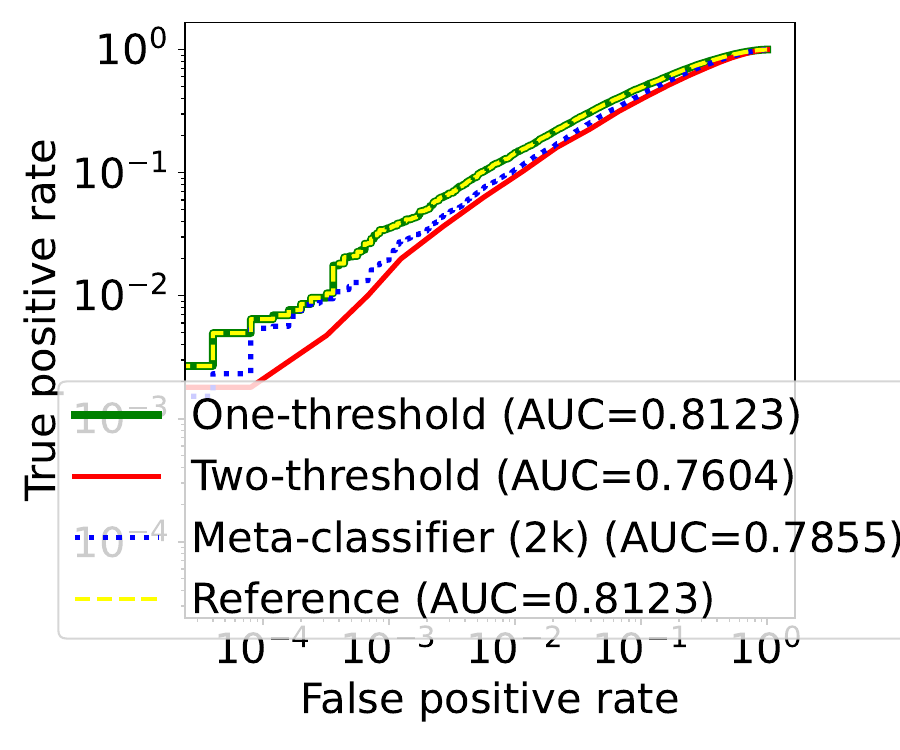}\label{subfig: metric-based-roc-pk-2k-gau}
    }
    \subfigure[Informed attacker with $200k$ shadow aggregates.]{
    \includegraphics[width=0.4\linewidth]{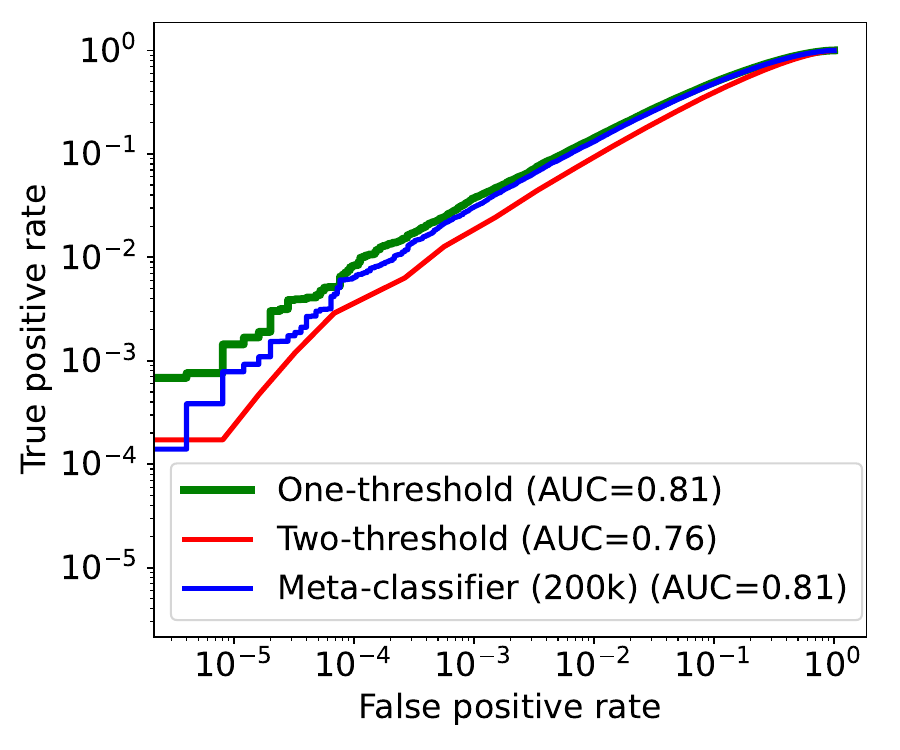}\label{subfig: metric-based-roc-fk-200k-gau}
    }
    \subfigure[Auxiliary attacker with $200k$ shadow aggregates.]{
    \includegraphics[width=0.4\linewidth]{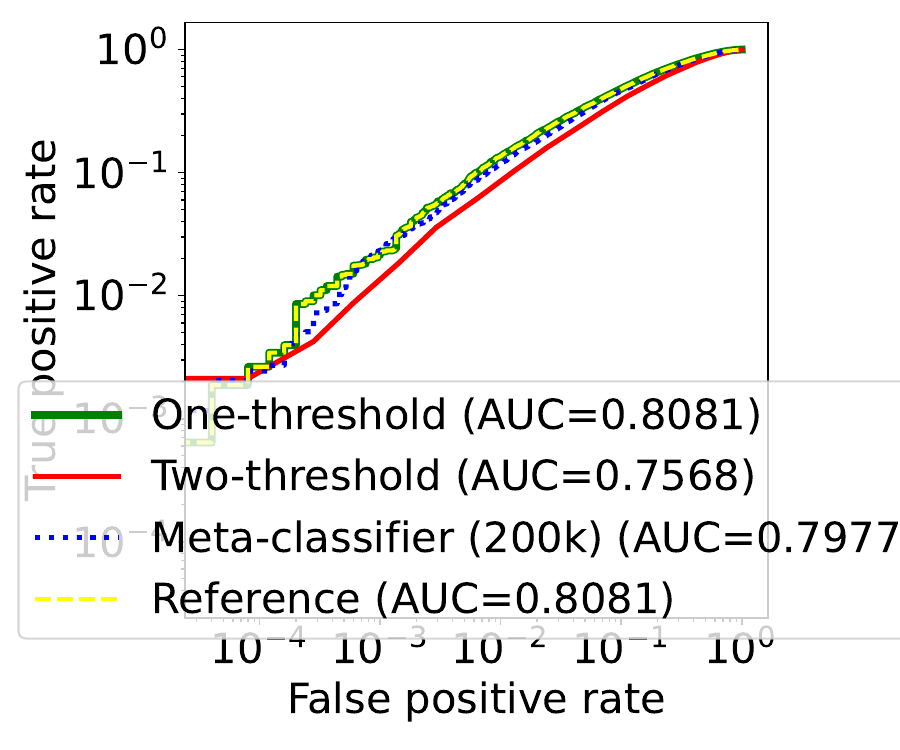}\label{subfig: metric-based-roc-pk-200k-gau}
    }
    \caption{Attack ROC with increasing number of positive observations under the Gaussian mechanism.}\label{fig: metric-based-roc-gau}
\end{figure}

In this section, we present the attack performance of all compared baselines in terms of accuracy, AUC, and AOC. As discussed in Section~\ref{subsubsec: comparison}, for the released aggregates perturbed by Gaussian noise, the one-threshold attack demonstrates a clear advantage over the two-threshold attack. In addition, as the MLP-based attack model makes it easier to learn the one-threshold rule compared to the two-threshold rule, it is thus easier to achieve the optimal performance in terms of the Gaussian noise perturbed aggregates with a relatively smaller training dataset size.

\section{Attack Performance with Varying Shadow Aggregates Number}\label{subsec: performance-with-varying-san}
\begin{figure}[htbp!]
    \centering
    \subfigure[Accuracy with informed attacker.]{
    \includegraphics[width=0.4\linewidth]{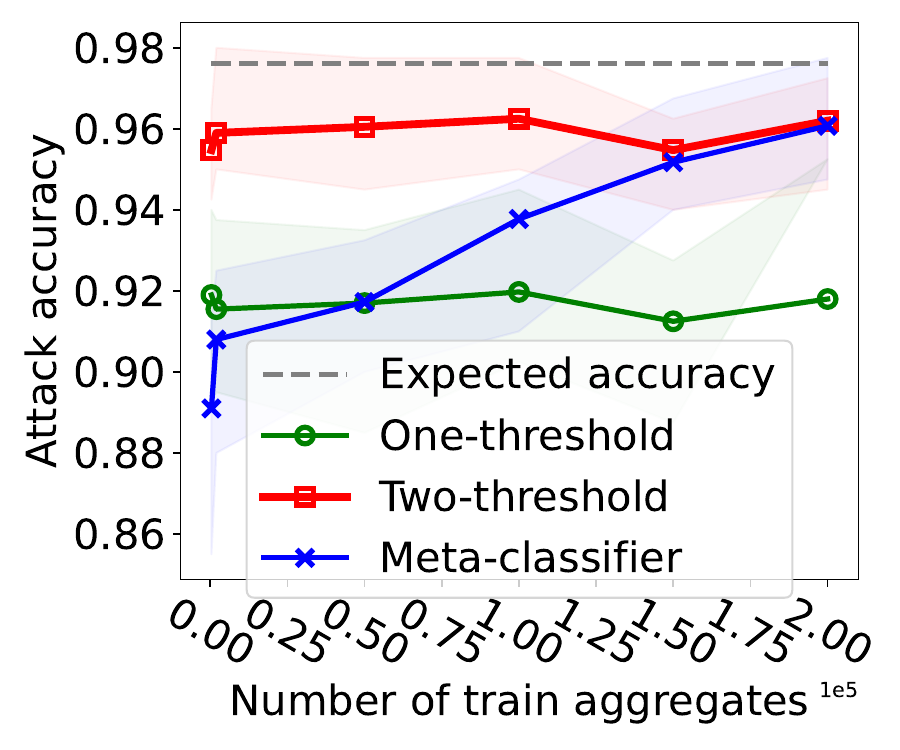}\label{subfig: metric-based-acc-varying-size-fk}
    }
    \subfigure[Accuracy with auxiliary attacker.]{
    \includegraphics[width=0.4\linewidth]{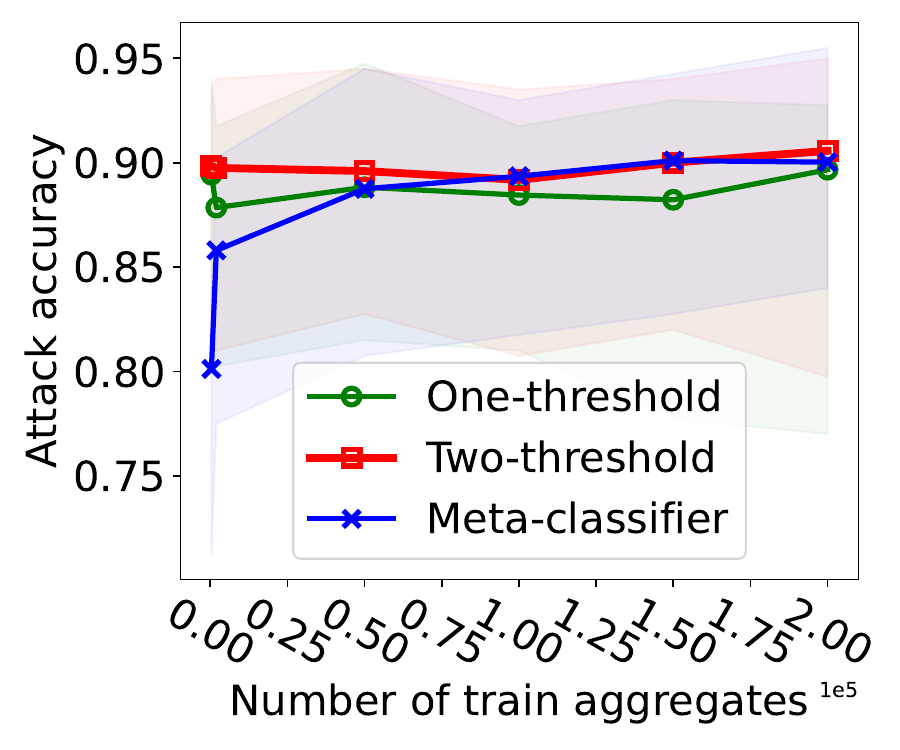}\label{subfig: metric-based-acc-varying-size-pk}
    }
    \subfigure[AUC with informed attacker.]{
    \includegraphics[width=0.4\linewidth]{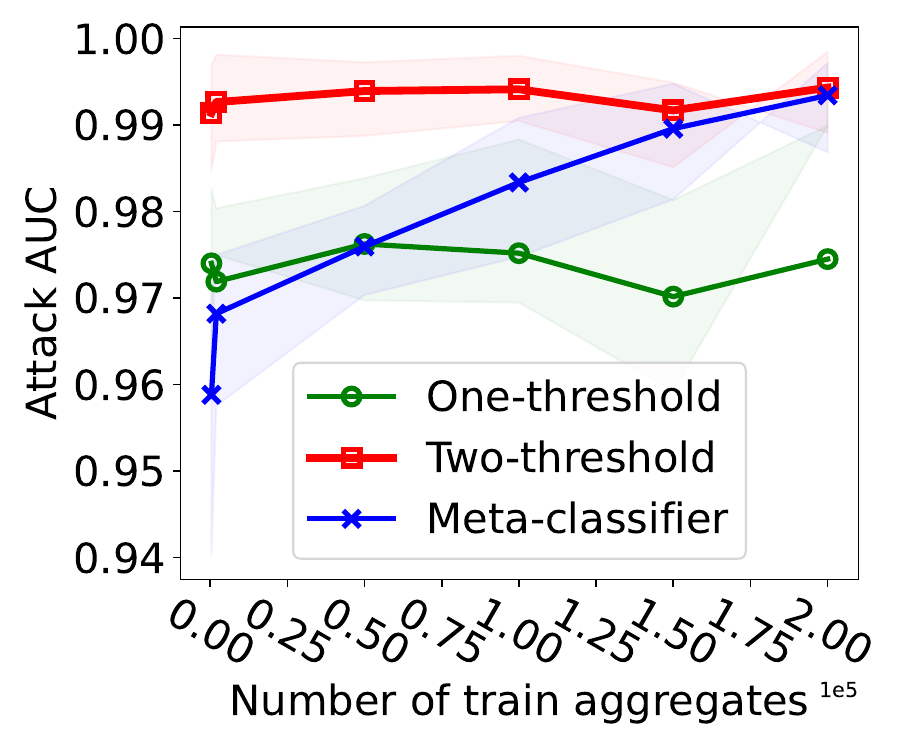}\label{subfig: metric-based-auc-varying-size-fk}
    }
    \subfigure[AUC with auxiliary attacker.]{
    \includegraphics[width=0.4\linewidth]{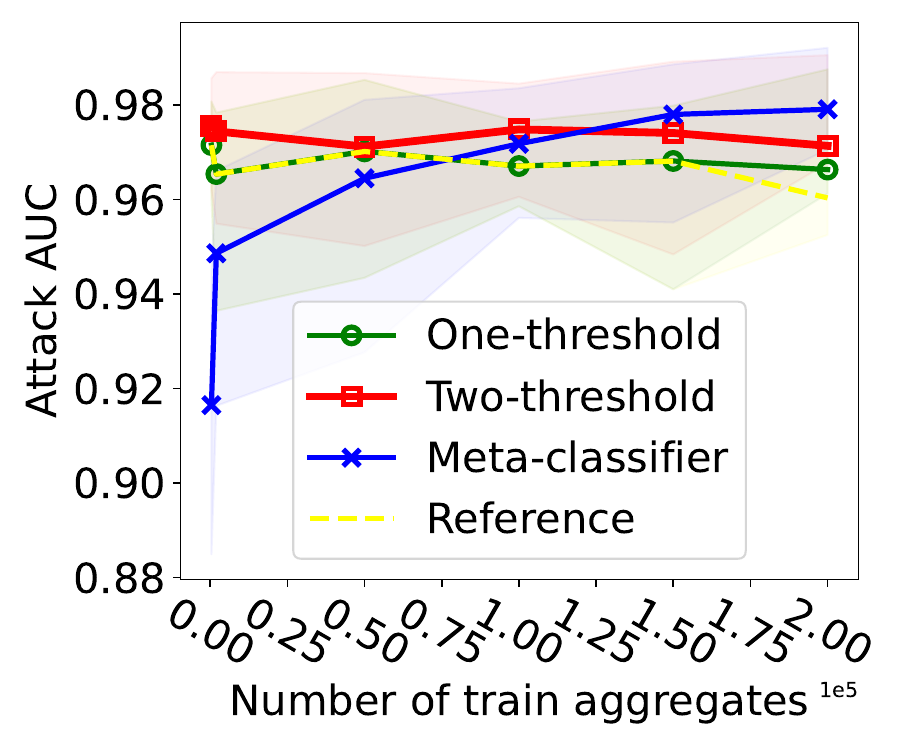}\label{subfig: metric-based-auc-varying-size-pk}
    }
    \caption{Attack performance with varying training shadow aggregates and $60$ positive observations under Laplace mechanism.}\label{fig: metric-based-varying-lap}
\end{figure}

In this section, we provide the attack accuracy and AUC with varying training shadow aggregates in Figure~\ref{fig: metric-based-varying-lap}.

The metric-based attack methods, including the one-threshold attack, two-threshold attack, and the reference attack, are not affected much by the number of shadow aggregates. On the other hand, the MLP-based attack results demonstrate a clear increasing trend with the increase of the number of training shadow aggregates, especially for the Laplace mechanism. At least $200k$ shadow aggregates are required for the MLP-based attack model to learn the two-threshold rule for informed attackers.

\section{Learned Weights for Auxiliary Attackers under Laplace mechanism.}\label{subsec: learned-weights-lap}
\begin{figure}[htbp!]
    \centering
    \includegraphics[width=0.45\linewidth]{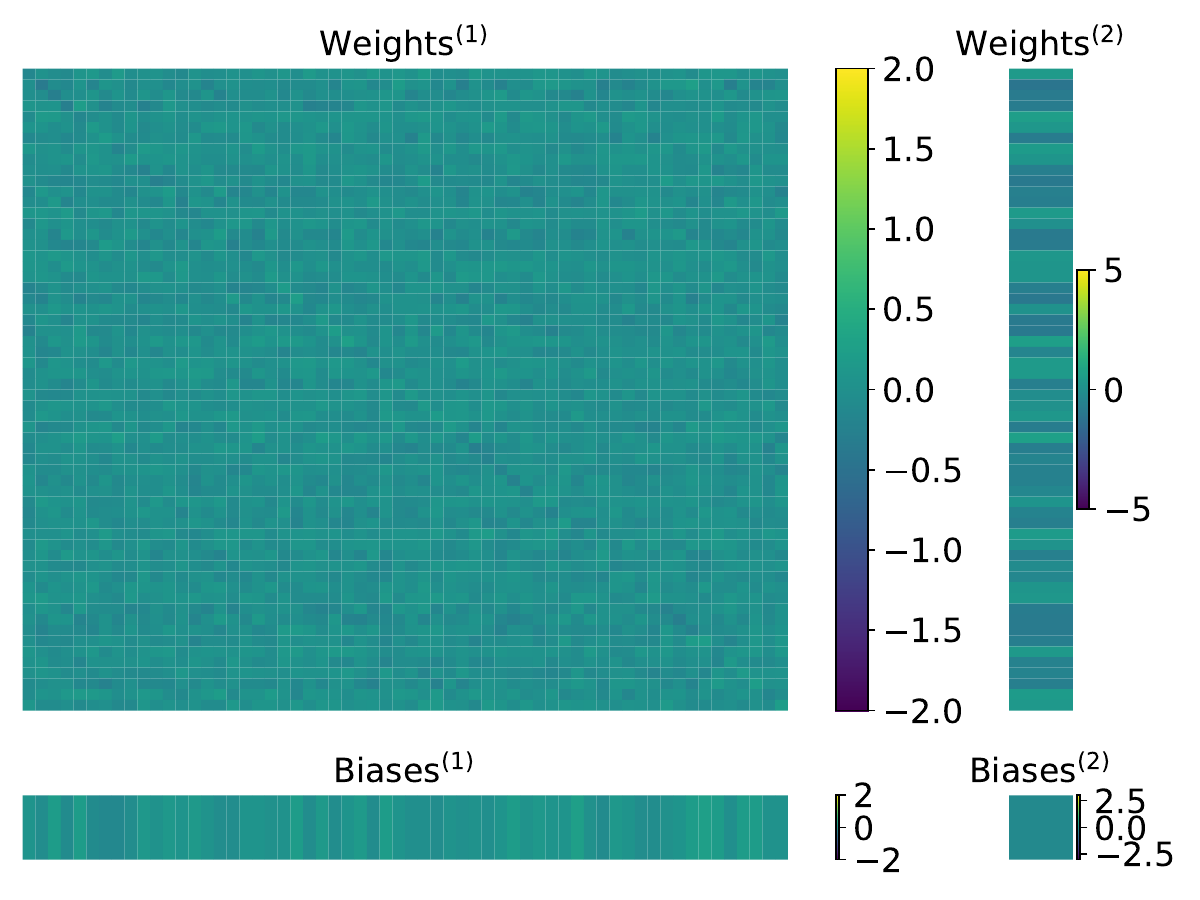}
    \caption{The learned weights of the MLP-based attacker model with an architecture as shown in Figure~\ref{subfig: mlp-multi-node} with $60$ input features. The MLP-based attack model is trained with $2k$ shadow aggregates under the auxiliary attacker.}
    \label{fig: lap_weights_2k_pk}
\end{figure}
\begin{figure}[htbp!]
    \centering
    \includegraphics[width=0.45\linewidth]{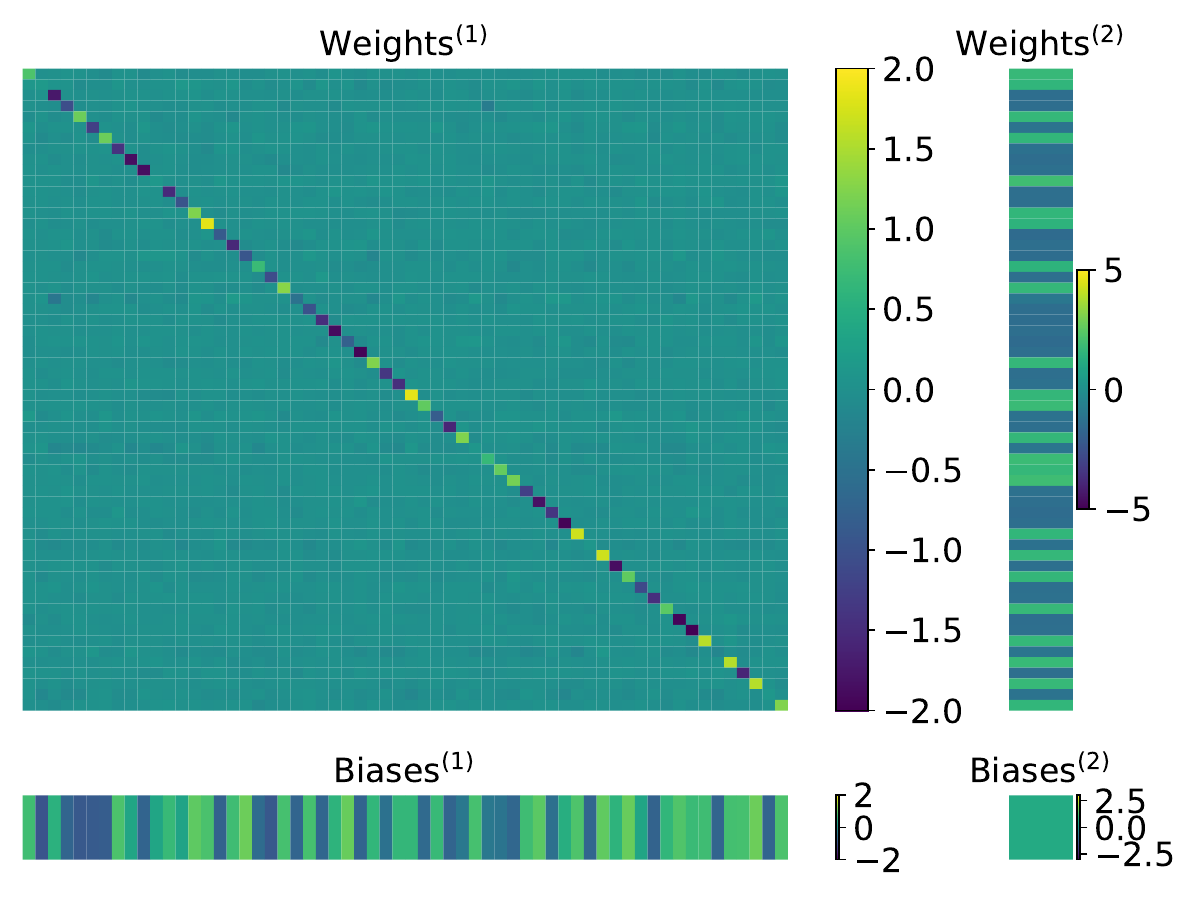}
    \caption{The learned weights of the MLP-based attacker model with an architecture as shown in Figure~\ref{subfig: mlp-multi-node} with $60$ input features. The MLP-based attack model is trained with $200k$ shadow aggregates under the auxiliary attacker.}
    \label{fig: lap_weights_200k_pk}
\end{figure}
In this section, we show the learned parameters of the MLP-based attack model with the architecture shown in Figure~\ref{subfig: mlp-multi-node} under auxiliary attackers. With the training dataset size equivalent to $2k$, the auxiliary attack learns only the one-threshold rule, which is sub-optimal for the Laplace noise perturbed aggregates. When the training shadow aggregate number increases to $200k$, the model finally learns the two-threshold rule, leading to a better attack performance in terms of the adopted metrics in this work. In particular, this finding agrees with the informed attackers as analyzed. 
\section{Attack Performance for MLP with Fixed Nodes Number}\label{subsec: fixed-nodes-num}
\begin{figure}[htbp!]
    \centering
    \includegraphics[width=0.45\linewidth]{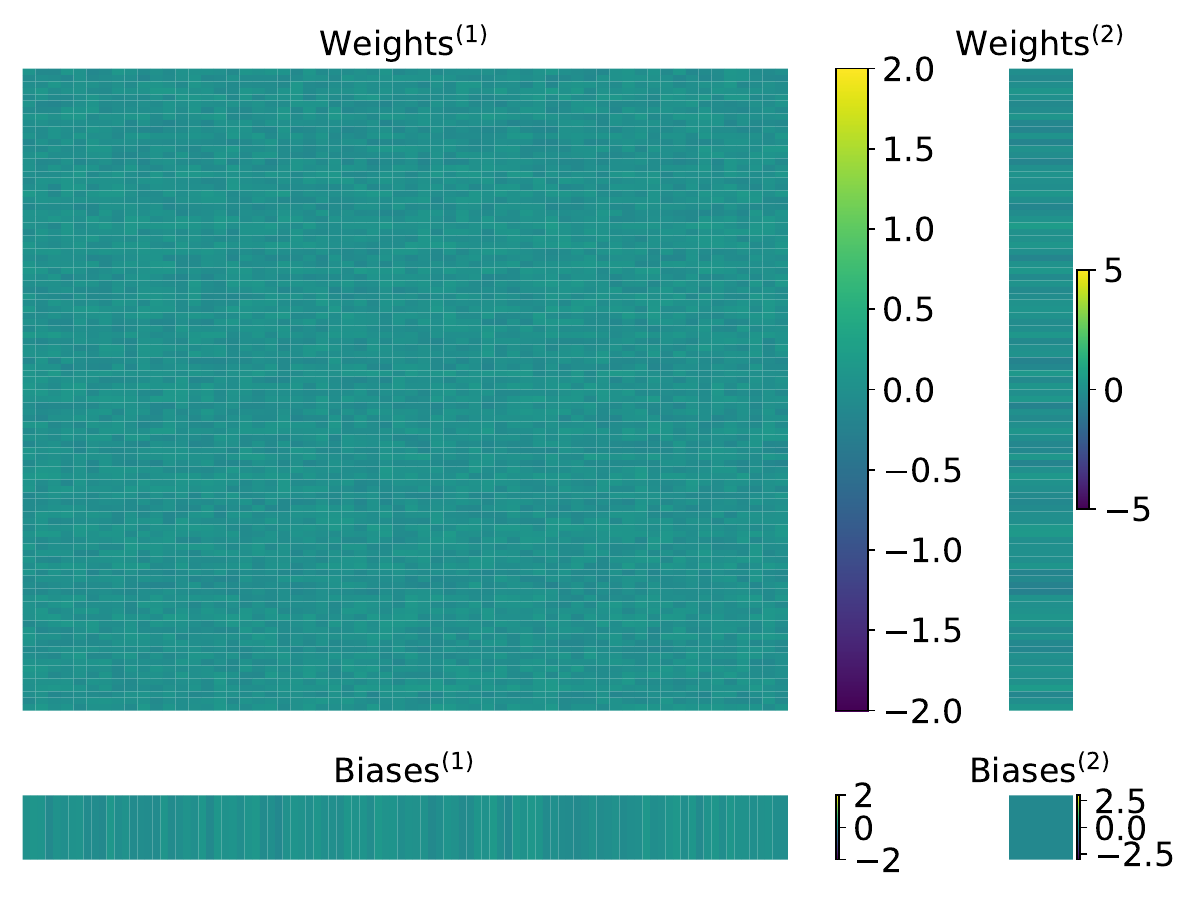}
    \caption{The learned weights of the MLP-based attacker model with $100$ nodes in one hidden layer and $60$ input features. The MLP-based attack model is trained with $2k$ shadow aggregates under the informed attacker.}
    \label{fig: lap_weights_2k_fk_100}
\end{figure}
\begin{figure}[htbp!]
    \centering
    \includegraphics[width=0.45\linewidth]{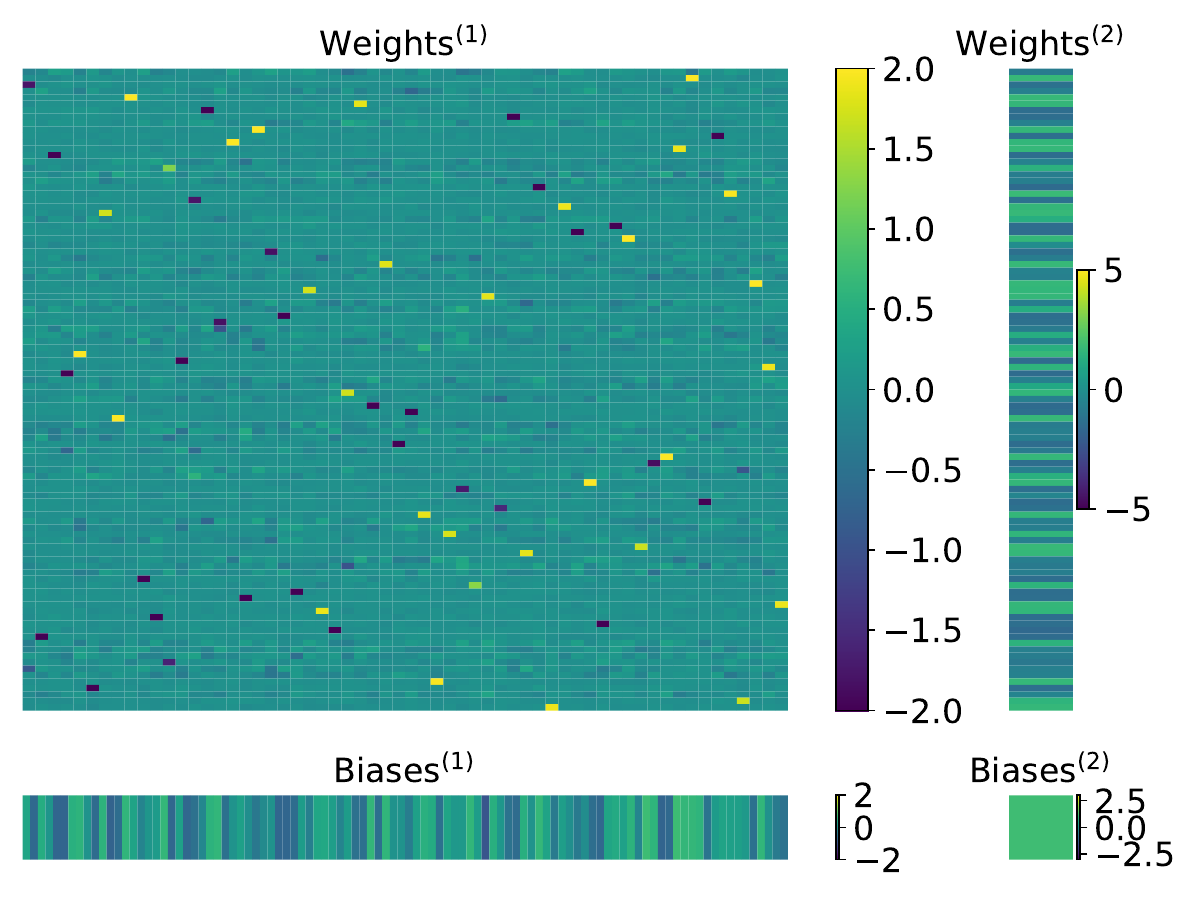}
    \caption{The learned weights of the MLP-based attacker model with $100$ nodes in one hidden layer and $60$ input features. The MLP-based attack model is trained with $200k$ shadow aggregates under the informed attacker.}
    \label{fig: lap_weights_200k_fk_100}
\end{figure}

\begin{figure}[tbp!]
    \centering
    \subfigure[$2k$ shadow aggregates.]{
    \includegraphics[width=0.4\linewidth]{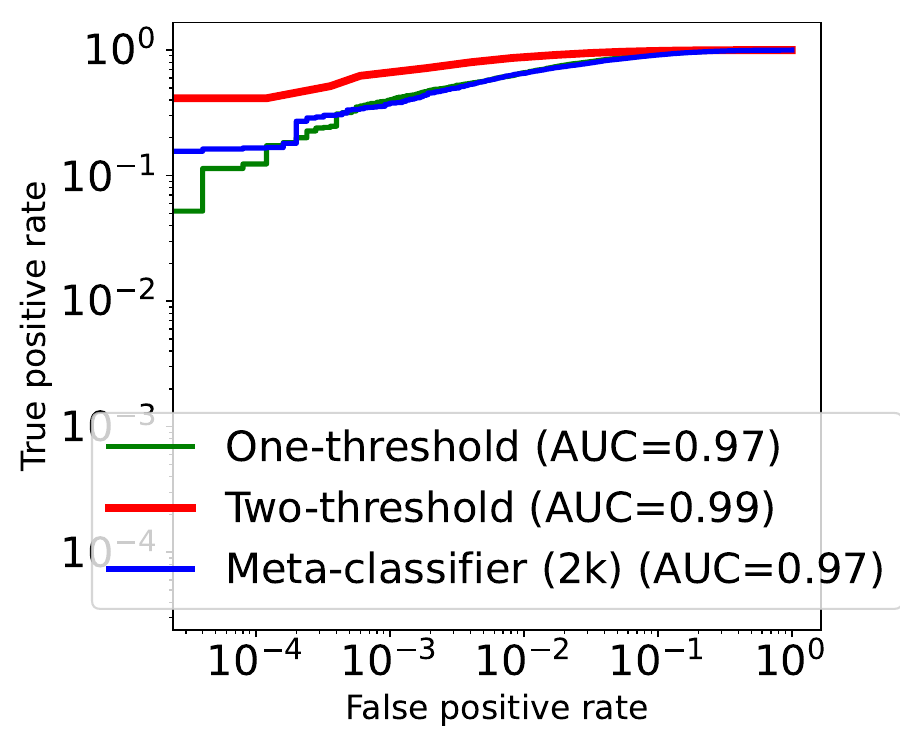}\label{subfig: roc-lap-2k-60-100}
    }
    \subfigure[$200k$ shadow aggregates.]{
    \includegraphics[width=0.4\linewidth]{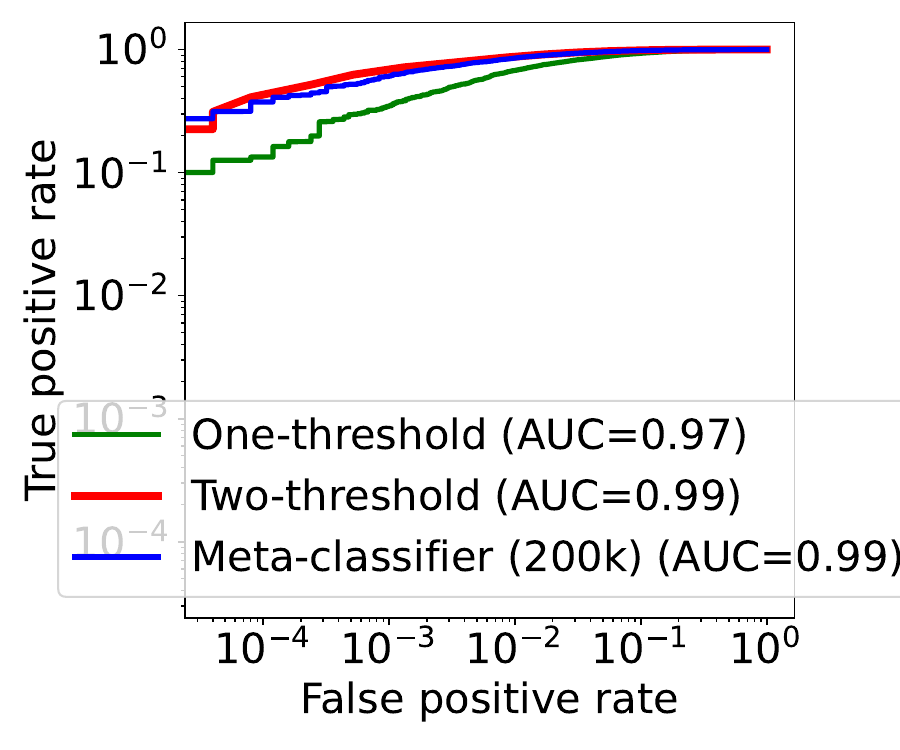}\label{subfig: roc-lap-200k-60-100}
    }
    \caption{Attack ROC $60$ positive observations under the Laplace mechanism. The meta-classifier-based attack model is an MLP model with $100$ nodes in one hidden layer.}\label{fig: roc-informed-100}
\end{figure}

In this section, we present both the attack ROC of the MLP-based attack model with one hidden layer and $100$ nodes for any number of input features in Figure~\ref{fig: roc-informed-100}, and the learned parameters in Figure~\ref{fig: lap_weights_2k_fk_100} and Figure~\ref{fig: lap_weights_200k_fk_100}. As demonstrated, its behavior is similar to the architecture adopted in Section~\ref{sec: evaluation}.



%



\end{document}